\RequirePackage{amsmath}
\documentclass{iopjournal}

\usepackage{cite}
\usepackage{amsfonts}
\usepackage{braket}
\usepackage{graphicx}
\usepackage{lipsum}
\usepackage[utf8]{inputenc}

\usepackage{xcolor}

\DeclareMathOperator{\Tr}{Tr}

\begin{document}

\articletype{paper} 

\title{Dissipation in Periodically Driven Quantum Systems: Partial Secularization and Thermodynamic Consistency}

\author{
Lu\'{\i}sa T. Tude$^{1,*}$\orcid{0000-0002-4058-2274},
Carlos Ortega-Taberner$^{2}$\orcid{0000-0002-3076-8526},
Roberta Zambrini$^{1}$\orcid{0000-0002-9896-3563} and
Gonzalo Manzano$^{1}$\orcid{0000-0003-1359-6344}
}

\affil{$^1$}{Instituto de F\'{\i}sica Interdisciplinar y Sistemas Complejos (IFISC, CSIC-UIB), Campus Universitat de les Illes Balears, E-07122 Palma de Mallorca, Spain}

\affil{$^2$}{School of Physics, Trinity College Dublin, Dublin 2, Ireland}

\email{luisa@ifisc.uib-csic.es}


\begin{abstract}

Periodically driven open quantum systems are central to quantum thermodynamics and quantum control. These systems are typically described using Floquet-Born-Markov master equations, derived with the use of a full secular approximation, and whose thermodynamic implications are often overlooked. In this context, we show that such a strong secular approximation may lead to unphysical predictions for steady state energy currents. We then demonstrate that a coarse-grained formulation of the master equation can regularize these issues while yielding completely positive dynamics and consistent energy currents. 
The coarse-graining time has a clear physical interpretation, as it defines the temporal resolution at which a Markovian master equation can describe the evolution of the periodically driven system.
We show the consistency and validity of our approach by comparing to an exact non-Markovian simulation in paradigmatic examples: a driven two-level system and a three-level maser coupled to hot and cold thermal reservoirs. Our work provides a practical framework for correctly applying the secular approximation in periodically driven-dissipative systems and for assessing the accuracy of master equations of the GKSL form.
\end{abstract}

\section{Introduction}
The development of quantum technologies increasingly relies on the precise control of open quantum systems, often achieved through external periodic driving. Such control can be designed in a variety of driven platforms, including optical systems, microwave and radio-frequency electromagnetic setups, and even mechanically driven architectures. Examples include trapped ions~\cite{martins_rydberg-ion_2023}, quantum dots~\cite{ramsay_review_2010}, or circuit and cavity QED systems~\cite{kavokin_microcavities_2017}, each enabling applications that range from quantum simulation~\cite{weimer_rydberg_2010, berloff_realizing_2017,aedo_analog_2018} to sensing~\cite{cabot_continuous_2024} and metrology~\cite{pavlov_quantum_2023}. To capture the relevant dynamical behavior and to accurately describe the energy exchanges of the driven system with the environment, it is constructive to adopt a microscopic description that allows the derivation of the time evolution using pertinent approximations. Such a formulation must also ensure both accuracy and thermodynamic consistency.

A typical approach for the microscopic derivation of the dynamics of open quantum systems follows from the Born–Markov approximation, which assumes weak system–environment coupling and negligible memory effects. In this context, the so-called secular (rotating wave) approximation leads to the derivation of a Gorini–Kossakowski–Sudarshan–Lindblad (GKSL) master equation (ME), characterized by jump operators that mediate transitions between the energy levels of the open system~\cite{breuer_theory_2002}. When the system possesses internal structure --- comprising interacting subsystems --- the interactions modify the energy-level structure and may induce degeneracies that are absent in the uncoupled (bare) system.
Depending on the approximations performed in the derivation, one can then obtain jump operators that act either in the basis of the bare subsystem states or in that of the dressed eigenstates of the interacting Hamiltonian. These two constructions are often known as the local and global approaches respectively. The local approach is valid when the internal coupling is weak enough that the energy shifts are negligible for the environmental action~\cite{trushechkin_perturbative_2016}, while for stronger couplings the global approach is required~\cite{gonzalez_testing_2017}, and may otherwise lead to thermodynamic inconsistencies~\cite{Levy_2014} if the extra costs for the coupling are not explicitly included~\cite{barra2015thermodynamic,de_chiara_reconciliation_2018}.
A critical aspect in the derivation of the GKSL equation is the application of the  secular approximation.
A solution guaranteeing consistency lies in performing only a partial secular approximation~\cite{farina_open-quantum-system_2019,cattaneo_local_2019}. Other approaches proposed phenomenological master equations~\cite{kirsanskas_phenomenological_2018}, or alternative regularizations of the Redfield equation to obtain a GKLS form~\cite{mccauley_accurate_2020,nathan_universal_2020,davidovic_completely_2020,dabbruzzo_time-dependent_2023,fernandez_de_la_pradilla_recovering_2024,potts_thermodynamically_2021,boubakour2026refined}. Moreover, it has been argued that the apparent loss of positivity in the Redfield equation is closely tied to the time dependence of the microscopic coefficients, which is commonly neglected when deriving the time-independent Redfield equation~\cite{whitney2008staying}.

In the presence of a periodic external driving, the energy levels of the system change cyclically, requiring the use of Floquet theory~\cite{Shirley65} alongside the Born-Markov approximation. In this case, the Floquet-interaction Hamiltonian introduces replicas of states (Floquet quasienergies) and induce energy shifts, which must be accounted for when defining the jump operators. 
When the quasienergies are sufficiently well separated with respect to the system-bath coupling, a full secular approximation can be performed to obtain a master equation in GKLS form preserving the periodicity of the drive~\cite{grifoni_driven_1998,kohler1998floquet,Breuer00,alicki_internal_2006,Hone09,szczygielski2013markovian,szczygielski_application_2014}. However, near-degeneracies and crossings in the quasienergy spectrum may often appear~\cite{kohler1998floquet}, specially when the dimension of the system increases~\cite{Hone09} or if the driving is not sufficiently fast compared with the system natural frequency~\cite{schnell_is_2020}. Moreover, the coupling to the reservoir might not be sufficiently weak, and higher orders in the system-bath coupling expansion are then required~\cite{Haddadfarshi15,Restrepo2016}. In this way, open periodically driven systems require similar attention to that given to open composite systems. 

In this context, Floquet-Redfield equations have been employed in a number of examples (see e.g. Refs.~\cite{nafari_qaleh_enhancing_2022,tude2024overcoming,kolisnyk_floquet_2024}). Although Redfield equations often show good agreement with the exact dynamics~\cite{jeske_bloch-redfield_2015,purkayastha_out_2016,murphy_laser_2022,Tello2024}, complete positivity of the evolution is not guaranteed anymore and hence a physical behavior cannot be ensured for every initial state and correlations with auxiliary systems~\cite{rivas2012open}.
The lack of complete positivity in the Redfield equation originates from fast oscillating terms that are often incompatible with the Markovian assumption~\cite{eastham_bath_2016,hartmann_accuracy_2020}. Notably, as we will shortly demonstrate, Floquet-Redfield equations may also be regularized through a partial secular approximation. 

Periodically driven systems that are weakly coupled to thermal reservoirs are ubiquitous in both stochastic and quantum thermodynamics, with important applications such as quantum heat engines and refrigerators~\cite{kosloff2013quantum,gelbwaser2013minimal,restrepo_quantum_2018,klatzow_experimental_2019}. While their energy currents have been examined in connection to the full secular approximation~\cite{szczygielski_markovian_2013,langemeyer2014energy,gasparinetti_heat-exchange_2014,cuetara_stochastic_2015} a consistent treatment of work and heat contributions in different regimes is not available yet, but highly desirable. In this context, we show from a first-principles approach that eliminating relevant oscillatory contributions via the full secular approximation leads to zero net power over a period from the drive in stationary conditions. This result cannot hold in general, indicating that the full secular approximation fails to accurately capture the system’s thermodynamic behavior.

To address together complete positivity and a consistent thermodynamic description, we employ a coarse-grained formulation of the Floquet–Redfield master equation, which enables partial secularization while ensuring complete positive dynamics, and hence GKLS form. Within this framework, energy currents can be defined consistently with the approximations underlying the derivation of the master equation. The relevant coarse-graining time interval has a clear physical interpretation by setting the temporal resolution at which a (time-divisible) Markovian description remains valid. Consequently, the resulting master equation provides a reliable description of the system’s evolution on time scales longer than the coarse-graining interval.

Building on this framework, we use a systematic procedure to compare the intrinsic dynamical timescales of the system with the minimal coarse-graining time required for complete positivity. This comparison yields a practical criterion to assess the validity of the Born-Markovian approximation and serves as a guideline for identifying regimes in which non-Markovian treatments become necessary. To support these considerations, we perform numerically exact non-Markovian simulations in both examples considered in the manuscript, which we use to benchmark the accuracy and limitations of the coarse-grain approach.

We stress that the resulting master equation and energy currents derived here have closed analytical forms that allow for a direct physical interpretation, and in some cases, as in the examples presented in this manuscript, they can be simplified further. The numerical implementation is simple and can be applied to more complex systems for which an exact simulation is impracticable.

This manuscript is organized as follows: In section~\ref{sec:ME}, we present a detailed analysis of the relevant time scales and approximations underlying the derivation of the Floquet–Redfield master equation and its secularization. We show that the full secular approximation leads to problematic definitions of energy currents (section~\ref{sec:thermodconsistency}) and, in section~\ref{sec:coarse-graining}, introduce an alternative approach based on a coarse-grained ME that allows these currents to be consistently evaluated (section~\ref{sec:thermoCG}). Our focus is on steady state thermodynamics, and we examine two fundamental examples which, when combined, serve as building blocks for more complex systems. In section \ref{sec:2ls} we provide a comprehensive analysis of a driven two-level system coupled with a thermal reservoir, we investigate the role of each oscillating term in the Redfield equation and its importance for different parameter regimes. We compare the steady state density matrix and energy currents obtained using the full secular approximation, the coarse-grain approach, the Redfield equation and an exact (non-Markovian) simulation. Next, in Section \ref{sec:3ls}, we investigate the role of the coherent terms of the steady state density matrix in ensuring the thermodynamic consistency of the three-level maser model, first proposed by Scovil~\cite{scovil_three-level_1959}. 

\section{Master equation of a periodically driven system}\label{sec:ME}

We consider a quantum system subjected to external periodic driving while coupled to a thermal bath. The total Hamiltonian of the system and bath reads
\begin{equation}
    H(t) = H_S(t) + H_B + H_{SB},\label{eq:eq1}
\end{equation}
where the time-dependent system Hamiltonian is periodic with period $T$, $H_S(t) = H_S(t + n T)$ with $n \in \mathbb{Z}$, $H_B$ is the Hamiltonian of the bath, and the system-bath interaction is generically described by $H_{SB} =  \lambda~ (A \otimes B)$, where $A$ and $B$ are Hermitian operators acting respectively on the system and the bath, and $\lambda$ explicitly quantifies the (small) coupling strength between them. 

The microscopic derivation of a Born-Markov master equation for a periodically driven open quantum system proceeds analogously to the case of a time-independent system Hamiltonian~\cite{breuer_theory_2002, farina_open-quantum-system_2019, cattaneo_local_2019}. The key difference lies in the use of a Floquet-interaction picture~\cite{szczygielski_application_2014}, where the reference frame rotates with the driving field. This is formally accomplished with a unitary operator of the form $U_{\rm int}(t) = U_S(t) \otimes e^{-i H_B t}$ where $U_S(t) \equiv \mathcal{T}e^{- i \int_0^t H_S(s) ds}$ acts only on the system and $e^{-i H_B t}$ on the bath. A plausible representation of $U_S(t)$ can be obtained from Floquet theory, which ensures its unique decomposition into two unitary operators: 
\begin{equation} \label{eq:U_S}
U_S(t) = P(t)~ e^{-i {H}_F t},    
\end{equation}
with a first periodic unitary operator $P(t + nT) = P(t)$ governing the fast dynamics of the system, and a second unitary operator involving the Floquet dressed Hamiltonian of the system ${H}_F \equiv \sum_k \epsilon_k \ket{\phi_k}\bra{\phi_k}$ (sometimes called the \emph{averaged Hamiltonian}~\cite{szczygielski_markovian_2013}) that is time independent and determines the slow system evolution over many periods. 

The Floquet Hamiltonian $H_F$ defines a set of quasienergies $\{\epsilon_k\}$ and the corresponding (Floquet) eigenvectors~$\{ \ket{\phi_k}\}$, which verify:
\begin{equation}
    U_S(t) \ket{\phi_k} = e^{-i \epsilon_k t} P(t) \ket{\phi_k} = e^{-i \epsilon_k t} \sum_{q \in \mathbb{Z}} e^{-i q \Omega t} \ket{\phi_k^{(q)}},
    \label{eq:floquetprop}
\end{equation}
where $\Omega = 2\pi/T$ is the driving frequency, and the set of states $\{\ket{\phi_k^{(q)}} = P(t) \ket{\phi_k}\}$, correspond to Fourier components (Floquet harmonics or Floquet modes), which form a complete basis of the system Hilbert space. As noticed in Ref.~\cite{alicki_internal_2006}, the number of Floquet modes (and hence the values of $q$ labeling the harmonics) is then limited by the dimension of the system Hilbert space. 

Using the properties of $U_S(t)$ and the Floquet-interaction picture, it is possible to decompose the system operators appearing in the interaction Hamiltonian $H_{SB} = \lambda (A \otimes B)$ into jumps over the states of the Floquet Hamiltonian $H_F$. In the following, we denote with a tilde operators in the Floquet-interaction picture, e.g. $\Tilde{O}$. We first perform a Fourier decomposition of $\Tilde{A}(t) = U^\dagger_S(t)~ A ~U_S(t) = \sum_q e^{i q \Omega t}  e^{i H_F t} ~ A_q ~ e^{-i H_F t}$, with $A_q$ transition operators associated to each Floquet mode $q$. Then we further decompose them into eigenoperators of $H_F$ to obtain:
\begin{equation}
   \Tilde{A}(t)= \sum_{\omega \in \mathbb{B}} \sum_{q \in \mathbb{Z}}~ e^{-i (\omega + q \Omega)t} ~{A}_{\omega,q} ,\label{eq:jumpops}
\end{equation}
where the operators $A_{\omega, q} \equiv \sum_{\epsilon_j, \epsilon_i} \delta(\epsilon_i - \epsilon_j - \omega)~ \ket{\phi_i}\bra{\phi_i} ~A_q~ \ket{\phi_j}\bra{\phi_j}$ describe jumps between the levels of the dressed system associated to the Floquet Hamiltonian $H_F$. As such, they verify $[H_F, A_{\omega, q}] = \omega~ A_{\omega, q}$ for all harmonics $q$, and $\omega$ in the set of Bohr quasifrequencies $\mathbb{B} \equiv \{\epsilon_j - \epsilon_k, \forall ~k \neq j\}$ comprising all possible differences of quasienergies in the spectrum of $H_F$. Moreover, since the system operators $\Tilde{A}(t)$ are Hermitian, they obey the extra symmetry property $A^\dagger_{-\omega, -q} =  {A}_{\omega, q}$. The above decomposition will be key to discuss the secular approximation below.

Many times it is convenient to split the system Hamiltonian into two parts, $H_S(t) = H_0 + V(t)$, with a static term $H_0 = \sum_k E_k \ket{k}\bra{k}$ and the time-periodic drive contribution $V(t) = V(t+ nT)$ with period $T$. A prototypical example of driving in this situation couples two states of the bare system Hamiltonian $H_0$, $\ket{i}$ and $\ket{j}$, with energies $E_j > E_i$ as 
\begin{equation}
    V(t) = g (e^{-i \Omega t} \ket{j}\bra{i} + e^{i \Omega t} \ket{i}\bra{j}),   \label{eq:V} 
\end{equation}
which represents e.g. a classical driving field acting on the two levels. 
In this case, the unitary transformation $U_S(t)$ in Eq.~\eqref{eq:U_S} can be explicitly calculated, giving a periodic unitary operator $P(t) = \exp{\left[- i t \Omega \big(\ket{j}\bra{j} - \ket{i}\bra{i}\big)/2\right]}$, 
and Floquet Hamiltonian:
\begin{equation} \label{eq:HFsplit}
    H_F = \sum_{k \neq i,j} E_k \ket{k}\bra{k} + (E_j - \Omega/2)\ket{j}\bra{j}  +(E_i + \Omega/2) \ket{i}\bra{i} + g( \ket{j}\bra{i} + \ket{i}\bra{j}).
\end{equation}
As can be explicitly appreciated, the effect of the driving is to both shift the energies of states $\ket{i}$ and $\ket{j}$ of the static term by an amount $\pm\Omega/2$, and induce a coherent exchange $g(\ket{j}\bra{i} + \ket{i}\bra{j})$ between them. In this case the quasienergy spectrum $\{ \epsilon_k \}$ matches the static energies $\{E_k\}$ except for states $\ket{i}$ and $\ket{j}$, for which $\epsilon_{i,j} = \frac{1}{2}(E_i + E_j \pm \Omega_R)$, where $\Omega_R = \sqrt{\Delta^2 + 4 g^2}$ is the Rabi splitting, and $\Delta = E_j - E_i - \Omega$ the detuning. Similarly, the Floquet eigenvectors are $\ket{\phi_k} = \ket{k}$ for $k \neq i,j$ while $\ket{\phi_i} \propto (\Delta + \Omega_R)\ket{i} + 2g \ket{j}$ and $\ket{\phi_j} \propto (\Delta - \Omega_R)\ket{j}  + 2g\ket{i}$. This result will be used in the examples of the next sections. However, we do not assume any specific form of $V(t)$ or $H_S(t)$ in the derivation below.

\subsection{Floquet-Redfield master equation}

In order to derive a master equation in the Floquet-interaction picture, we proceed as usual by integrating the Liouville-von Neumann equation for the compound evolution of system and bath up to second order in the system-bath coupling $\lambda$. Taking the trace over the bath degrees of freedom we obtain an integro-differential equation for the system density operator $\Tilde{\rho}$:
\begin{equation}
    \dot{\Tilde{\rho}} (t) = - \int_0^t ds \Tr_B\big[\Tilde{H}_{SB}(t), \big[\Tilde{H}_{SB}(s), \Tilde{\rho}_{\rm tot}(s)\big]\big] + \mathcal{O}(\lambda^3),\label{eq:voneummann}
\end{equation}
where $\Tilde{\rho}_{\rm tot}(s)$ denotes the global density operator of system and bath at time $s$ in the Floquet-interaction picture.  
Under the Born approximation, the system-bath interaction is considered to be weak, $\lambda \ll \omega$ so that terms $\mathcal{O}(\lambda^3)$ are neglected and $\Tilde{\rho}_{\rm tot}(s) \approx \Tilde{\rho}(s) \otimes \Tilde{\rho}_B$ on the right hand side of the equation. Here $\Tilde{\rho}_B = \rho_B = e^{- \beta H_B}/Z_B$ is the density operator of the bath, which is assumed to be in a thermal state at inverse temperature $\beta$.

The Markov approximation can be taken when the time scale at which correlations decay in the bath is much smaller than the time scale at which the system evolves $\tau_B \ll \tau_S$. This effectively means that the bath does not retain information on the system dynamics and allows us to replace $\Tilde{\rho}(s)$ by $\Tilde{\rho}(t)$ inside the integral, and further $H_{SB}(s)$ by $H_{SB}(t - s)$ while taking the limit $t\rightarrow \infty$.  
Making these approximations and introducing the explicit form of $\Tilde{H}_{SB}(t)$, one finds after some algebra:
\begin{equation}
    \dot{\Tilde{\rho}} (t) = \sum_{\omega, \omega^\prime \in \mathbb{B}} \sum_{q , q^\prime \in \mathbb{Z}} \Gamma(\omega + q \Omega) e^{i(\omega^\prime - \omega)t + i(q^\prime-q)\Omega t } \Big( {A}_{\omega,q} \Tilde{\rho}(t) {A}^\dagger_{\omega^\prime,q^\prime} - A_{\omega^\prime,q^\prime}^\dagger A_{\omega, q} \Tilde{\rho}(t) \Big) + \text{ h.c.},\label{eq:bm1}
\end{equation}
where we have used the decomposition in Eq.~\eqref{eq:jumpops} together with $A_{-\omega^\prime,-q^\prime} = A^\dagger_{\omega^\prime,q^\prime}$, and introduced the one-side Fourier transform of the bath correlation function $\Gamma(\omega + q \Omega) \equiv \lambda^2 \int_0^\infty d s e^{i (\omega + q \Omega)s} \Tr_B\big[\Tilde{B}(s) \Tilde{B}(0) \rho_B\big]$~\footnote{Here we have also used that for a thermal state the bath correlation function is  homogeneous in time, that is, $\Tr_B\big[\Tilde{B}(t) \Tilde{B}(t-s) \rho_B\big] = \Tr_B\big[\Tilde{B}(s) \Tilde{B}(0) \rho_B\big]$ for all $t$.}.

We refer to the above equation as the Floquet-Redfield master equation. A more compact form of the master equation can be obtained by using combined indices $\alpha = \omega + q \Omega$ and $\alpha^\prime = \omega^\prime + q^\prime \Omega$ and rearranging the Fourier transforms above 
so that Eq.~\eqref{eq:bm1} becomes:
\begin{align}
\dot{\Tilde{\rho}}(t) =& -i \sum_{\alpha, \alpha^\prime \in \mathbb{F}} e^{i(\alpha^\prime - \alpha)t} \chi(\alpha, \alpha^\prime) [A^\dagger_{\alpha^\prime} A_\alpha, \Tilde{\rho}(t)] \nonumber \\
& + \sum_{\alpha, \alpha^\prime \in \mathbb{F}} e^{i(\alpha^\prime - \alpha)t}~\gamma(\alpha,\alpha^\prime) \Big(A_{\alpha} ~\Tilde{\rho}(t) A^\dagger_{\alpha^\prime} - \frac{1}{2} \{ A^\dagger_{\alpha^\prime} A_{\alpha} , \Tilde{\rho}(t)\}\Big),
\label{eq:ME}
\end{align}
where we used the short-hand notation $A_{\alpha} \equiv A_{\omega, q}$ and introduced the combinations of Fourier transforms $\chi(\alpha, \alpha^\prime) = [\Gamma(\alpha) - \Gamma^\ast(\alpha^\prime)]/(2i)$, and $\gamma(\alpha, \alpha^\prime) = \Gamma(\alpha) + \Gamma^\ast(\alpha^\prime)$. Note also that the sums in $\alpha$ and $\alpha^\prime$ run along all possible combinations of Bohr quasifrequencies and harmonics (denoted compactly as the set $\mathbb{F}$).  

The form of the master equation, in Eq.~(\ref{eq:ME}), resembles the GKLS form, but with extra terms that couple jump operators with different Floquet-Bohr frequencies ($\alpha \neq \alpha^\prime$). The term in the first line is reminiscent of a Lamb shift, inducing a small perturbation to the system Hamiltonian (of order $\lambda^2$). Indeed, it can be written in the form $-i[H_{\mathrm{LS}}(t),\Tilde{\rho}(t)]$, with
\begin{equation}
    H_{LS} = \sum_{\alpha, \alpha^\prime \in \mathbb{F}} e^{i(\alpha^\prime - \alpha)t} \chi(\alpha, \alpha^\prime) A^\dagger_{\alpha^\prime} A_\alpha .\label{eq:Hls}
\end{equation}
Using the symmetry relation $\chi(\alpha,\alpha')=\chi^*(\alpha',\alpha)$, it follows that $H_{\rm LS}(t)$ is Hermitian and, consequently, does not require any further regularization to preserve complete positivity~\cite{trushechkin_unified_2021}.
On the other hand, the second line of Eq.~(\ref{eq:ME}) is reminiscent of a GKLS dissipator can be expressed as elements of the so-called Kossakowski matrix, with elements $K_{\alpha \alpha^\prime}(t) = \gamma(\alpha, \alpha^\prime) e^{i (\alpha^\prime - \alpha)t}$. The dissipator can be then written as 
\begin{equation}
        \mathcal{D}[\Tilde{\rho}] \equiv  \sum_{ \alpha, \alpha^\prime \in \mathbb{F}} K_{\alpha \alpha^\prime}(t) \Big({A}_{\alpha} \Tilde{\rho}(t) {A}^\dagger_{\alpha^\prime} - \frac{1}{2} \Big\{{A}^\dagger_{\alpha^\prime} {A}_{\alpha} , \Tilde{\rho}(t)\Big\}\Big),\label{eq:BRdissipatorK}
\end{equation}
which is guaranteed to generate a completely positive and trace-preserving (CPTP) dynamics provided that the Kossakowski matrix is positive semi-definite~\footnote{Note that the functions $\gamma(\alpha, \alpha^\prime)$ can be incorporated into either the Kossakowski matrix or inside the jump operators, keeping the form of Eq.(\ref{eq:BRdissipatorK}). Here, we adopted the convention used in~\cite{gorini_completely_1976,agredo_kossakowski_2022}, where the rates are absorbed into the Kossakowski matrix, so that  $\Tr[A_\alpha A^\dagger_{\alpha^\prime}] = \delta_{\alpha,\alpha^\prime}$.}. However, in order to guarantee that the Kossakowski matrix $K(t)$ is positive semi-definite at all times, further approximations are required.

\subsection{Weak driving limit}\label{sec:weakdriving}
 
An important case where the Floquet-Redfield master Eq.~(\ref{eq:ME}) can be transformed into a simple GKLS form is the limit of weak driving. In this regime, the effect of the drive is perturbative with respect to the static Hamiltonian, and in analogy with the local approach for multipartite systems, the dissipative terms in the master equation reduce to those of the bare system~\cite{trushechkin_perturbative_2016,manzano_quantum-enhanced_2023}. 

Consider a Hamiltonian of the system of the form
\begin{equation}
    H_S = H_0 + g V(t),
\end{equation}with $V(t + T) = V(t)$ a small periodic perturbation to the static Hamiltonian $H_0 = \sum_k E_k \ket{k}\bra{k}$, and $g \ll \omega_{jk}$ for all energy gaps $\omega_{jk} = E_k - E_j$. The Floquet Hamiltonian hence reads $H_F = H_0 + g V$, which is time-independent as mentioned before. 
In the weak-driving limit, the periodic operator $P(t)$ can be expanded perturbatively in $g$. Substituting the decomposition $U_S(t) = P(t)e^{-iH_F t}$ 
into the equation $i\partial_t U_S = H_S(t) U_S$ and expanding
\begin{equation}
P(t) = P^{(0)}(t) + g P^{(1)}(t) + \mathcal{O}(g^2),
\end{equation} 
we find that the zeroth order term $P^{(0)}(t) = 1 $, reflecting the absence of Floquet replicas when $g=0$. 
Using perturbation theory up to second order in $g$, the quasienergies take the form $\epsilon_n = \epsilon_n^{(0)} + g \epsilon_n^{(1)} + g^2 \epsilon_n^{(2)}$, where $\epsilon_n^{(0)}= E_n$ are the energies of $H_0$. Analogously, the Floquet eigenstates read $\ket{\phi_n}= \ket{\phi_n^{(0)}} + g \ket{\phi_n^{(1)}} + g^2 \ket{\phi_n^{(2)}}$,
with $\ket{\phi_n^{(0)}} = \ket{n}$ the eigenvectors of $H_0$. Using this expansion, the jump operators in Eq.~\eqref{eq:jumpops} can also be expanded perturbatively to obtain
\begin{equation}
A_{\omega} = A_{\omega}^{(0)} + g A_{\omega}^{(1)} + g^2 A_{\omega}^{(2)},
\end{equation}
where $A_{\omega,q}^{(0)} = \sum_{E_i, E_j} \delta(E_i - E_j - \omega) \ket{i}\bra{i} A \ket{j}\bra{j}$ represent jumps in the $H_0$ ladder associated to the Bohr frequencies $\omega$ in the set $\mathbb{B}= \{E_j - E_i, \forall i \neq j \}$, that is, calculated from the differences in the energies of the static Hamiltonian. As a consequence, the dissipator of the Floquet-Redfield equation~\eqref{eq:BRdissipatorK} can be expanded as
    $\mathcal{D}[\rho] = \mathcal{D}^{(0)}[\rho] + g \mathcal{D}^{(1)}[\rho] + g^2\mathcal{D}^{(2)}[\rho]$.

Turning back to the Schr\"odinger picture, the master Eq.~(\ref{eq:ME}) becomes:
\begin{equation} \label{eq:MEweak}
    \dot{{\rho}}(t) =-i [H_0 + g V(t) + H_{LS}, {\rho}(t)] + \mathcal{D}^{(0)}[\rho(t)] + \mathcal{O}(\lambda^2 g)
\end{equation}
where we neglected terms $\mathcal{O}(\lambda^2 g) \sim \mathcal{O}(\lambda^3)$, which amounts to keep only the zeroth order dissipator, reading: 
\begin{equation}
    \mathcal{D}^{(0)}[\rho] = \sum_{\omega} \gamma_{\omega} \left( L_\omega \rho L_{\omega}^\dagger - 1/2 \big\{L_{\omega}^\dagger L_{\omega}, \rho \big\} \right).\label{eq:localME}
\end{equation}
Therefore in the weak-driving limit a GKSL master equation can be consistently derived.
Here the Lindblad operators $L_{\omega}= A_\omega^{(0)}$ correspond to jumps among the states of the undriven Hamiltonian (bare basis) and verify $[H_0, L_\omega] = - \omega L_{\omega}$. We obtained Eq.~(\ref{eq:localME}) by applying the 
standard secular approximation~\cite{breuer_theory_2002} for time-independent systems, which is controlled entirely by the static spectrum of $H_0$. That is, it is well justified whenever the energies of the static Hamiltonian are well separated, $|E_k - E_l| \gg \lambda^2$ for all $k,l$. The corresponding rates $\gamma_\omega$, verify the local detailed balance relation $\gamma_{-\omega} = e^{-\beta \omega} \gamma_{\omega}$ with $\omega \in \mathbb{B}$ computed as the differences in energies of the static energy levels. 

The weak-driving master equation~\eqref{eq:MEweak} is valid when the driving strength $g$ is comparable to the system--bath coupling $\lambda$, and terms of order $\mathcal{O}(\lambda^2 g)$ can be treated as terms $\mathcal{O}(\lambda^3)$ and be neglected. As a consequence the dissipator verifies a local detailed balance relation w.r.t. the bare Hamiltonian which implies a thermal fixed point $\mathcal{D}^{(0)}[e^{- \beta H_0}/Z] = 0$. However, the presence of the perturbation $V$ in the Hamiltonian part of Eq.~\eqref{eq:MEweak}, where generically $[H_0, V] \neq 0$, prevents the system to reach a thermal equilibrium state in the $H_0$ basis, and leads instead to a non-equilibrium steady state in the interaction picture $\Tilde{\pi}$ with coherences in the $H_0$ basis, verifying $-i[V,\Tilde{\pi}] + \mathcal{D}^{(0)}[\Tilde{\pi}] = 0$, but $-i[V,\Tilde{\pi}] \neq 0$ and $\mathcal{D}^{(0)}[\Tilde{\pi}] \neq 0$.

\subsection{Full secular approximation} \label{sec:FSA}
 
A commonly employed way of ensuring the positivity of the Kossakovski matrix $K(t)$ in Eq.~\eqref{eq:BRdissipatorK} is to perform the full secular approximation~\cite{alicki_internal_2006,Hone09,szczygielski2013markovian,szczygielski_application_2014}, in which all terms where $\alpha \neq \alpha^\prime$ in the  Floquet-Redfield master Eq.~\eqref{eq:ME} are neglected, even when going beyond the weak-driving limit discussed above. In that case, the master equation in the Floquet-interaction picture reads:
\begin{equation}
    \dot{\Tilde{\rho}}(t) = - i [\Tilde{H}_{LS}, \Tilde{\rho}(t)] + \mathcal{D}[\Tilde{\rho}(t)]
    \label{eq:FullsecularME}
\end{equation}
with $\Tilde{H}_{LS} = \sum_\alpha \chi(\alpha,\alpha) A_\alpha^\dagger A_\alpha$ the Lamb-shift Hamiltonian (which becomes diagonal in the $H_F$ basis), and the dissipator in Eq.~\eqref{eq:BRdissipatorK} yielding
\begin{equation}
        \mathcal{D}[\Tilde{\rho}(t)] =  \sum_{ \alpha} K_{\alpha \alpha} \Big({A}_{\alpha} \Tilde{\rho}(t) {A}^\dagger_\alpha - \frac{1}{2} \Big\{{A}^\dagger_\alpha {A}_{\alpha}. \Tilde{\rho}(t)\Big\}\Big),\label{eq:FullseculardissipatorK}
\end{equation}
where only the (time-independent) diagonal elements of the Kossakowski matrix survive, $K_{\alpha, \alpha} = \gamma(\alpha,\alpha) = \lambda^2 \int_{-\infty}^\infty ds e^{i \alpha s} \Tr[B(s) B(0) \rho_B] \geq 0 $. These elements can then be interpreted as proper rates associated to jumps $A_\alpha$ among quasienergy levels (which become the Lindblad jump operators), and Eq.~\eqref{eq:FullsecularME} can be written in the form of differential (Pauli-type) equations for the system populations~\cite{Breuer00,Hone09,Voberg13}.

For $\rho_B$ a thermal state, the rates appearing in the dissipator  Eq.~\eqref{eq:FullseculardissipatorK} satisfy a generalized version of local detailed balance~\cite{Breuer00,kohn2001periodic}, coming from the Kubo-Martin-Schwinger condition (valid for arbitrary bath operators $\Tilde{B}$ in the interaction Hamiltonian)~\cite{alicki_internal_2006,szczygielski2013markovian}:  
\begin{equation} \label{eq:LDB}
K_{-\alpha, -\alpha} = e^{-\beta \alpha}  ~ K_{\alpha,\alpha},    
\end{equation}
where we recall that $\alpha = \omega + q \Omega$ contains both the differences in quasienergies of the Floquet spectrum $\omega$  and a contribution from the driving harmonics $q \Omega$. The above relation implies that the rate of jumps $A_{-\omega, -q}$ transferring an excitation $\omega + q \Omega$ to the bath, are exponentially more probable than the complementary jumps $A_{\omega, q}$ absorbing $\omega + q \Omega$ from it.

Turning back to the Sch\"odinger picture, the full secular master equation takes the form:
\begin{equation} \label{eq:full-sec-Schrodinger}
    \dot{\rho}(t) = -i [H_S(t) + H_{LS}(t), \rho(t)] + \mathcal{D}_t[\rho(t)],
\end{equation}
with time-dependent dissipator $\mathcal{D}_t[\rho] \equiv U_S(t) \mathcal{D}[U^\dagger_S(t) ~\rho~ U_S(t)] U_S^\dagger(t)$, which is periodic (see demonstration in~\cite{szczygielski_application_2014}) 
with the drive period $T$. Since the (interaction-picture) master equation possesses a unique steady state solution $\Tilde{\pi}$ verifying $-i[\Tilde H_{LS}, \Tilde\pi] + \mathcal{D}[\Tilde \pi] = 0$, the above equation reaches in the long time run a periodic steady state (or limit cycle) solution $\pi(t) = U_S(t) \Tilde{\pi} U_S^\dagger(t)$ with $\pi(t + nT) = \pi(t)$ for all $n \in \mathbb{Z}$~\cite{szczygielski_application_2014}. This is an important property that we will use later to discuss the thermodynamic features predicted by this equation.

The full secular approximation can be consistently assumed when all off-diagonal terms of the Kossakovski matrix $K_{\alpha, \alpha^\prime}(t)$ oscillate fast enough~\cite{alicki_internal_2006}, so that the matrix becomes diagonal at longer stroboscopic times, ensuring a CPTP dynamics. In particular, it requires $|\alpha - \alpha^\prime| \gg \lambda^2$ for all $\alpha \neq \alpha^\prime$, which requires that the quasienergies are well separated~\cite{Hone09}, together with a sufficiently fast driving.  Otherwise, an appropriate secular approximation can only eliminate terms that satisfy $|\alpha - \alpha^\prime| \gg 1/\tau_S$, where $\tau_S$ is a suitable time scale in which the system changes appreciably (in the interaction picture)~\cite{farina_open-quantum-system_2019,cattaneo_local_2019}.

We note that a stronger form of the full secular approximation can be made by performing the secularization not with respect to the combined Bohr-Floquet quasifrequencies $\alpha=\omega+q\Omega \in \mathbb{F}$, but by retaining only the terms with both equal quasifrequencies $\omega = \omega^\prime \in \mathbb{B}$ and harmonics $q=q^\prime\in \mathbb{Z}$ independently.
This approach, known as the ultrasecular approximation, provides a more simplified master equation with respect to the full secular treatment. However, its predictions may in some cases deviate more strongly from the exact dynamics than those obtained within the full secular approximation presented above~\cite{hotz_coarse-graining_2021}.

In the following sections, we show that the master equation~\eqref{eq:full-sec-Schrodinger} is quite restrictive for the description of the dissipation in periodically-driven systems, which often depart from the required conditions that guarantee the application of the full secular approximation.
We then analyze partial secularization via coarse-graining in time, where we focus on its physical implications and interpretation, and compare this approach with the Floquet-Redfield master equation. In Appendix~\ref{sec:otherreg} we also analyse other alternative methods commonly employed in the literature.

\section{Thermodynamic limitations of the full secular approximation}\label{sec:thermodconsistency}

In order to show the limited thermodynamic scope of the full-secular master equation~\eqref{eq:full-sec-Schrodinger} we adopt a first-principles approach to identify mechanical work in driven open systems. Let $\rho_{\rm tot}(t)$ denote the system and bath density operator, whose evolution verifies the Liouville-von Neumann equation (in Schr\"odinger picture):
\begin{equation} \label{eq:vNeq}
    \dot{\rho}_{\rm tot}(t) = -i [H(t), \rho_{\rm tot}(t)],
\end{equation}
with $H(t)$ the total Hamiltonian in Eq.~\eqref{eq:eq1}. Since the system and bath compound is an isolated system governed by a time-dependent Hamiltonian, the mechanical work exerted by the drive can be generically identified with the total change in energy $W(t) = \Tr[H(t) \rho_{\rm tot}(t)] - \Tr[H(0) \rho_{\rm tot}(0)]$, or in differential form:
\begin{align} \label{eq:power}
    \dot{W}(t) &= \Tr[\dot{H}(t) \rho_{\rm tot}(t)] + \Tr[{H}(t) \dot{\rho}_{\rm tot}(t)] \nonumber \\ 
    &= \Tr[\dot{H}(t) \rho_{\rm tot}(t)] - i \Tr[{H}(t)[H(t), \rho_{\rm tot}(t)]] = \Tr_S[\dot{H}_S(t) \rho(t)].
\end{align}
where to reach the second line we used Eq.~\eqref{eq:vNeq}, which cancels the second term in the r.h.s. In the last equality we performed the trace over the bath and used that the only time-dependence in the total Hamiltonian comes form the system part, and hence $\dot{H}(t)= \dot{H}_S(t)$. This derivation is general and valid for any driven open system with Hamiltonian not necessarily periodic (the only requirement is that the bath and interaction Hamiltonians are time-independent).

We now calculate the mechanical power $\dot{W}(t)$ using the expression above under the full secular approximation master equation~\eqref{eq:full-sec-Schrodinger}. In particular we focus in the power delivered along a cycle $T$ in the long-time limit, where the system density operator reaches the periodic steady state $\pi(t + nT) = \pi(t)$ (see Sec.~\ref{sec:FSA} above):
\begin{equation} \label{eq:Wcycle}
  W_{\rm cycle} =  \int_t^{t +T}ds~ \Tr[\dot{H}_S(s) \pi(s)] = 
  \Tr[H_S(s) \pi(s)]\Big|_{s=t}^{s=t+T} - \int_t^{t +T}ds~ \Tr[H_S(s) \dot{\pi}(s)].
\end{equation}
where the first term vanishes because $H_S(t)$ and $\pi(t)$ are periodic with same period $T$. To show that the second term also vanishes we evaluate $\dot{\pi}(s)$ using the full-secular master equation~(\ref{eq:full-sec-Schrodinger}). Using the (interaction-picture) stationary state $\Tilde{\pi}$, the evolution in the Schr\"odinger picture is 
$\dot{\pi}(t) = -i [H_{S}(t), \pi(t)]$,
which generates the limit cycle solution discussed in the previous section. Substituting back to Eq.~(\ref{eq:Wcycle}) and using the cyclic property of the trace we obtain
\begin{equation} \label{eq:WcycleSec}
  W_{\rm cycle} =   i \int_t^{t +T}ds~ \Tr[H_S(s) ~ [H_{S}(s), \pi(s)]] = 0.
\end{equation}
The above result implies that any periodically driven system described by the full secular master equation does not directly consume power in the (periodic) steady state. This result might be expected for systems driven far from resonance, where energy exchanges are typically weaker and may be mediated by other mechanisms (e.g. Coulomb repulsion in electronic systems). Examples include high-frequency far-off-resonance optical trapping~\cite{Miller1993,kaufman2021}, dispersive qubit readout for quantum non-demolition measurements in circuit QED~\cite{blais2021}, or, more generally, any very slow (quasi-static) driving, where the system follows the instantaneous equilibrium state at any time~\cite{Cavina2017}. However, in most driven-dissipative systems, that is not the case, pointing to a systematic failure of the full-secular master equation to describe driven dissipative thermodynamics. As a matter of fact, typically $\dot{W} \neq 0$ in the steady state for weak-driving fields near resonance~\cite{Geva94,Boukobza07,Hofer16,mitchison_realising_2016,klatzow_experimental_2019,AlmanzaMarrero2025certifyingquantum}.

On the other hand, under the full secular approximation, the diagonal form of the Kossakowski matrix in Eq.~\eqref{eq:FullsecularME}, together with the generalized local detailed balance relation in \eqref{eq:LDB}, implies the absorption and emission of discrete energy quanta $\alpha = \omega + q \Omega$ into the bath, mediated by jump operators $A_{\alpha}$ in the Floquet Hamiltonian ladder, c.f. Eq.~\eqref{eq:jumpops}. As a consequence, heat can be identified with the transitions in the quasienergy spectrum mediated by the drive harmonics~\cite{szczygielski_markovian_2013,langemeyer2014energy,gasparinetti_heat-exchange_2014,cuetara_stochastic_2015}. 
In terms of the Floquet Hamiltonian and full secular dissipator, the heat current from the bath can be expressed as~\cite{szczygielski_markovian_2013,kosloff2013quantum}:
\begin{equation}
    \dot{Q}(t) 
    = \sum_{\omega \in \mathbb{B}} \sum_{q \in \mathbb{Z}} \left( \frac{\omega + q \Omega}{\omega} \right) K_{\omega + q \Omega, \omega + q \Omega} \Tr \Big[H_F \Big({A}_{\omega, q} \Tilde{\rho}(t) {A}^\dagger_{\omega, q} - \frac{1}{2} \Big\{{A}^\dagger_{\omega, q} {A}_{\omega, q},
\Tilde{\rho}(t)\Big\}\Big) \Big],\label{eq:heatnormalized}
\end{equation}
which is consistent with the second law of thermodynamics~\cite{kosloff2013quantum,cuetara_stochastic_2015}. Using the commutation relation $[H_F, A_{\omega, q}] = \omega~ A_{\omega, q}$ this expression reduces to 
\begin{equation}
    \dot{Q}(t) = \sum_{\omega \in \mathbb{B}} \sum_{q \in \mathbb{Z}} ( \omega + q \Omega) K_{\omega + q \Omega, \omega + q \Omega} \Tr\Big[{A}_{\omega, q} \Tilde{\rho}(t) {A}^\dagger_{\omega, q} \Big].\label{eq:heatnormalized_simplified}
\end{equation}
Despite the full secular master equation leads to zero work in a cycle in the long time run, that is not the case for the heat exchanged with the environment.  Indeed, because of the prefactors $(\omega + q\Omega)/\omega$ in Eq.~\eqref{eq:heatnormalized}, the above expressions do not lead in general to zero flux in the steady state $\tilde{\pi}$, and the system can sustain non-zero heat current during a cycle, $Q_{\rm cycle} \neq 0$, through transitions between Floquet quasienergy states. This mismatch between the expected work and heat during a cycle in the long-time run points to a breakdown of the first law.

It is also worth pointing out that despite the above predicted absence of steady-state mechanical work from the drive in a cycle, $W_{\rm cycle} = 0$, the system does not typically reach a (periodic) thermal equilibrium state~\cite{Breuer00,Shirai15}. This follows from the presence of the extra contribution $q \Omega$ from the Floquet modes (harmonics) in the local detailed balance relation \eqref{eq:LDB}, which prevents, in general, thermalization~\cite{kohn2001periodic,Hone09}. The non-thermal nature of the periodic steady-state can be explained by interpreting such extra terms as a non-conservative force, leading to an extra work contribution associated to jumps between quasienergy levels [c.f. Eq.\eqref{eq:LDB}], as considered in Refs.~\cite{cuetara_stochastic_2015,elouard_thermodynamics_2020}. Such non-conservative work would compensate for the non-vanishing heat current, thus saving the first law. It is identified as the exchange of integer multiples of the driving frequency $q\Omega$ during each microscopic transition, namely, $\dot{W}_{\rm nc} \equiv - \sum_{\omega \in \mathbb{B}} \sum_{q \in \mathbb{Z}} q \Omega~ K_{\omega + q \Omega, \omega + q \Omega} \Tr\Big[{A}_{\omega, q} \Tilde{\rho}(t) {A}^\dagger_{\omega, q} \Big]$. From Eq.~\eqref{eq:Wcycle} it is clear that such non-conservative work contribution is not directly exchanged by the drive, but it is a nonequilibrium effect induced by the drive on the bath whose origin remains, however, obscure. These issues suggest that the full secular master equation~\eqref{eq:full-sec-Schrodinger} may not provide in general an adequate tool to study the thermodynamics of periodically driven-dissipative systems.

We now show that, in the case of the drive introduced in Eq.~(\ref{eq:V}), the property of zero power in the steady state can be traced back to the absence of coherences in the system density operator $\Tilde{\rho}$ as expressed in the Floquet basis $\{\ket{\phi}_k\}$. In that case, the instantaneous power reads: 
\begin{equation}
     \dot{W}(t) =  - 2 g \Omega ~\mathbb{I}\text{m}[\rho_{j,k}(t) e^{i \Omega t}], \label{eq:powerV}
\end{equation}
and therefore a diagonal steady-state does not consume power from the drive \footnote{Note that in Eq.~\eqref{eq:powerV} $\rho_{j, k}$ corresponds to the system density matrix elements in the Schrodinger picture and in the (bare) basis of $H_0$.}. Under the full secular approximation, the dynamics of the diagonal and off-diagonal elements of $\Tilde \rho(t)$ become decoupled, and the coherences vanish in the long time limit (see Sec.\ref{sec:FSA} and appendix~\ref{appendix:rhodiag}).

If, on the other hand, the density matrix in the long time limit retains coherences in the Floquet basis, the instantaneous work is in general non zero. This is the case of the partial secularization, in which the master equation in the interaction picture does not have a fixed point (so that in general $\dot{\pi}(t) \neq -i[H_S(t), \rho(t)]$ anymore) and the state of the system in the long time limit is not diagonal. A similar situation is obtained in the weak driving limit (see Sec.~\ref{sec:weakdriving}), where there exists a steady state in the interaction picture, but it shows coherences in the Floquet Hamiltonian basis.

\section{Coarse-graining approach and master equation} \label{sec:coarse-graining}

While the full secular approximation may neglect important oscillatory terms responsible for steady-state coherences, the Floquet–Redfield equation retains rapidly oscillating contributions that can be unphysical, leading to a lack of complete positivity. To address these issues, we now turn to a coarse-grain approach, which allows us to examine and reconcile the limitations of both the Floquet–Redfield equation and the full secular approximation even when going beyond the weak driving limit. 

To coarse-grain the dynamics, we average the evolution of $\Tilde{\rho}$ over a time interval $\Delta t$. Importantly, such an interval must be chosen accordingly with the previous approximations taken to derive Eq.~(\ref{eq:ME}). These approximations impose a sandwich inequality of the form 
\begin{equation}
    \tau_B \ll \Delta t \ll \tau_S, \label{eq:CGcondition}
\end{equation}
where $\tau_B$ is the typical time-scale for the decay of the bath correlations, and $\tau_S = \min(\tau_{\rm int},\tau_{H_F})$, with $\tau_{\rm int}\sim 1/\lambda^2$ and $\tau_{H_F}$ the time scale of changes in the system due to the interaction with the bath and the Floquet Hamiltonian, respectively. The latter depends on the quasienergy splittings and therefore on the driving parameters, including the driving frequency $\Omega$. The left inequality is a consequence of Markovianity, that is, the time scale at which correlations decay in the bath is much smaller than the time scale at which the master equation is coarse-grained. On the other hand, the right inequality guarantees that the coarse-graining can accurately resolve the dynamics of the system in the interaction picture. The coarse-grained approach can only be consistently used when there exists a $\Delta t$ satisfying these conditions.
 
The coarse-grained master equation is derived by averaging the Floquet-Redfield Eq.~(\ref{eq:ME}) in the Floquet-interaction picture
over an interval $\Delta t$: 

\begin{align}
    \frac{1}{\Delta t}  \int_{t- \Delta t/2}^{t + \Delta t/2} \dot{\Tilde{\rho}}(s) ds =& -i \frac{1}{\Delta t}  \int_{t- \Delta t/2}^{t + \Delta t/2}  ds\sum_{\alpha, \alpha^\prime \in \mathbb{F}} e^{i(\alpha^\prime - \alpha)s} \chi(\alpha, \alpha^\prime) [A^\dagger_{\alpha^\prime} A_\alpha, \Tilde{\rho}(s)]\label{eq:intred}  \\
& + \frac{1}{\Delta t}  \int_{t- \Delta t/2}^{t + \Delta t/2}  ds\sum_{\alpha, \alpha^\prime \in \mathbb{F}} e^{i(\alpha^\prime - \alpha)s}~\gamma(\alpha,\alpha^\prime) \Big(A_{\alpha} ~\Tilde{\rho}(s) A^\dagger_{\alpha^\prime} - \frac{1}{2} \{ A^\dagger_{\alpha^\prime} A_{\alpha} , \Tilde{\rho}(s)\}\Big). \nonumber 
\end{align}
When the coarse-grain interval is much shorter than the characteristic time scale of the system’s density matrix evolution $\Delta t \ll \tau_S$ [right inequality in Eq.~\eqref{eq:CGcondition}] the system density operator is not significantly modified within the interval $\Delta t$ and integration above only modifies the oscillatory phase terms in the r.h.s.~\cite{farina_open-quantum-system_2019}:
\begin{equation}
    \frac{1}{\Delta t}  \int_{t- \Delta t/2}^{t + \Delta t/2}  e^{i(\alpha^\prime - \alpha)s} = e^{i(\alpha^\prime - \alpha)t} \text{sinc}\left[\frac{(\alpha^\prime- \alpha) \Delta t}{2}\right],
\end{equation}where $\text{sinc}(x) = \sin(x)/x$ is the cardinal sinus function. Therefore, the Floquet-Redfield master equation~\eqref{eq:ME} becomes
\begin{equation} \label{eq:MECG}
    \dot{\Tilde{\rho}}(t) = -i [H_{LS}^{\rm c.g.}, \Tilde{\rho}(t)] + \mathcal{D}^{\rm c.g.}[\Tilde{\rho}(t)],
\end{equation}
with a coarse-grained Lamb-shift term $H_{LS}^{c.g.} = \sum_{\alpha, \alpha^\prime \in \mathbb{F}} e^{i(\alpha^\prime - \alpha)t} \text{sinc}[(\alpha^\prime - \alpha) \Delta t/2] \chi(\alpha, \alpha^\prime) A^\dagger_{\alpha^\prime} A_\alpha$, and a coarse-grained dissipator 
\begin{align}
     \mathcal{D}^{\rm c.g.}[\Tilde{\rho}] =  \sum_{ \alpha^\prime, \alpha \in \mathbb{F}} K^{\rm c.g.}_{\alpha^\prime \alpha}(t) \Big(A_{\alpha^\prime} \Tilde{\rho}(t) A^\dagger_\alpha - \frac{1}{2} \Big\{A^\dagger_\alpha A_{\alpha^\prime}, \Tilde{\rho}(t)\Big\}\Big),\label{eq:BRdissipatorKdt}
\end{align}
with associated (coarse-grained) Kossakowski matrix with elements given by
\begin{equation}
    K_{\alpha^\prime \alpha}^{\rm c.g.}(t) = \gamma_{\alpha^\prime, \alpha} \text{sinc}\left[\frac{(\alpha^\prime- \alpha) \Delta t}{2}\right] e^{i (\alpha^\prime- \alpha)t}.\label{eq:Kcoarsegrain}
\end{equation}
The coarse-grained Kossakowski matrix above has same diagonal elements as the original one, $K_{\alpha, \alpha}^{\rm c.g.} = K_{\alpha, \alpha}$, but introduces a non-trivial damping of the off-diagonal elements via the sinc functions. Indeed, by varying $\Delta t$ the sinc function in Eq.~\eqref{eq:Kcoarsegrain} acts as a frequency filter, introducing unequal weighting among the off-diagonal components of the Kossakowski matrix.

\begin{figure}
    \centering
    \includegraphics[scale=0.85]{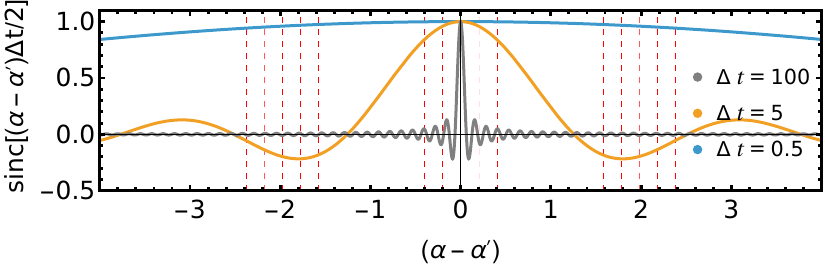}
    \caption{Illustration of the effect of coarse-graining on the Kossakowski matrix for different averaging intervals. Three cases are displayed corresponding to different time intervals $\Delta t$ compatible with full secular approximation (gray), the Floquet-Redfield equation (blue) and an intermediate regime (orange). The sinc function leads to a damping of some of the off-diagonal elements $K_{\alpha, \alpha^\prime}$ when the frequency difference $\alpha - \alpha^\prime$ becomes large. As an example, we show the resulting frequency differences $\alpha -\alpha^\prime$ of the oscillating terms in the Kossakowski matrix (dashed red vertical lines) for the driven two-level system studied in Sec.~\ref{sec:2ls}, with Hamiltonian Eq.~(\ref{eq:hs}). We use units such that $\omega_0 =1$ serves as the reference frequency, and set $g = 0.1\omega_0$ and $\Delta = 0.01\omega_0$.}
    \label{fig:sincs}
\end{figure}

In Fig.~\ref{fig:sincs} we illustrate how the sinc functions in Eq.~\eqref{eq:Kcoarsegrain} may damp the off-diagonal terms $K_{\alpha, \alpha^\prime}^{\rm c.g.}$ associated to the set of frequency differences $\alpha - \alpha^\prime$ 
in the case of a driven-dissipative two-level system (dashed red vertical lines). A more thorough discussion of this model is reserved for Sec.~\ref{sec:2ls}, here we simply use it to illustrate the role of $\Delta t$ in shaping the terms of the Kossakowski matrix. In the case of large coarse-graining intervals, as $\Delta t \rightarrow \infty$ the sinc function approaches a $\delta$ function, and all off-diagonal elements are washed out, leading to the same result as full secular approximation (gray line), see Sec.~\ref{sec:FSA}.
This ensures the positivity of the Kossakowski matrix $K^{\rm c.g.}$, but it is only justified if a sufficiently large $\Delta t$ exists to suppress all the off-diagonal terms associated to frequencies $\alpha - \alpha^\prime$ without violating the inequality $\Delta t \ll \tau_S$. If such a condition is not fulfilled, important oscillations might be averaged out, leading to an erroneous dynamics and steady state. On the other hand, for small intervals $\Delta t \rightarrow 0$, the sinc function approaches a flat distribution around $1$ over the relevant frequencies (blue line) and all off-diagonal elements of the Kossakowski matrix are almost no weighted. This case corresponds to no coarse-graining being applied at all, which naturally recovers the Floquet-Redfield equation~\eqref{eq:ME}. Such a small $\Delta t$, however, can violate the inequality $\tau_B \ll \Delta t$ in Eq.~\eqref{eq:CGcondition}, which breaks down the Markovian assumption and possibly leads to non physical solutions. For intermediate values of $\Delta t$ (orange line) only some of the off-diagonal terms $K_{\alpha, \alpha^\prime}^{\rm c.g.}$ will be damped.  Intuitively, such a coarse-grained master equation may neglect the contribution from terms that oscillate fast and are responsible for the loss of complete positivity, but keep oscillations that are important for an accurate and thermodynamically consistent evolution. 

Here we argue that in systems where the time scales are such that there exists a $\Delta t$ satisfying both conditions in Eq.~\eqref{eq:CGcondition}, a coarse-gain time interval can be chosen to satisfactorily regularize the Floquet-Redfield equation~\eqref{eq:ME}. Our prescription consists of selecting the smallest coarse-graining time $\Delta t_{\rm min}$ that guarantees the positivity of the coarse-grained Kossakowski matrix $K^{\rm c.g.}(t)$ at any time $t$. This can be done by decomposing it as $K^{\rm c.g.}(t)=D(t)K^\prime D^\dagger(t)$, where $D(t)$ is a unitary diagonal matrix with elements $D_{\alpha, \alpha}(t) = e^{-i \alpha t}$ and $K^\prime$ is a time-independent matrix with elements
\begin{equation} \label{eq:Ksinc}
K^\prime_{\alpha^\prime\alpha}=\gamma_{\alpha^\prime\alpha}
\text{sinc}\left[\frac{(\alpha^\prime-\alpha)\Delta t}{2}\right].
\end{equation}
Increasing $\Delta t$ progressively suppresses the off-diagonal elements $K^\prime_{\alpha^\prime\alpha}$ through the sinc filter, until all their eigenvalues become non-negative. 
Since the spectrum of $K^\prime$ equals the one of $K^{\rm c.g.}(t)$, ensuring a $\Delta t$ for which $K^\prime$ is positive semidefinite implies also that $K^{\rm c.g.}(t)$ is positive semidefinite at all times $t$. 

The advantage of this approach is that the coarse-graining interval $\Delta t$ can be tuned to reflect the relevant time scales of the dynamics. Indeed explicitly introducing $\Delta t$ effectively sets a temporal (and frequency) resolution of the master equation. In our approach, $\Delta t$ is chosen as the smallest interval for which the Kossakowski matrix becomes positive semidefinite. In this way, it has a clear physical interpretation: the finest resolution at which the evolution can still be consistently described by a (time-divisible) Markovian master equation with a completely positive generator. 

We notice that different coarse-graining approaches, with $\Delta t$ often being dynamically varied, have already been used in the literature to enhance the accuracy of master equations, mainly in the case of systems with time-independent Hamiltonians~\cite{schaller_preservation_2008,majenz_coarse_2013,farina_open-quantum-system_2019}. In Ref.~\cite{hotz_coarse-graining_2021} coarse-graining in the master equations was applied to cases of periodically-driven systems to obtain time-independent Markovian equations. In contrast, in this work we are interested in obtaining a time-dependent master equation in GKLS form, valid for generic periodically-driven-dissipative systems in the weak coupling regime.

Furthermore, other approaches for regularization used in systems with time-independent Hamiltonians can be extended to regularize the Floquet-Redfield master equation~\cite{kirsanskas_phenomenological_2018,mccauley_accurate_2020,nathan_universal_2020,davidovic_completely_2020,soret_thermodynamic_2022,boubakour2026refined,dabbruzzo_time-dependent_2023,potts_thermodynamically_2021,trushechkin_unified_2021}. In the Appendix~\ref{sec:otherreg} we summarize these methods and compare with the Floquet-Redfield to the examples below.

In the weak-driving limit, the coarse-grained approach introduced above reduces in the interaction picture to
\begin{equation}
    \mathcal{D}^{(0)}[\rho] = \sum_{\omega,\omega^\prime \in \mathbb{B}_0} \gamma_{\omega, \omega^\prime} e^{i(\omega - \omega^\prime)t} \text{sinc}\left[\frac{(\omega - \omega^\prime) \Delta t}{2}\right] \left( L_\omega \rho L_{\omega^\prime}^\dagger - 1/2 \big\{L_{\omega^\prime}^\dagger L_{\omega}, \rho \big\} \right).\label{eq:localMEsinc}
\end{equation}
where $\omega$ and $\omega^\prime$ are taken from the Bohr frequencies of the bare (static) Hamiltonian, $\mathbb{B}_0= \{E_k - E_l~;k \neq l\}$. When the system does not exhibit near degeneracies, the off-diagonal contributions with $\omega \neq \omega^\prime$ in Eq.~\eqref{eq:localMEsinc} contain rapidly oscillating phases $e^{i(\omega-\omega^\prime)t}$. These terms average out over the dissipative timescale $\sim \lambda^{-2}$ whenever the standard secular condition $|E_k-E_l| \gg \lambda^2$ for all $k,l$ is satisfied by the spectrum of $H_0$. Neglecting these contributions directly recovers the weak-driving dissipator in Eq.~\eqref{eq:localME}, characterized by a diagonal and positive Kossakowski matrix. In this limit, complete positivity is ensured independently of the coarse-graining procedure, and the limit $\Delta t \to 0$ can therefore be consistently taken.

\section{Energetics of driven-dissipative systems}\label{sec:thermoCG}

The Floquet theory used in the derivation of the master equations discussed above introduces a new picture for the structure of transitions associated with energy exchanges. The Floquet Hamiltonian $H_F$ describes the dressed Hamiltonian of the system with a spectrum of quasienergies $\{ \epsilon_k \}$. Each quasienergy state $\{\ket{\phi_k}\}$ is associated with a set of replicas  given by the different Floquet modes (harmonics), separated by integer multiples of the driving frequency $\Omega$. 
As shown in Sec.~\ref{sec:FSA}, under the full secular approximation, the mechanical work performed by the drive vanishes in the periodic steady state. Furthermore, this picture leads to a heat current expressed in terms of independent transitions between Floquet sidebands, leading to the prefactors appearing in Eq.~\eqref{eq:heatnormalized_simplified}, which lead to non-vanishing heat currents in the long time run. Consequently, the first law can only be recovered by introducing an additional non-conservative work contribution.

As we show below, with a proper partial secular approximation (as given by the coarse-grained master equation) the coherences present in the periodic steady state yield a finite mechanical power, which verifies the first law. Importantly, the mechanical work contribution is non-zero even in parameter regimes where the full secular treatment accurately reproduces the populations and the associated heat currents. This suggests that the coherences neglected by the full secular approximation can play a key thermodynamic role, and that when they are properly taken into account, make unnecessary the postulation of the non-conservative work.

Without the full secular approximation, the expression for the heat in Eq.~(\ref{eq:heatnormalized}) is no longer valid, requiring a more general formulation.
In particular, the dissipator of the master equation cannot be decomposed into independent channels labeled by a single pair of indices $(\omega,q)$ (or alternatively $\alpha$), but instead contains contributions that couple different virtual transitions, characterized by pairs of indices $(\omega,q)$ and $(\omega^\prime,q^\prime)$ (or alternatively $\alpha$ and $\alpha^\prime$). 
A consistent definition of heat can be obtained using the framework of full counting statistics~\cite{esposito_nonequilibrium_2009,gasparinetti_heat-exchange_2014}, which can be applied beyond the secular approximation~\cite{Friedman_2018,kilgour_coherence_2018,liu_coherences_2021,murphy_laser_2022}. The procedure consists of re-deriving the master equation while introducing (one or several) auxiliary variables, referred to as the counting fields. These fields track the energy exchanged with the reservoirs (see Appendix~\ref{appendix:A} for a full derivation of the method). The counting fields applied to the coarse-grained approach yield the general expression for the heat current \begin{equation}
    \dot{Q}(t) = - \sum_{\alpha, \alpha^\prime \in \mathbb{F}} \mathbb{R}\text{e}\left[\alpha \Gamma(\alpha) + \alpha^\prime \Gamma(\alpha^\prime) \right] e^{-i(\alpha-\alpha^\prime)t} \text{sinc}\left[\frac{(\alpha-\alpha^\prime)\Delta t}{2}\right] \Tr \big[ A_{\alpha} \Tilde{\rho}(t) A^\dagger_{\alpha^\prime}  \big],\label{eq:Qcountingfields}
\end{equation} 
where we used the approximation as in the case studies (See Eq.~(\ref{eq:smoothbathapprox})). 

We note that, among the two terms in the dissipator of Eq.~\eqref{eq:BRdissipatorKdt}, only the first one appears above in within the trace. Since the operators $A_{\alpha}$ denote transitions operators associated to given quasi-frequencies and Floquet modes $\alpha = \omega + q \Omega$, the terms with $\alpha = \alpha'$ can be interpreted as heat due to the actual emission and absorption of quanta with such energy into the environment. 
However, in contrast to Eq.~(\ref{eq:heatnormalized}), the above expression does not rely on a decomposition into independent transitions, and instead incorporates also contributions that couple different channels for terms such that $\alpha \neq \alpha^\prime$. As such, it captures quantum effects arising from coherence between them, which are absent in the full secular description. For example, as we see in the Section~\ref{sec:3ls}, these nonsecular terms can give rise to interference between energy currents from different reservoirs. As in Eq.\eqref{eq:BRdissipatorKdt}, the $\text{sinc}$ function acts as a filter, mitigating the action of these terms accordingly with the condition of complete positivity. In the secular regime, where the driving frequency is large compared to the relaxation timescale and the coarse-graining interval satisfies $\Delta t \gg \Omega_R^{-1}$, the filter function is sharply peaked and well approximated by a delta function, $\delta(\alpha -\alpha^\prime)$. Cross-frequency terms are thus suppressed in this limit, and the counting-fields expression for the heat current Eq.~(\ref{eq:Qcountingfields}) reduces to the full secular form in Eqs.~\eqref{eq:heatnormalized} and~\eqref{eq:heatnormalized_simplified}.

In the presence of multiple (independent) reservoirs, an equation analogous to Eq.~\eqref{eq:ME} can be derived for each reservoir separately. The heat exchanged with reservoir $r = 1, 2, ...$, denoted by $Q_r$, is given by Eq.~\eqref{eq:Qcountingfields}, but with bath correlation function, $\Gamma^{(r)}(\alpha)$ now depending on the parameters of the corresponding reservoir and system-bath coupling operators $A_\alpha^{(r)}$ which might be different for each reservoir. 
Accordingly, the first law for a driven open system coupled to multiple reservoirs reads
\begin{equation}
\dot{E} = \dot{W} + \sum_r \dot{Q}_r,
\end{equation}
with $\dot{E} = (d/dt)\Tr[H_S(t)~\rho(t)]$ the rate of energy change in the system, and for the case of a drive $V(t)$ as introduced previously in Eq.~(\ref{eq:V}), the power is given by Eq.~(\ref{eq:powerV}). For a single period of the driving $T$ in the long time run, we have $\Delta E = 0$ since the diagonal elements of $\rho(t)$ are periodic with period $T$. However now we obtain 
\begin{equation}
W_{\rm cycle} = \int_{t}^{t+T} ds~ \dot{W}(s) = - \sum_r\int_{t}^{t+T} ds  ~\dot{Q}_r(s) \neq 0,
\end{equation}
which becomes generically non-zero when the steady state of $\Tilde{\rho}$ is not time-independent, as it is the case for the coarse-grained master equation~\eqref{eq:MECG} and hence $\pi(t)$ does not follows a unitary limit-cycle orbit.

We note that the first law above does not follow directly from the full-counting-statistics formalism. The work is evaluated as $\Tr[\dot{H}\rho]$, whereas the heat is obtained from the counting-field statistics, which track the change in the energy of the reservoirs. Since these two quantities are defined independently, their sum does not close into the first law at the level of the exact expressions. Nevertheless, the first law is recovered in the relevant limits. In the secular limit, our expressions reduce to the standard definition in Eqs.~\eqref{eq:heatnormalized_simplified}, for which the first law holds, and the same is true in the weak driving limit discussed below. Beyond these regimes, non-secular terms make the standard definitions ill-defined, so a direct comparison is not possible. However, we do a numerical analysis of its consistency in the examples studied in the manuscript.

In the weak driving limit, the above expression for heat reduces to   
\begin{equation}
    \dot{Q}_\text{weak} = \Tr[\mathcal{D}^{(0)}[\Tilde{\rho}] H_0] = \sum_{\omega \in \mathbb{B}} \omega \gamma_{\omega} \Tr\Big[{L}_{\omega} \Tilde{\rho}(t) {L}^\dagger_{\omega} \Big],
\end{equation}
while the expression for $\dot{W}(t)$ in Eq.~\eqref{eq:powerV} remains valid in the case of weak driving. Therefore we recover a smooth transition in the thermodynamic quantities from the weak to strong driving regimes, contrary to the case of the full secular approximation. 

Finally, we remark that the above expressions for heat and work for the coarse-grained master equation are also consistent with the second law of thermodynamics, which can be expressed for generic (bipartite) open quantum system dynamics as~\cite{esposito2010entropy}
\begin{equation} \label{eq:secondlaw}
   {S}_{\rm tot}(t) = \Delta S_{\rm sys}(t) - \beta Q(t) = S(\rho_{\rm tot}(t) || \rho(t) \otimes \rho_B ) \geq 0,
\end{equation}
where $\Delta S_{\rm sys}(t)$ is the change in von-Neumman entropy of the system, $Q(t) = - \Tr[H_B (\rho_B(t) - \rho_B)]$ is the energy exchange with the thermal bath, and $S(\rho || \sigma)$ is the relative entropy between generic density operators $\rho$ and $\sigma$. The inequality above follows from the non-negativity of relative entropy $S(\rho || \sigma)\geq 0$, which becomes zero if and only if $\rho =  \sigma$, corresponding to reversibility conditions.

The above expression for the entropy production in Eq.~\eqref{eq:secondlaw} can be derived from first principles. It is valid for generic (non-Markovian) open system dynamics in contact with a bath that starts the dynamics in thermal equilibrium, $\rho_B = e^{-\beta H_B}/Z_B$, and is reset to it afterwards~\cite{manzano2018quantum}. In the present case,  Eq.~\eqref{eq:secondlaw} can be applied to every stroboscopic coarse-grain time $\Delta t$ used to derive the master equation~\eqref{eq:MECG}. As a consequence of Markovianity, we can assume that after an interval of time $\Delta t$ the bath correlation functions have decayed ($\Delta t \gg \tau_B$), and hence the bath can be considered to be effectively (re)thermalized, and the (previously build) correlations with the system do not affect the subsequent system evolution. Therefore Eq.~\eqref{eq:secondlaw} can be written for the coarse-grained master equation~\eqref{eq:MECG} in dynamical form as:
\begin{equation} \label{eq:EPrate}
    \dot{S}_{\rm tot}(t) = \dot{S}_{\rm sys}(t) - \beta \dot{Q}(t) \geq 0
\end{equation}
where the heat exchanged with the bath can be identified from the full counting statistic approach to be given by Eq.~\eqref{eq:Qcountingfields}.

\section{Case Studies}

The coarse-graining procedure presented here, allows us to work with analytical expressions and to connect coarse graining directly to a physically meaningful resolution scale governing the validity of Markovian and thermodynamic descriptions, as we will discuss in the following in more detail for specific examples. In particular, in the examples presented in subsections~\ref{sec:2ls} and~\ref{sec:3ls}, we compare the coarse-grained approach against the Floquet–Redfield, weak driving, and full secular master equations, in order to assess its validity across different dynamical regimes. We emphasize that the examples considered here involve relatively simple Floquet structures that allow for analytical treatment, however, the general framework developed in the previous sections applies to arbitrary sets of Floquet harmonics. More complex driving protocols may make analytical solutions more difficult to obtain, but the dynamics and heat currents can still be readily implemented numerically.

For completeness, we also consider a rotating-wave approximation (RWA) in the system--bath coupling. This master equation again provides an averaged description of the dynamics (with frequency $2\Omega$), but the averaging is performed at the Hamiltonian level, before the master equation is derived. The resulting equation reproduces the dynamics with accuracy comparable to the Redfield equation, although it does not guarantee complete positivity. The RWA results are presented in Appendix~\ref{sec:RWA}.

An approximation that we will use in the following examples is to retain only the real part of $\gamma(\alpha,\alpha^\prime)$ and the imaginary part of $\chi(\alpha,\alpha^\prime)$. This is justified by the smoothness of the bath spectral response over the system frequencies under the Born-Markov assumption, so that $\mathbb{R}\text{e}[\Gamma(\alpha) - \Gamma(\alpha^\prime)] \approx \mathbb{I}\text{m}[\Gamma(\alpha) - \Gamma(\alpha^\prime)] \approx 0$ for the frequency differences $\alpha - \alpha^\prime$ considered. Under this approximation, the prefactors in Eq~(\ref{eq:ME}) reduce to
\begin{equation} \begin{aligned}
    \chi(\alpha, \alpha') &\approx\frac{1}{2} (\mathbb{I}\text{m}[\Gamma(\alpha)] + \mathbb{I}\text{m}[\Gamma(\alpha^\prime)]), \\
    \gamma(\alpha, \alpha') &\approx \mathbb{R}\text{e}[\Gamma(\alpha)] + \mathbb{R}\text{e}[\Gamma(\alpha^\prime)],\label{eq:smoothbathapprox}
\end{aligned}\end{equation}
which separates the unitary and dissipative contributions. We stress that this approximation is not required for the consistency of the formalism, but is introduced to simplify the analytical treatment. We have verified numerically in the examples that it has a negligible impact on the results predicted by the master equation.

In both examples the environment is composed by one or two bosonic baths. The bath operators appearing in the interaction Hamiltonian $H_{SB}= \lambda (A \otimes B)$ contain a sum over the environmental modes, $B = \sum_l \mu_l (b_l +b_l^\dagger)$ with $[b_l,b_l^\dagger]=\mathbb{I}$. When computing the Fourier transforms of the bath correlation function $\Gamma(\alpha)$, in Eq.~\eqref{eq:bm1}, the sum can be replaced by an integral over a spectral density $J(\omega)= \lambda^2 \sum_l \mu_l^2\delta(\omega-\omega_l)$, resulting in 
\begin{align} \label{eq:bosonicGamma}
    \Gamma(\alpha) &= \pi  (N(\alpha)+1) J(\alpha) \Theta(\alpha) + \pi N(\alpha)J(\alpha)(1-\Theta(\alpha)) \nonumber \\ 
    &~~+ i \mathcal{P} \int_0^\infty d\omega^\prime J(\omega^\prime) \left( \frac{N(\omega^\prime) + 1}{\alpha - \omega^\prime} + \frac{N(\omega)}{\alpha + \omega^\prime} \right)
\end{align}
where $\Theta(\omega)$ is the Heaviside function, $N(\omega)= (e^{\beta \omega} -1 )^{-1}$ the Plank distribution, and $\mathcal{P}$ denotes the principal value (the derivation is analogous to the case of time-independent Hamiltonians~\cite{breuer_theory_2002}). We assume an Ohmic spectral density with exponential cutoff $J(\omega) = \xi~ \omega~ e^{-\omega/\omega_c}$, and apply the approximation in Eq.~(\ref{eq:smoothbathapprox}).

\subsection{Example I: Driven-dissipative two-level system}\label{sec:2ls}

We examine a periodically driven qubit coupled to a bosonic thermal reservoir. The thermodynamics of this system has been previously studied in~\cite{grifoni_driven_1998,geva_relaxation_1995,szczygielski_markovian_2013,gasparinetti_heat-exchange_2014,juan-delgado_first_2021} with the use of the full secular approximation. A coarse-grained average was also employed in~\cite{elouard_thermodynamics_2020,soret_thermodynamics_2025}, albeit with a different interpretation. Here, we concentrate on the role of coherences and assess analytically the validity of the secular approximation from a thermodynamic standpoint analytically, comparing our findings with numerically exact reduced dynamics of the system-Bath Hamiltonian Eq.~\eqref{eq:eq1} obtained using the PT-TEMPO algorithm~\cite{strathearn2018efficient,link2024open}. The time-dependent system Hamiltonian is
\begin{equation}
    H_S(t) =  \frac{\omega_0}{2} \sigma_z + g \left(\sigma_- e^{i \Omega t} + \sigma_+ e^{-i \Omega t}\right),\label{eq:hs}
\end{equation} 
where the rotating wave approximation has been performed in the interaction with the drive. The system Hamiltonian can be split into the static term $H_0 =  \frac{\omega_0}{2} \sigma_z$, with spectrum $E_\mp=\{-\omega_0/2, \omega_0/2\}$, and the drive term $V(t) = g \left(\sigma_- e^{i \Omega t} + \sigma_+ e^{-i \Omega t}\right)$, whose period is $T= 2\pi/\Omega$. 

As mentioned in Sec.~\ref{sec:ME}, the periodic unitary operator for this type of driving reads $P(t) = \exp{\left[- i t \Omega \sigma_z/2\right]}$, which leads to a single pair of harmonics ($q=\pm1$). The Floquet Hamiltonian, defined in Eq.~(\ref{eq:U_S}), is then of the form in Eq.~\eqref{eq:HFsplit}, namely, $H_F =  \frac{\Delta}{2} \sigma_z + g\sigma_x$, with quasienergies $\epsilon_{\mp} = \mp \Omega_R/2$, where we introduced the detuning $\Delta= \omega_0 - \Omega$, and the Rabi frequency $\Omega_R= \sqrt{\Delta^2 + 4 g^2}$. The set of Bohr quasifrequencies is $\mathbb{B} = \{\pm \Omega_R,0\}$ corresponding, respectively, to transitions from the upper to the lower quasienergy levels (and vice-versa), and no transitions.

\begin{figure}
    \centering
    \includegraphics[scale=0.75]{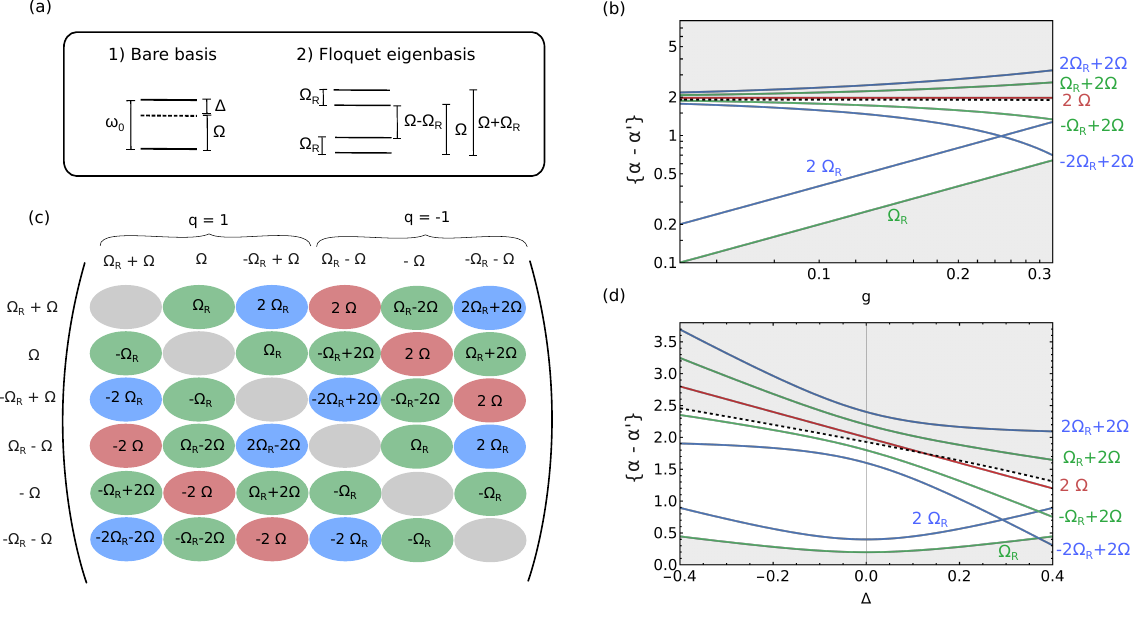}
    \caption{{\bf a.} Energy diagram of the driven two-level system in the bare basis and in the Floquet eigenbasis. The Floquet eigenstates are separated by $\Omega_R$, and the Floquet replicas are shifted by $\Omega$. {\bf b.} Maximum coarse-grain frequency resolution $\omega_\text{max} = 2 \pi/\Delta t_\text{min}$ as a function of $g$ is depicted in black dashed. The minimum $\Delta t$ is the finest time resolution for a Markovian master equation to yield a positive Kossakowski matrix. Off-diagonal oscillation frequencies $\{\alpha - \alpha^\prime\}$ are color-coded as in (c). The gray shaded regions represent the limits of validity of the coarse grain interval~(\ref{eq:CGcondition}) {\bf c.} Kossakowski matrix, with the oscillation frequency of each term indicated at its position and colors denoting the type of contribution: gray for secular terms, blue for coherence oscillations, green for coupling between coherences and populations, and red for terms oscillating at frequency $2\Omega$. {\bf d.} Maximum coarse-grain frequency resolution and off-diagonal oscillation frequencies as a function of detuning.}
    \label{fig:kmatrix}
\end{figure}

Figure \ref{fig:kmatrix}(a) shows a diagram with the energy levels of the bare two-level system (left) and of its Floquet eigenbasis when combined with the harmonics (right). The first represents the static energies of the two-level system (i.e. when the drive is turned off), but with the driving frequency $\Omega$ sketched for comparison. In the Floquet picture, we obtain instead the quasienergy levels of the Floquet Hamiltonian, together with replicas spaced by integer multiples of the drive frequency. These virtual levels form distinct Floquet zones (here we have two zones) separated by $\Omega$. The quasienergies of each zone are given by the Floquet Hamiltonian $H_F$. The emission spectrum of this system features three different peaks of frequency corresponding to the extended set of Bohr quasifrequencies combined with the harmonics $\mathbb{F} = \{\Omega +\Omega_R,\Omega, \Omega-\Omega_R\}$ given by transitions between distinct Floquet zones, known as the Mollow triplet~\cite{mollow1969power}.

The corresponding jump operators defined in Eq.~(\ref{eq:jumpops}), assuming a system-bath coupling of the form $H_{SB} = \lambda A \otimes B$ with $A= \sigma_x$, take the form:
\begin{equation}
    \begin{aligned}
        A_{-\Omega_R,\pm1}= \left(\frac{\Delta - \Omega_R}{2 \Omega_R} \right)\sigma_{\pm}, \quad
        A_{0,\pm1}= \frac{g}{\Omega_R} \sigma_z, \quad
        A_{\Omega_R,\pm1}= \left(\frac{\Delta + \Omega_R}{2 \Omega_R} \right)\sigma_{\mp},
    \end{aligned}
\end{equation}
where we recall that $A_{\omega,q} = A_{-\omega,-q}^\dagger$ with $q=\pm1$. In the following, we will absorb the numerical prefactors into the Kossakowski matrix and consider the jump operators as simply $A_{-\Omega_R,\pm1} = \sigma_{\pm}$, $A_{0, \pm 1}=\sigma_z$ and $A_{\Omega_R, \pm1} =\sigma_{\pm}$. 

To better understand the effect of the non-secular terms in the Floquet-Redfield master equation~\eqref{eq:ME}, we obtain analytical expressions showing how the dissipative contribution [Eq.~\eqref{eq:BRdissipatorK}] enters the coarse-grain master equation. We construct the Kossakowski matrix with indices $(\alpha, \alpha^\prime)$ ordered as $\{ \Omega_R +\Omega, \Omega, - \Omega_R + \Omega ,\Omega_R -\Omega, -\Omega, - \Omega_R - \Omega \},$ and show all its elements in the Appendix~\ref{appendix:B}. We show a schematic representation of the matrix elements in Fig.~\ref{fig:kmatrix}(c), where different colors highlight their distinct contributions in the system's dynamics. The oscillation frequency associated with each element is given by the difference $\alpha - \alpha^\prime = (\omega + q \Omega) - (\omega^\prime + q^\prime \Omega)$, indicated at its corresponding position in the matrix. We calculate analytically the effect of each term in the sum in Eq.~(\ref{eq:BRdissipatorK}) and categorize the elements in four contributions: 
\begin{equation}
     \mathcal{D}[\Tilde{\rho}] =   \mathcal{D}[\Tilde{\rho}] \Big |_\text{sec}  +  \mathcal{D}[\Tilde{\rho}]\Big |_\text{coherence} +  \mathcal{D}[\Tilde{\rho}]\Big |_\text{coupling} +  \mathcal{D}[\Tilde{\rho}]\Big |_{2 \Omega},\label{eq:termsdisspator}
\end{equation}
associated, respectively, to the secular terms (grey), terms leading to extra coherence oscillations (blue), terms that couple populations and coherences (green), and fast terms oscillating at $2\Omega$ (red). A detailed discussion of the role of these different contributions is provided in Appendix~\ref{appendix:B}.

In Figures~\ref{fig:kmatrix}(b) and (d), we compare the oscillation frequencies of the off-diagonal elements of the Kossakowski matrix $K$ with the minimum coarse-grain time, $\Delta t_\text{min}$ (obtained numerically) for which the $K$ matrix remains positive. The different frequencies are displayed with the corresponding color of their matrix element in panel~\ref{fig:kmatrix}(c), and the dashed line corresponds to the cutoff frequency $\omega_\text{max} \equiv 2 \pi/\Delta t_\text{min}$ in our approach. The shaded region above this frequency indicates the values of $\Delta t$ for which the dynamics is not positive, and therefore inequality~(\ref{eq:CGcondition}) is not verified from below. This condition also has an upper limit, $2\pi/\tau_S$, to ensure an accurate resolution of the ME, shown by the shaded region at the bottom of the figures. 

\begin{figure}
    \centering
    \includegraphics[scale=0.48]{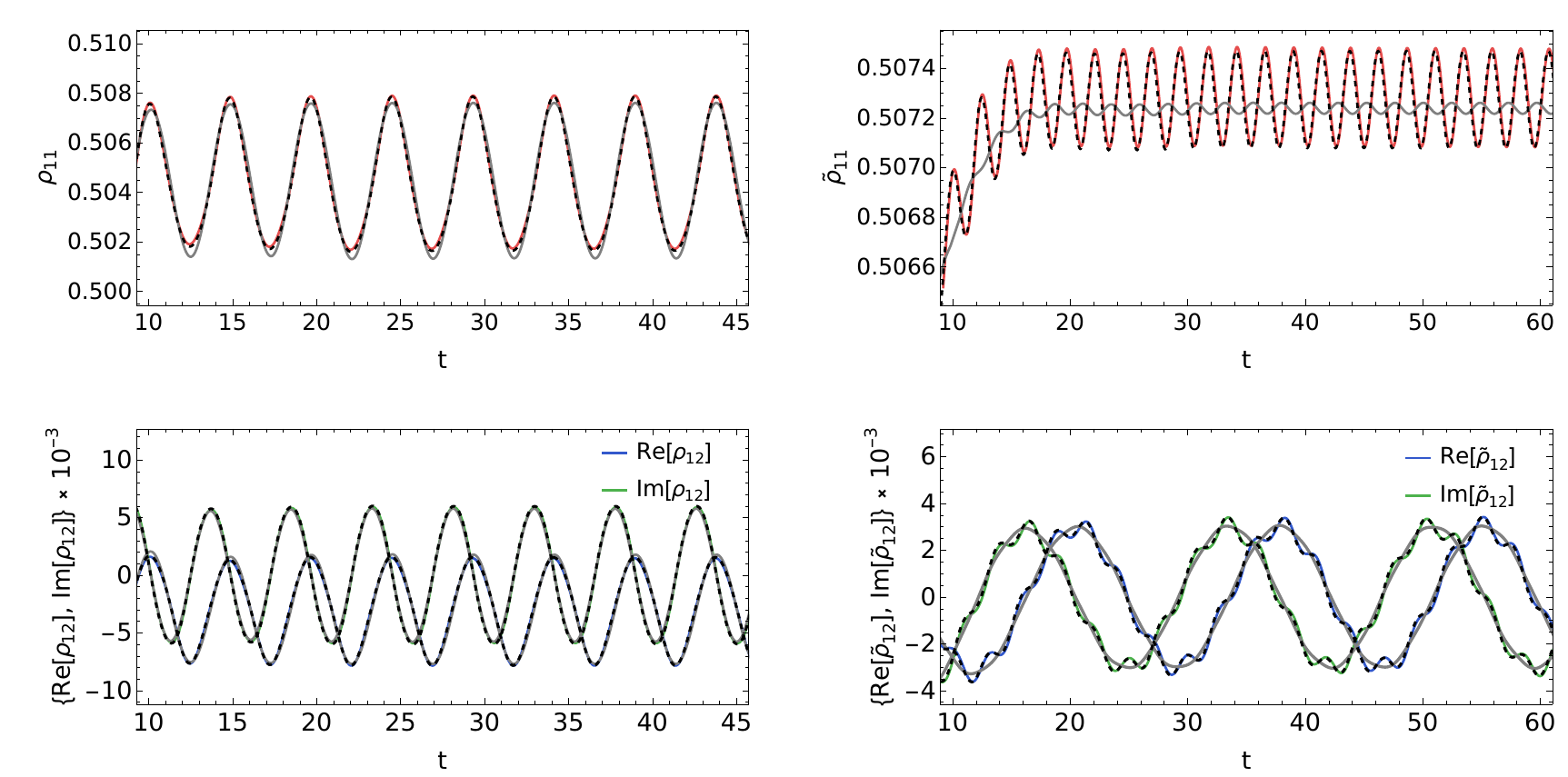}
    \caption{Exact dynamics of the driven two-level system in the Schr\"odinger picture (left), and interaction picture (right). The top row has the populations and the bottom row coherences. We show curves calculated with the Floquet-Redfield (black dotted), and coarse-grained (solid gray) master equations, together with exact non-Markovian simulations (colored lines). In the long-time limit, the system exhibits oscillations with frequency $\Omega$ in the Schr\"odinger picture, which are well captured by the three approaches. In the interaction picture, we observe population oscillations with frequency $2 \Omega$ that are well captured by the Floquet-Redfield equation, but present reduced amplitude in the coarse-grained approach to ensure completely positive dynamics. The coherences contain oscillatory components at different frequencies. The coarse-graining procedure filters out the high-frequency oscillations, resulting in a periodic function with period $2\pi/\Omega_R$. The state is shown in the dressed basis of the Floquet Hamiltonian, and the color scheme is the same in all panels: $\rho_{11}$ in red, the real part of the coherences in blue, and the imaginary part in green. We use $\omega_0 = 1$ as the reference frequency, and set $g = 0.1 \omega_0$, $\Delta = - 0.3 \omega_0$, $\omega_c = 20 \omega_0$, $\xi = 10^{-3}$, and $k_b T=30 \omega_0$. Axis intervals differ for visualization purposes.} 
    \label{fig:Nonmarkov}
\end{figure}

The cutoff frequency also corresponds to the first zero of the cardinal sinus function appearing in Eq.~\eqref{eq:Kcoarsegrain} [or analogously Eq.~\eqref{eq:Ksinc}] and illustrated in Fig.~\ref{fig:sincs}. This means that terms oscillating with frequencies near $\omega_{\mathrm{max}}$ are suppressed. From Figs.~\ref{fig:kmatrix}(b) and (d), we observe that the numerically obtained values of $\omega_{\mathrm{max}}$ are approximately $2\Omega$, implying that these contributions are strongly filtered. This suggests that terms oscillating at frequency near $2\Omega$ constitute the dominant contribution responsible for the breakdown of positivity in the Floquet-Redfield equation. 

For relatively weak driving, as in the regime shown in Fig.~\ref{fig:kmatrix}(b), all terms oscillating at frequencies $\pm \{2 \Omega, 2 \Omega \pm \Omega_R\}$ are near $\omega_{\mathrm{max}}$ and hence filtered out in the coarse-graining. As $g$ increases, the Rabi splitting $\Omega_R$ becomes larger, thus reducing the spectral separation between $\{\Omega_R, 2 \Omega_R\}$ and the other frequencies. As a result, $\{\Omega_R, 2 \Omega_R\}$ are increasingly filtered out together with the other oscillatory contributions. An analogous effect is observed at larger (positive) detunings, when the frequencies $\{\Omega_R, 2 \Omega_R\}$ approach $\omega_{\mathrm{max}}$. These suggest that the coarse-graining approximation performs most accurately in the regime of relatively weak driving and negative detuning. Furthermore, we point out that in the limit of very strong coupling, the coarse-graining approach with $\Delta t_\text{min}$ filters out all off-diagonal oscillations, recovering the full secular ME. 

We now demonstrate the validity of our approach by comparing it with the dynamics of the system using an exact (non-Markovian) simulation. The exact simulations are performed using a variant of the PT-TEMPO algorithm~\cite{strathearn2018efficient,link2024open}, implemented using the OQuPy framework~\cite{fux2024oqupy}, which uses tensor-network techniques to compress the influence functional of the bath, allowing for numerical calculation of the system dynamics. 

In Figure~\ref{fig:Nonmarkov}, we compare the dynamics obtained from the exact simulation with those predicted by the coarse-grained (with $\Delta t_\text{min}$) and Floquet-Redfield master equations in the Schr\"odinger and interaction picture. In the interaction picture, the phase factors $e^{-i(\alpha-\alpha^\prime)t}$ in the master Eq.~\eqref{eq:ME} give rise to oscillations at frequencies $\alpha-\alpha^\prime$. As a result, the populations exhibit long-lived oscillations at frequency $2\Omega$, while the coherences contain contributions oscillating at different frequencies. These oscillations are accurately captured by the Floquet-Redfield master equation (black dashed), which agrees closely with the exact simulation (colored). In contrast, the coarse-grained master equation (gray) suppresses the faster oscillations. In particular, for the coherences, the coarse-grained approach filters out the high-frequency components, generating an averaged dynamics with oscillation frequency $\Omega_R$. Although the coarse-graining procedure can suppress oscillations seen in the interaction picture, this effect is not readily apparent when the dynamics are viewed in the Schr\"odinger picture. 

In Appendix~\ref{appendix:positivity}, we show the dependence of the minimum eigenvalue, $\lambda_{\text{min}}$, of the Kossakowski matrix on the coarse-graining interval $\Delta t$. The results demonstrate that the value of $\Delta t_{\text{min}}$ is very sensitive to the chosen tolerance for the negativity of the minimum eigenvalue. Here, we present results obtained with no tolerance, i.e., by imposing $\lambda_{\text{min}} = 0$. Allowing for a small negative tolerance is expected to yield even better agreement with the exact method.

\begin{figure}
    \centering
    \includegraphics[scale=0.4]{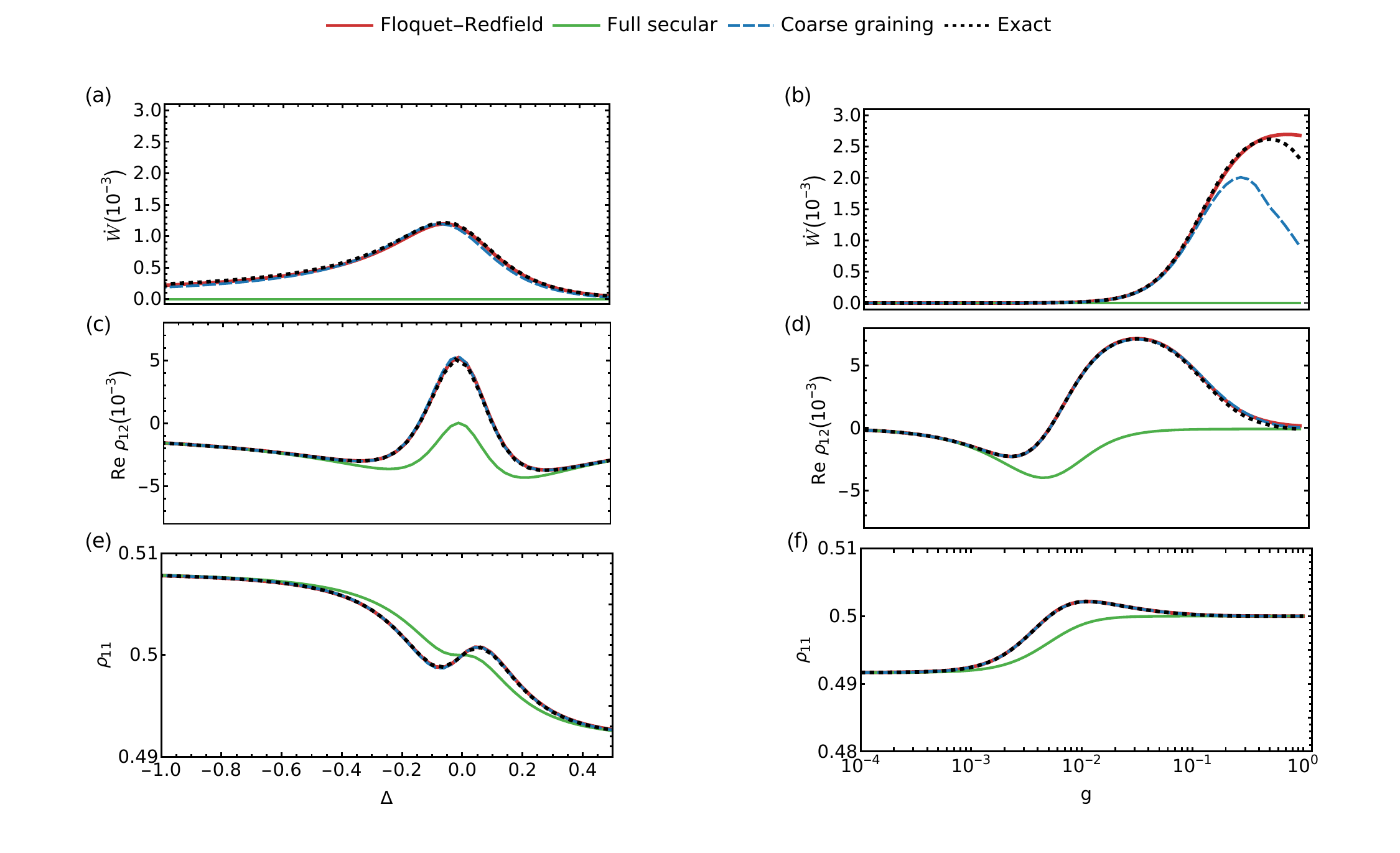}
    \caption{Comparison of the averages over a period of the drive of work ({\bf a.} and {\bf b.}),  steady state coherence ({\bf c.} and {\bf d.}), and population ({\bf e.} and {\bf f.}) given by the coarse-grained, the Floquet-Redfield, and the full secular master equations, and an exact non-Markovian simulation (see legend) as a function of detuning (left) and driving strength (right). The full secular approximation can describe the populations and real part of coherences in the off-resonant regime, but fails to provide physical values of the imaginary part of coherences and therefore work current. The coarse-grained and Floquet-Redfield master equations agree with the exact simulation for the parameter inside its validity range Eq.~(\ref{eq:CGcondition}). The density matrix is shown in the dressed basis of the Schr\"odinger picture. The bath parameters are the same as in Fig.~\ref{fig:Nonmarkov}, and when not varied, we take $g=0.1\omega_0$, $\Delta = 0.01\omega_0$.}
    \label{fig:comparison2ls}
\end{figure}

Next, we calculate the averaged steady states over a period of the drive ($ 2 \pi /\Omega$) of the exact, Floquet-Redfield, and coarse-grained master equation simulations and compare them with the full secular one. The results of steady-state populations and coherence are presented in the Schr\"odinger picture and shown in Fig.~\ref{fig:comparison2ls}. The full secular approximation is expected to correctly capture the populations to zeroth order~\cite{tupkary_fundamental_2022}, and indeed, we find good agreement with the exact solution, except when the drive is close to resonance (panels (e) and (f)). For $\Delta =0.01~\omega_0$, deviations start occurring with driving strengths of $10^{-3}~\omega_0$ and increase with $g$. However, for off-resonant values of detuning, the agreement improves and is maintained across the entire range of driving strengths considered. Nevertheless, the full secular approximation predicts incorrect off-diagonal density-matrix elements. In particular, it fails to reproduce the behavior of the real part of the coherences near resonance (panels (c) and (d)). Moreover, the work input from the drive (panels (a) and (b)), $\dot{W}(t) =  - 2 g \Omega ~\mathbb{I}\text{m}[\rho_{12}(t) e^{i \Omega t}]$ vanishes across all parameter regions considered. 

On the other hand, the Floquet-Redfield and coarse-grained master equations exhibit good agreement with the exact results for both populations and the real part of coherences. Deviations in the input power are observed for stronger driving strengths, which can be attributed to the Rabi frequency approaching the cutoff frequency of the approximation, $\omega_{\max}$. In this limit, the cardinal sinus function filters the Rabi oscillations ($\tau_S^{-1}$) induced by the system Hamiltonian (see bottom right panel of Fig.~\ref{fig:Nonmarkov}), leading to a loss of accuracy. Furthermore, we observe that in the limit of weak driving , all three approaches (Floquet-Redfield equation, coarse-grained master equation and exact non-Markovian simulation) recover the predictions from the weak driving master equation~(\ref{eq:MEweak}). In this regime, the Kossakowski matrix is diagonal, and the Lindblad operators are eigenoperators of the static Hamiltonian, satisfying local detailed balance~\cite{manzano_quantum_2022}. We also point out that when no drive is considered, i.e., $g=0$ exactly, the steady-state density matrix corresponds to a Gibbs thermal equilibrium state at the reservoir's temperature.

\begin{figure}[t]
    \centering
    \includegraphics[scale=0.4]{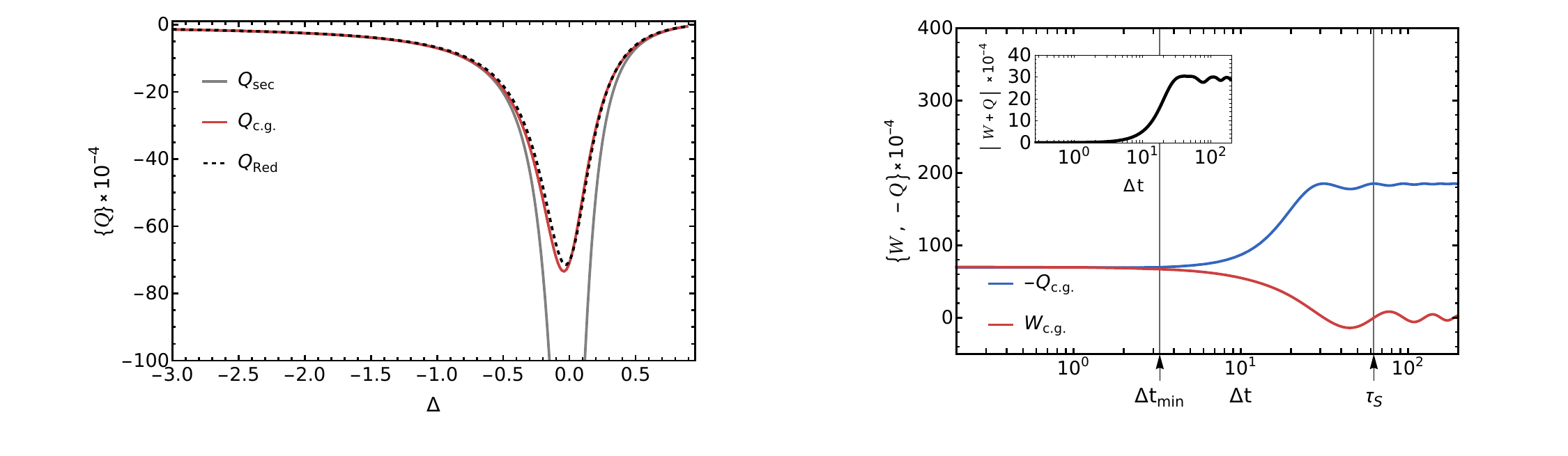}
    \caption{Averaged steady state and heat per cycle as a function of the  detuning (left) and coarse-grain time interval (right). The energy currents are consistent with the first law of thermodynamics for small $\Delta t$. The inset of the right panel shows the sum $W + Q$, representing the accuracy with which the first law of thermodynamics is obtained. 
    }
    \label{fig:1stlaw2ls}
\end{figure}

In the left panel of Fig.~\ref{fig:1stlaw2ls}, we compare the steady-state heat current per cycle [Eq.~(\ref{eq:Qcountingfields})] into the reservoir obtained from the Floquet-Redfield, coarse-grained, and full secular master equations. As it can be appreciated, the full secular approximation fails to correctly capture the heat current near resonance (inside the interval $\Delta \in [-1/2, 1/2]$), whereas the coarse-grained and Floquet-Redfield master equations yield similar results across all regimes. Far from resonance, however, all three approaches produce nearly identical heat currents (i.e. for very fast or very slow driving). As can be appreciated, all the three master equations predict a positive heat current in all regimes. This is consistent with the second law of thermodynamics, which predicts a non-negative entropy production rate, Eq.~\eqref{eq:EPrate}. When applied to a cycle of the periodic steady-state, it leads to $\Sigma_{\rm cycle} = - {Q}_{\rm cycle}/T \geq 0$, which constrains the direction of heat flow to dissipate into the environment. 

It is important to stress that any type of secular approximation, including the full secular, coarse-grained, and any other partial secularization, will affect the steady-state populations and coherence differently. This has a direct impact on the energy currents and therefore on the thermodynamic consistency of the master equation. The power, given by Eq.~(\ref{eq:powerV}), is proportional to the imaginary part of $\rho_{ 12}(t)$, while the heat is predominantly influenced by the populations of $\rho$. In this way, the stronger the secularization, the less accurate the first law of thermodynamics. This fact is illustrated in the right panel of Fig.~\ref{fig:1stlaw2ls}, where we show the averaged steady-state power and heat for various coarse-grain intervals. For small coarse-grain intervals, which include both the Floquet-Redfield equation ($\Delta t = 0$) and the minimum coarse-grained approach ($\Delta t = \Delta t_{\rm min}$), energy conservation is verified up to fourth order corrections in the system-reservoir coupling $\lambda$, i.e. $ 10^{-6} \omega_0 $ and $10^{-5} \omega_0 $, respectively, in this parameter regime. In contrast, in the limit of large $\Delta t$ where the ME approaches the full secular limit, it predicts non-zero heat exchange but zero work, making necessary the postulation of extra non-conservative work contributions. 
We also show, for comparison, the upper timescale limit on the right-hand side of inequality~\eqref{eq:CGcondition}, which in this case is given by $\tau_S = 2\pi/\Omega_R$. Any coarse-graining interval that does not satisfy $\Delta t \ll \tau_S$ cannot properly resolve the system oscillations and hence fails to correctly predict the mechanical work.

As mentioned before, in the context of thermodynamics, the full secular approximation has been analyzed in the literature, where the absence of mechanical work was compensated by the inclusion of a non-conservative work contribution. In our approach, no redefinition of work is required whenever the coarse-grain interval respects the upper bound $\Delta t \ll \tau_S$, as the mechanical work is naturally captured through the coherences in the periodic steady state. This difference reflects the fact that time averaging and time evolution do not, in general, commute. In other words, evolving the state with a full secular ($\Delta t \rightarrow \infty $) master equation is generally not equivalent to taking the time average of the state obtained from the underlying dynamics over long times.

Taken together, these results suggest that it is difficult to identify a universally optimal Markovian master equation for driven-dissipative systems. In the parameter regime considered here, the Floquet-Redfield master equation provides the most accurate description of the dynamics and hence of the power. However, this accuracy comes at the cost of losing complete positivity, which, in principle, can lead to unphysical behavior. Conversely, the coarse-grained master equation preserves complete positivity by construction, but it loses accuracy in the description of the coherences, and therefore in the power. The choice of approximation must therefore be made on a case-by-case basis, balancing the desired level of accuracy in the dynamics against the need for consistency. When neither requirement can be satisfactorily met within a Markovian description, more sophisticated approaches, including non-Markovian simulations, may be required.

Additionally, the alternative approach of applying the rotating wave approximation (RWA) in the system-bath coupling could also be applied. In this case it consists in neglecting oscillations of frequency $2 \Omega$ in the interaction picture
\begin{equation}
    \Tilde{H}_{SB}= \sum_l \mu_l (b_l e^{i\Omega t}+b_l^\dagger e^{-i\Omega t})(\sigma_+ e^{-i\Omega t}+\sigma_- e^{i\Omega t}) \simeq  \sum_l \mu_l  (b_l \sigma_++b_l^\dagger \sigma_-).
\end{equation}
This procedure is equivalent to applying a coarse-graining interval of $\Delta t = 2 \pi/2 \Omega$ directly to the master equation in a rotated frame of the Schrödinger picture, and results in a time-independent block-diagonal Kossakowski matrix. By completely averaging out the oscillatory components of the steady state, it significantly simplifies the calculations. The resulting steady state remains in good agreement with the Redfield prediction; however, this simplification does not guarantee complete positivity yet. A more detailed discussion of this approach is provided in Appendix~\ref{sec:RWA}.

\subsection{Example II: Three-level  maser}\label{sec:3ls}

In this section, we analyze the model of a three-level maser acting as a heat engine, as first proposed in Refs.~\cite{scovil_three-level_1959,geusic_quantum_1967}. This paradigmatic setup consists of a three-level system coupled to hot and cold thermal reservoirs, where the energy absorbed from the hot bath is partially converted into work in the form of coherent radiation, while the remaining energy is dissipated into the cold bath. When operated in reverse, the same configuration functions as a refrigerator~\cite{geusic1959three,Palao2001}.

The thermodynamic properties of this model have been extensively investigated within the framework of open quantum systems and continuous heat engines~\cite{Geva94,geva_quantum_1996,Boukobza07,kosloff2014quantum,Uzdin2015,mitchison_realising_2016}, with more recent works addressing optimal performance, coherence effects, and fluctuations~\cite{Scully2017,singh_three-level_2019,kalaee_violating_2021,bayona-pena_thermodynamics_2021,AlmanzaMarrero2025certifyingquantum}. Beyond its theoretical relevance, the three-level maser model has been successfully applied to model a wide range of physical platforms, including superconducting quantum circuits~\cite{aamir_thermally_2025}, nitrogen-vacancy centers in diamond~\cite{klatzow_experimental_2019}, and polariton condensation~\cite{toledo_tude_quantum_2024}.

\begin{figure}[t]
    \centering
    \includegraphics[scale=0.76]{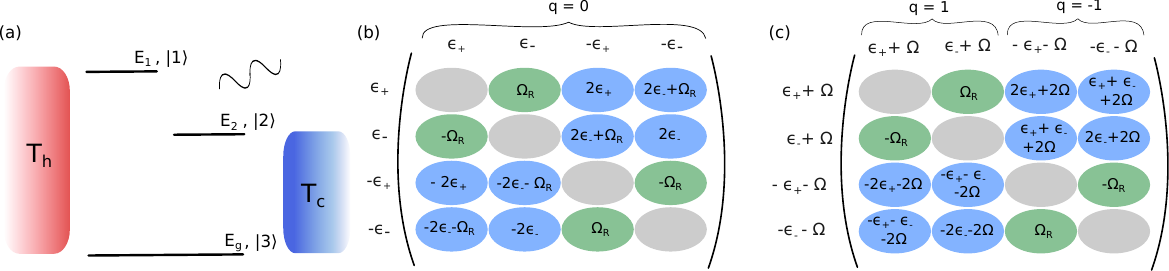}
    \caption{{\bf a.} Schematic of the three-level maser showing the energy levels coupled to a hot and a cold reservoir and driven by an external periodic field. {\bf b.} and {\bf c.} Kossakowski matrix of the cold (b) and hot (c) baths, with the oscillation frequency of each element indicated at its position. Colors denote the type of contribution: gray for secular terms, blue for affecting only coherence, and green for terms coupling populations and coherences. The Kossakowski elements lie within a single Floquet zone ($q = 0$) for the cold bath, but span two Floquet zones ($q = \pm 1$) for the hot  bath.}
    \label{fig:diagrams3ls}
\end{figure}

The model, illustrated in Fig.~\ref{fig:diagrams3ls}(a), consists of a three-level system with states labeled $\ket{3}, \ket{2}, \ket{1}$. The system interacts with two thermal reservoirs: a hot bath weakly couples to transition $\ket{3}\leftrightarrow\ket{1}$, while a cold bath weakly couples to transition $\ket{3}\leftrightarrow\ket{2}$. In addition, a periodic driving field of arbitrary strength $g$ is applied to transition $\ket{1}\leftrightarrow\ket{2}$. The system Hamiltonian reads:
\begin{equation}
\begin{aligned}
      H_S &= E_1 \ket{1}\bra{1} + E_2\ket{2}\bra{2} + E_g \ket{3}\bra{3} + g (e^{i \Omega t} \ket{1}\bra{2} + e^{-i \Omega t} \ket{2}\bra{1}), \label{eq:H3ls}
\end{aligned}
\end{equation}
where in the following we set $E_g = 0$ without loss of generality. The first terms of Eq.~\eqref{eq:H3ls} correspond to the static Hamiltonian term, $H_0 = E_1 \ket{1}\bra{1} + E_2\ket{2}\bra{2}$, while $V(t) = g (e^{i \Omega t} \ket{1}\bra{2} + e^{-i \Omega t} \ket{2}\bra{1})$ is a driving term of the form in Eq.~\eqref{eq:V}, with period $T= 2 \pi/\Omega$. We consider the periodic unitary $P(t) = \exp{\left(- i t \Omega \ket{1}\bra{1}\right)}$ \footnote{The operator $P(t)$ is not uniquely defined, and other choices such as $P(t) = \exp{\left[- i t \Omega \big(\ket{1}\bra{1} - \ket{2}\bra{2}\big)/2\right]}$ would also be possible.}, which leads to a rotating frame that cancels out the time dependence of $H_S$ by transferring it to the coupling operator with the hot bath. This choice of transformation leads to the Floquet modes (harmonics) $q = \{0, \pm 1\}$.


The Floquet Hamiltonian is of the form in Eq.~\eqref{eq:HFsplit}, which in the computational basis reads: 
\begin{equation}
    H_{F}=\begin{pmatrix}
    E_1 + \Delta && g && 0 \\
    g && E_1 && 0\\
     0 && 0 && 0
\end{pmatrix},
\end{equation}
where we introduced the detuning $\Delta = (E_1 -E_2) - \Omega$. The Floquet Hamiltonian can be diagonalized, giving the Floquet eigenvectors $\ket{\phi_\pm} = (\Delta \pm \Omega_R)\ket{1} + 2 g \ket{2}$ and $\ket{\phi_0} = \ket{0}$. These dressed states are obtained from a rotation of the bare basis by an angle $\theta/2$, where the mixing angle is $\tan\theta = \frac{2g}{\Delta}$. The eigenvalues of $H_F$ give us the set of Floquet quasi-energies $\epsilon_{\pm}= (2 E + \Delta \pm \Omega_R)/2$, where we introduced the Rabi splitting $\Omega_R =\sqrt{\Delta^2 + 4 g^2}$, and ground state energy $\epsilon_0 = 0$.

The system-bath interaction consist of two terms $H_{SB} = g_{\rm h} (\ket{3}\bra{1} + \ket{1}\bra{3}) \otimes B_{\rm h} + g_{\rm c} (\ket{3}\bra{2} + \ket{2}\bra{3}) \otimes B_{\rm c} $, where $B_{\rm h}$ and $B_{\rm c}$ are bath operators corresponding to the hot and cold reservoirs, respectively, and $g_{\rm h}$ and $g_{\rm c}$ are their corresponding strengths. Again, we model the baths as collections of harmonic oscillators, so that $B_{r} = b_{r} + b_{r}^\dagger$, for $r={\rm c, h}$, and consider the approximation in Eq.~(\ref{eq:smoothbathapprox}). The transition operators $A_\alpha^{(r)}$ in Eq.~(\ref{eq:ME}) for reservoirs $r= {\rm c, h}$ are then:
\begin{equation}
    \begin{aligned}
        A^{\rm (c)}_{\epsilon_-, 0} = \sqrt{\frac{\Omega_R + \Delta}{2 \Omega_R}} \ket{3}\bra{2} = \cos(\theta/2) \ket{3}\bra{2} &\qquad
        A^{\rm (c)}_{\epsilon_+, 0} = \sqrt{\frac{\Omega_R - \Delta}{2 \Omega_R}} \ket{3}\bra{1}  = \sin(\theta/2) \ket{3}\bra{1}\\
        A^{\rm (h)}_{\epsilon_-, 1} = \sqrt{\frac{\Omega_R - \Delta}{2 \Omega_R}} \ket{3}\bra{2}= \sin(\theta/2) \ket{3}\bra{2} &\qquad
        A^{\rm (h)}_{\epsilon_+, 1} = \sqrt{\frac{\Omega_R + \Delta}{2 \Omega_R}} \ket{3}\bra{1} = \cos(\theta/2) \ket{3}\bra{1},
    \end{aligned}\label{eq:As3ls}
\end{equation} together with the complementary transitions given by ${A^{(r)}}_{-\omega, -q} =  {A^{(r)}}^{\dagger}_{\omega, q}$.  The corresponding set of Bohr quasifrequencies is now $\mathbb{B} = \{\pm \epsilon_+, \pm \epsilon_- , \pm (\epsilon_+ - \epsilon_-), 0\}$, which correspond to transitions between the ground state and the two non-zero Floquet quasienergies, together with no transitions at all . When we combine the Bohr quasifrequencies with the harmonics, we obtain the extended set $\mathbb{F} = \{\pm \epsilon_+, \pm \epsilon_-, \pm (\epsilon_+ - \epsilon_-), 0,\pm \epsilon_+ \pm \Omega, \pm \epsilon_- \pm \Omega, \pm (\epsilon_+ - \epsilon_-) \pm \Omega, \pm \Omega\}$ which also contains transitions between distinct Floquet zones. However, not all quasifrequencies in $\mathbb{F} $ correspond to reservoir-induced transitions. Only those associated with the nonvanishing transition operators in Eq.~\eqref{eq:As3ls} contribute to the dissipative dynamics. That is, the hot reservoir couples different transitions within the first harmonic $\pm\{\epsilon_+ \pm \Omega,  \epsilon_- \pm \Omega\}$, while the cold reservoir couples transitions within the zeroth harmonic $\pm\{\epsilon_+,  \epsilon_-\}$ (see Fig.~\ref{fig:diagrams3ls}).

As in the previous example, the dissipators corresponding to the hot and cold reservoirs can be decomposed into distinct contributions according to their action on the density matrix. Panels (b) and (c) of Fig.~\ref{fig:diagrams3ls} provide a schematic representation of the elements associated with the Kossakowski matrix of the cold and hot reservoir, respectively. The different colors highlight their respective roles in the system dynamics. The oscillation frequency associated with each element in the interaction picture is indicated at its corresponding position in the matrix. The secular (diagonal) terms are shown in gray, the terms responsible for pure effects on coherences are indicated in blue, and those that couple populations and coherences are marked in green. The full expression of the dissipator is given in Appendix~\ref{appendix:C}. An analysis of the positivity of the Kossakowski matrix as a function of the coase graining interval is presented in Appendix~\ref{appendix:positivity}.

Figure~\ref{fig:dyn3levels} shows the evolution of the population of the intermediate level and of the absolute value of the coherence in the upper levels, in the Schr\"odinger picture, in the frame rotating with the driving frequency (not to be confused with the interaction picture), using the full secular, coarse-grained, and Floquet-Redfield master equations.  We assumed Ohmic baths, and system energies $E_1 = 2~E_2$, with detuning $\Delta = 0.1~E_2$. While the population dynamics coincide for the three approaches (and similarly the populations of the other levels), the coherence show significant differences. The Floquet-Redfield and coarse-grained equations both display oscillations, which are slightly suppressed in the coarse-grained case. The full secular approximation, on the other hand, shows no oscillations and yields very different transient values of the coherences. The inset of Fig.~\ref{fig:dyn3levels}(b) shows the long-time limit, where the three approaches appear to converge to similar values. However, as shown in Fig.~\ref{fig:3levels}, the full secular approximation gives again vanishing coherence in the steady state, while the other two do not.

\begin{figure}
    \centering
    \includegraphics[scale=0.4]{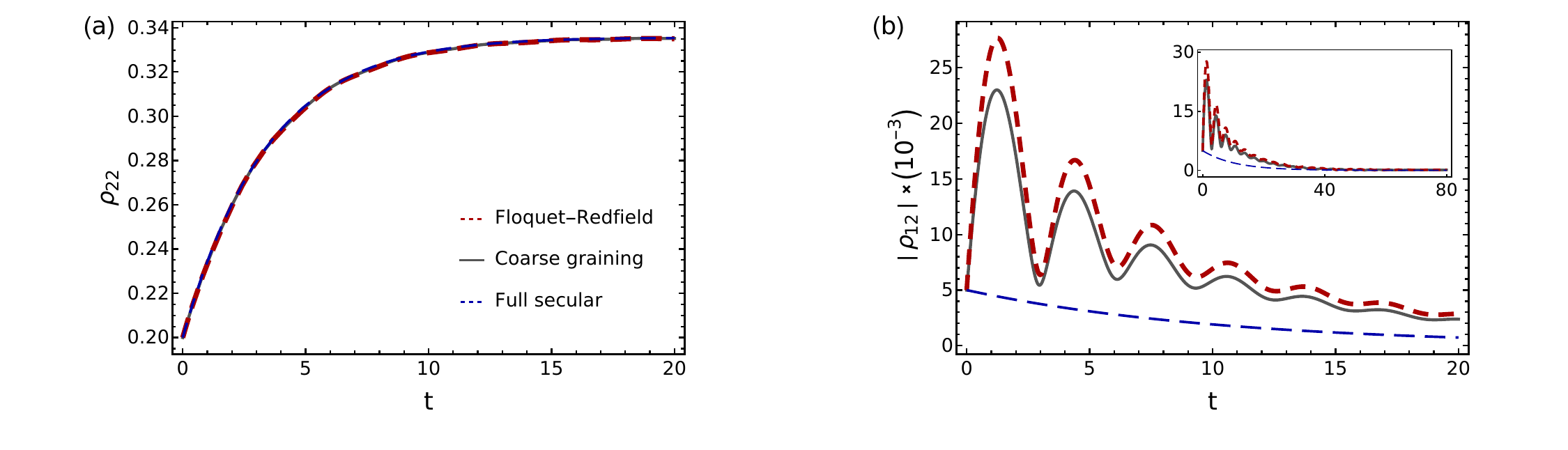}
    \caption{Dynamics of the three-level maser model, obtained with the Floquet-Redfield, coarse-grained, and full secular master equations. (a) Population of the intermediate level, $\rho_{22}$. (b) Coherence $\rho_{12}$ (in units of $10^{-3}$). The inset shows the long-time evolution of $\rho_{12}$. We take $\Delta = 0.1 E_2$, and use the same bath parameters as in the last section with temperatures $k_b T_h= 200 E_2$ and $k_b T_c= 10 E_2$ and $g=1$.}
    \label{fig:dyn3levels}
\end{figure}

In this case, the baths act on the system in such a way that the dynamics reach a time-independent steady state. Figure~\ref{fig:3levels} presents some of the steady state density matrix elements of the three-level maser when it operates as a heat engine ($\dot{W}<0$, $\dot{Q}_{\rm h}>0$, $\dot{Q}_{\rm c} < 0$). Panels (a) and (b) compare steady-state density matrix elements obtained from the coarse-graining approach, the full secular master equation, the weak-driving master equation, and an exact simulation. As can be observed, the coarse grain equation agrees well with the exact simulation, while the weak-driving master equation performs accurately within its domain of validity ($g/\Delta \ll 1$), but departing from the other curves when the driving strength becomes larger. The full secular master equation reproduces the populations at large coupling, but predicts vanishing coherences in all the range of parameters. We also notice that when the drive is turned off, $g = 0$, all approaches converge to the thermal equilibrium distribution imposed by the reservoirs. The Floquet-Redfield equation yields steady states visually indistinguishable from those of the coarse-grained approach (although not exactly equal), so we omit them to avoid overcrowding the figure.

\begin{figure}
    \centering
    \includegraphics[scale=0.49]{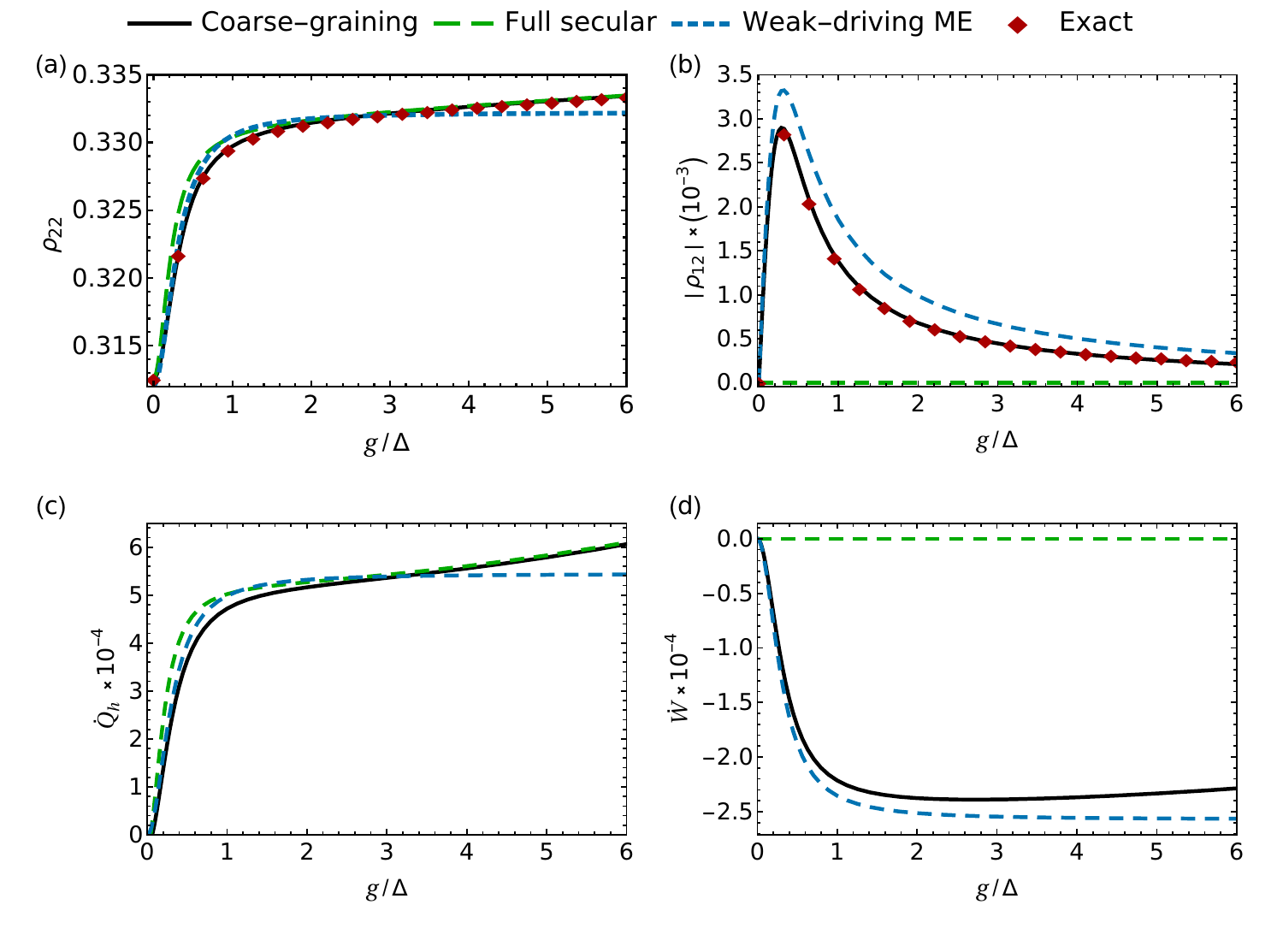}
    \caption{Steady state density matrix elements and thermodynamic fluxes of the three-level maser operating as a heat engine. The different curves are obtained using the coarse-graining approach, the full secular master equation, the weak driving master equation, and an exact simulation (see legend). {\bf a}). Population of state $\rho_{22}$ in the interaction picture. {\bf b}). Absolute values of the coherence between the excited levels, $|\rho_{12}|$. {\bf c}). Instantaneous heat currents from the hot reservoir. {\bf d}) Instantaneous power (negative when delivered by the heat engine). 
    In panels (a) and (b), the results of the exact simulation are shown as red diamonds. We take the same parameters as in Fig.~\ref{fig:dyn3levels} and vary $g$.}
    \label{fig:3levels}
\end{figure}

It is instructive to look at the expression of the heat currents for the Floquet-Redfield equation. By taking $\Delta t= 0$, and substituting the jump operators Eqs.~(\ref{eq:As3ls}), the heat currents, given by Eq.~(\ref{eq:Qcountingfields}) can be simplified to 
\begin{equation}
\begin{aligned}
 \dot{Q}_{\rm c} &= - \epsilon_{+}\Bigg\{(1-\cos\theta) \Big(\Gamma^{\rm (c)}(\epsilon_{+})\rho_{11} -\Gamma^{\rm (c)}(-\epsilon_{+})\rho_{33}\Big) + \sin\theta \,\Gamma^{\rm (c)}(\epsilon_{+})\mathrm{Re}\rho_{12}
\Bigg\} \\
& -\epsilon_{-}\Bigg\{(1+\cos\theta)\Big(\Gamma^{\rm (c)}(\epsilon_{-})\rho_{22}-\Gamma^{\rm (c)}(-\epsilon_{-})\rho_{33}\Big)  + \sin\theta\,\Gamma^{\rm (c)}(\epsilon_{-}) \mathrm{Re}\rho_{12}\Bigg\}.
\end{aligned}\label{eq:3lsQc}
\end{equation} for the cold bath, and 
\begin{equation}
\begin{aligned}
 \dot{Q}_{\rm h} &= -(\epsilon_{+}+\Omega)\Bigg\{ (1+\cos\theta)\Big(\Gamma^{\rm (h)}(\epsilon_{+}+\Omega)\rho_{11}-\Gamma^{\rm (h)}(-\epsilon_{+}-\Omega)\rho_{33}\Big) + \sin\theta\, \Gamma^{\rm (h)}(\epsilon_{+}+\Omega)\mathrm{Re}\rho_{12}\Bigg\} \\
&- (\epsilon_{-}+\Omega)\Bigg\{(1-\cos\theta)\Big(\Gamma^{\rm (h)}(\epsilon_{-}+\Omega)\rho_{22}-\Gamma^{\rm (h)}(-\epsilon_{-}-\Omega)\rho_{33}
\Big) + \sin\theta\, \Gamma^{\rm (h)}(\epsilon_{-}+\Omega) \mathrm{Re}\rho_{12}
\Bigg\}.
\end{aligned}\label{eq:3lsQh}
\end{equation}for the hot bath. Where $\Gamma^{\rm (c)}$ and $\Gamma^{\rm (h)}$ refer to the cold and hot bath correlation functions, respectively, and $\rho$ in the rotating frame.  

The above expressions contain contributions corresponding to two different quanta of energy being exchanged with the reservoirs: $\epsilon_\pm$ for the cold bath, and $\epsilon_\pm + \Omega$ for the hot bath. Each of these contributions has, in addition, two terms. The first one is proportional to the populations, which comes from the secular contribution, and is similar to that obtained from a semiclassical approach such as Fermi's golden rule. It has been suggested that these rates may be interpreted as transition rates for two fictitious thermodynamic cycles that operate in opposite directions, competing with each other~\cite{geva_quantum_1996,kosloff2014quantum}. In the first cycle, acting clockwise, the relevant transitions are those associated with $\epsilon_+$ and $\epsilon_+ + \Omega$, together with the driving frequency $\Omega$. A second cycle, running counter-clockwise, is obtained by replacing $\epsilon_+$ with $\epsilon_-$, which acts as a refrigerator. Their relative contributions depend on the driving strength, and their competition determines the net heat currents and power produced by the engine. 
On the other hand, the second terms in the two lines Eqs.\eqref{eq:3lsQc} and \eqref{eq:3lsQh} depend on the real part of the coherence $\rho_{12}$, which comes from non-secular corrections to the dissipator, and have no classical counterpart~\cite{toledo_tude_quantum_2024}. They correspond to interference terms between the currents of both cycles, and are stronger for higher mixing angles, i.e, strong drive or near resonance. These terms can lead to substantial differences in the heat currents, as appreciated for moderate values of $g/\Delta$ in Fig.~\ref{fig:3levels}(c).

Different limits corresponding to different master equations can be recovered by examining the effect of the mixing angle $\theta$ in the heat currents \eqref{eq:3lsQc} and \eqref{eq:3lsQh} of the Floquet-Lindblad equation. In the weak driving limit, when $g/\Delta\ll 1$, the mixing angle is such that $\sin \theta \approx 0$ and $\cos\theta \approx 1$. Therefore, the heat currents reduce to a rate equation in which only the bare transition energies are exchanged. In the opposite limit, $g/\Delta \gg 1$, the off-diagonal elements of the density matrix are suppressed [see Fig.~\ref{fig:3levels}(b)], and Eqs.~(\ref{eq:3lsQc}) and~(\ref{eq:3lsQh}) recover the result of the full secular master equation. Finally, for the coarse-grained master equation, the sinc function filtering the non-secular terms of the dissipator leads to a suppression of the interference terms in Eqs.~\eqref{eq:3lsQc} and~\eqref{eq:3lsQh}. Because each non-secular term in the dissipator sum is weighted by a different filter value, the resulting expressions cannot be cast in a compact form and are not given here.

The power, defined in Eq.(\ref{eq:power}) and shown in panel (d) is given by the imaginary part of the coherence $\rho_{12}$. Under the coarse-graining approach (the absolute value of) the power produced by the engine for moderate values of $g/\Delta$ is lower than the one predicted from the weak-driving master equation. It also cannot be described by the full secular approximation, which always predicts zero power in the long-time run (as discussed in Sec.~\ref{sec:thermodconsistency}). 

The results presented here are genuinely consistent with the first law of thermodynamics (again with errors of order $\lambda^4$)
\begin{equation}
   \dot{U} =  \dot{Q}_{\rm c} + \dot{Q}_{\rm h}  + \dot{W},
\end{equation} where $\dot{U}  = 0$ in the periodic steady state and $\dot{Q}_{\rm c}$, $\dot{Q}_{\rm h}$, and $\dot{W}$ are calculated independently. In contrast, previous works analyzing the three-level maser as a heat engine using the full secular master equation (for a review see Ref.~\cite{kosloff2014quantum}) defined work by directly enforcing the first law above. The steady state entropy production is in this case
\begin{equation}
    \dot{S}_{\rm tot} = - \frac{\dot{Q}_{\rm c}}{T_{\rm c}} -\frac{\dot{Q}_{\rm h}}{T_{\rm h}} \geq 0 
\end{equation} as established by the second law of thermodynamics, which is consistent in all the regimes. In particular, it guaranties an efficiency $\eta \equiv - \dot{W}/\dot{Q}_{\rm h}$ bounded by Carnot's efficiency, $\eta \leq \eta_c = 1 - T_{\rm c}/T_{\rm h}$.

\begin{figure}[t]
    \centering
    \includegraphics[scale=0.4]{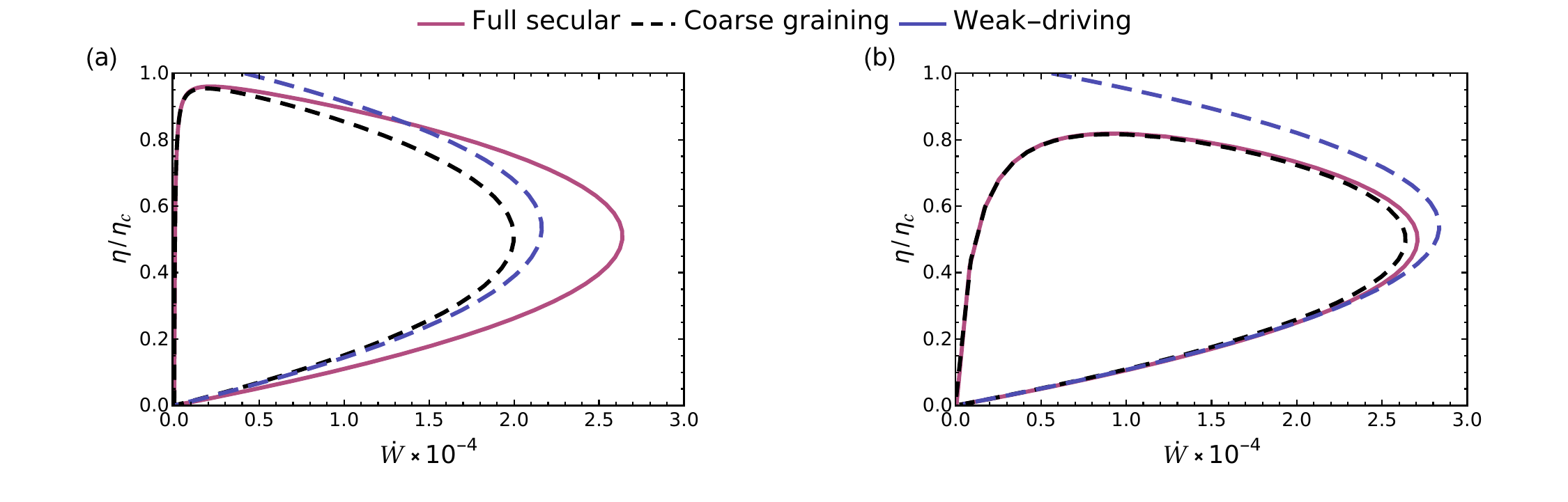}
    \caption{Efficiency as a function of the output power of the three-level heat engine for {\bf a.} $g = 0.02~E_1$ and $\Delta = 0.01~E_1$, and {\bf b.} $g = 0.1~E_1$ and $\Delta = 1\times 10^{-3}~E_1$. The curves are obtained by varying $E_2$, while keeping the detuning fixed. All other parameters are the same as in Fig.~\ref{fig:3levels}.}
    \label{fig:PvsEff}
\end{figure}

Figure~\ref{fig:PvsEff} shows how the predictions for the performance of the three-level  heat engine, in terms of power and efficiency, can differ when using the coarse-grain approach with respect to the predictions from the full secular master equation. As mentioned above, since the full secular master equation predicts vanishing mechanical work, the efficiency is instead obtained from the first law, $\eta = (Q_h + Q_c)/Q_h$, as is typically done in the previous literature. We obtain parametric power-efficiency curves by varying $E_2$ in the range $[E_1,E_3]$ and consider two situations: a relatively weak driving strength $g = 0.02~E_1$ and detuning $\Delta = 0.01~E_1$ (panel a) and stronger drive $g = 0.1~E_1$, but closer to resonance $\Delta = 1\times 10^{-3}~E_1$ (panel b). For the weaker driving, the weak-driving master equation predictions agree better with the master equation obtained from the coarse-grained approach, even though $g$ already lies outside its formal regime of validity, while the full secular master equation largely fails to predict the correct power-efficiency curve. As the driving strength increases, the weak-driving master equation predictions become more inaccurate, while the full secular approximation approaches the coarse-grained results more closely. However, it still fails to capture the performance close to resonance, especially near the maximum output power: it predicts nearly the same efficiency at maximum power, yet systematically overestimates the power itself.

\section{Conclusions}

In this work, we have addressed in detail the derivation of master equations for periodically driven systems in contact with a thermal environment. We showed that the indiscriminate application of the full secular approximation on the Floquet-Redfield equation, typically employed in the literature, can yield to incorrect predictions that challenge thermodynamic consistency. On the other hand, the Floquet-Redfield equation without further approximations is not guaranteed to be completely positive, and therefore can lead to nonphysical predictions. To address this gap, we proposed a robust coarse-graining approach to regularize the Floquet-Redfield equation, guaranteeing complete positivity, while retaining the essential oscillatory features of the dynamics. Such a partial secularization based on coarse-graining offers the advantage of interpolating between the Floquet-Redfield and the full secular master equation with a tunable parameter that determines the extent to which non-secular terms are filtered, and carries a clear physical interpretation: the time resolution.

Comparing the coarse-graining time interval with the other relevant time scales of the system and bath [see Eq.~(\ref{eq:CGcondition})] reveals the potential issues of both the Floquet-Redfield equation and a full secular approximation,  and defines the regime in which a coarse-graining approach can be safely adopted. We argue that choosing the coarse-graining interval $\Delta t$ to be the minimum value required to ensure that the Kossakowski matrix is positive semi-definite yields a master equation with the highest possible resolution while guaranteeing a completely positive dynamics. Using counting-field statistics we show that this approach guarantees also thermodynamic consistency, both at the level of the first and second laws of thermodynamics. A remarkable feature of this approach is that, contrary to the full secular master equation, it predicts non-zero average power in the periodic stationary state (which is no longer a unitary orbit), without the need of postulating further sources of non-conservative work, while guaranteeing non-negative entropy production.

We tested our approach in two minimal but relevant examples of driven-dissipative systems which capture the essential physical mechanisms while also serving as building blocks for the study of more complex driven-dissipative systems. In the first example, we investigated the role of the non-secular terms in the evolution of a driven two-level system connected to a heat bath. We show how each term of the Kossakowski matrix precisely affects the evolution and conclude that the minimal coarse-grain interval to ensure complete positivity must be approximately half of the driving period. We benchmark the dynamics and the time-averaged steady state predicted by the Floquet-Redfield equation, the full secular master equation, and the coarse-grained master equation against non-Markovian exact simulations. The results demonstrate that the coarse-grained master equation accurately describes the system within its regime of validity. In contrast, the full secular approximation fails to capture steady-state coherences across all parameter regimes and gives inaccurate populations in the near-resonant regime. 

Computing the time-averaged steady-state heat currents using counting-field statistics, we show that both the Floquet-Redfield and coarse-grained master equations predict stationary power dissipated into the environment, satisfying the first and second laws of thermodynamics. Moreover, we find that even in regimes where the full secular approximation provides an accurate description of the heat currents, it fails to correctly capture the work performed by the drive. Consequently, it does not provide a thermodynamically consistent description and should therefore not be used for thermodynamic analyses.

Next, we investigated the master equation predictions, the heat currents and the power of a three-level maser in a configuration acting as a heat engine. In this case, the system is led to a time-independent steady state in the the rotated frame of the Schr\"odinger picture. The coarse-grained master equation yields transient oscillations with reduced amplitude compared with the Floquet-Redfield equation (see Fig.~\ref{fig:dyn3levels}), and its steady state agrees with the exact simulation. Remarkably, the use of our approach unveils substantial differences in the power-efficiency curves of the three-level maser as a heat engine for moderate and strong driving fields with respect to predictions based on the full secular master equation with (by-hand) enforcement of the first law. These results call for a more comprehensive analysis of the performance of the three-level maser heat engine in the strong driving regime, including also power and heat fluctuations~\cite{kalaee_violating_2021,bayona-pena_thermodynamics_2021}, in the search of genuine quantum enhancements~\cite{AlmanzaMarrero2025certifyingquantum}. 

The analysis of both examples confirms that the loss of positivity originates from rapidly oscillating terms and therefore mainly affects oscillatory dynamics. When the steady state is periodic, the coarse-grained equation accurately describes the cycle-averaged steady state and energy currents. When the system instead reaches a time-independent steady state, the coarse-graining provides an averaged effect on the transient dynamics, leading to correct predictions of steady-state density matrix and currents.

In summary, our results show that the full secular approximation is generally not applicable for describing the thermodynamics of driven-dissipative systems. The non-secular terms of the Kossakowski matrix couple the dynamics of the diagonal and off-diagonal elements of the density matrix, thereby sustaining steady-state coherences that are essential for a consistent thermodynamic description. In contrast, the coarse-grained master equation provides an accurate framework for describing the thermodynamics of these systems. While the Floquet-Redfield equation can yield a accurate description of the coherence dynamics in the long time run, the coarse-grained approach guarantees complete positivity and therefore prevents possible nonphysical behavior for any initial state. The choice between these approaches must therefore be made on a case-by-case basis, balancing the advantages and drawbacks of each approach, or by deciding to resort to non-Markovian methods. 

The results presented here can serve as a practical and physically transparent guide for applying the secular approximation to driven-dissipative systems, as well as assessing the accuracy of Markovian master equations. Our framework is general and can be extended to more complex systems and structured drivings. While the analytical solutions may become more involved in such cases, the resulting expressions can still be evaluated numerically with similar cost as the Floquet-Redfield equation. These results have potential applications to various driven-dissipative platforms such as lasers~\cite{ferioli2023non, kirton_introduction_2019}, superconducting qubits~\cite{magazzu2018probing, mezzacapo_digital_2014,lamata_digital-analog_2017,benhayoune2025kerr}, or dissipative time crystals~\cite{Iemini18, kongkhambut2022observation, carraro2024solid, jiao2025observation}.


\ack{We thank Juan M.R. Parrondo for enlightening discussions.}

\funding{We acknowledge financial support from the CoQuSy
project (Grants No. PID2022-140506NB-C21 and C22) funded by
MCIU/AEI/10.13039/501100011033, from the QuantERA QNet project
CoQuaDis (Grant No. PCI2024-153446) funded by MCIU/AEI/10.13039/501100011033 and co-financed by European Union; and from the María de Maeztu Grant for excellent R$\&$D units, No. CEX2021-001164-M funded by MCIU/AEI/10.13039/501100011033 and European Union ERDF, EU. L.T.T. acknowledges Juan de la Cierva Postdoctoral Fellowship (JDC2025-057716-I) funded by MICIU/AEI/10.13039/501100011033 and the European Social Fund Plus (FSE+).}
\roles{G.M. conceptualized the research. L.T.T. performed the main derivations leading to the research results, together with the formal analysis and visualization, with feedback from C.O., R.Z. and G.M. C.O. contributed to numerical simulations and visualization. All authors participated in the writing and revision of the original manuscript. R.Z. and G.M. supervised the research and participated in funding acquisition and project administration tasks.}

\data{The code used to generate the data that support the findings of this study is openly available in Zenodo at https://doi.org/10.5281/zenodo.21836541.}


\section*{Appendix}

\appendix

\section{Diagonal steady states of GKLS generator}\label{appendix:rhodiag}
We show that, in the energy eigenbasis of a non-degenerate Hamiltonian, a GKLS generator with jump operators that are either diagonal or induce transitions between eigenstates leads to decay of all coherences. The GKLS dissipator is 
\begin{equation}
\mathcal{D}[\rho] = \sum_\alpha \gamma_\alpha \left( L_\alpha \rho L_\alpha^\dagger - \frac{1}{2} \{L_\alpha^\dagger L_\alpha, \rho\} \right),
\end{equation}
with rates $\gamma_\alpha \ge 0$.

Let's start by considering a jump operator that corresponds to transitions between energy eigenstates, $L = \ket{i}\bra{j}$, $i \neq j$
\begin{equation}
\begin{aligned}
        \mathcal{D}[\rho] &= \sum_{i,j} \gamma_{i,j} \Big(\ket{i}\bra{j} \rho \ket{j}\bra{i} - \frac{1}{2} \ket{j}\bra{i} \ket{i}\bra{j} \rho- \frac{1}{2} \rho \ket{j}\bra{i} \ket{i}\bra{j}\Big)\\
        &= \sum_{i,j} \gamma_{j,j} \rho_{ij} \ket{i}\bra{i} - \frac{1}{2} \sum_k \gamma_{i,j} \Big( \rho_{jk} \ket{j}\bra{k}  + \rho_kj \ket{k}\bra{j}\Big).
\end{aligned}
\end{equation}
Looking to each element in the eigenbasis separately, the evolution of populations corresponds to the semi-classical rate equation. And the off-diagonal elements decay exponentially.

On the other hand, for diagonal jump operators
\begin{equation}
\begin{aligned}
        \mathcal{D}[\rho] &= \sum_{i} \gamma_{i} \Big(\ket{i}\bra{i} \rho \ket{i}\bra{i} - \frac{1}{2} \ket{i}\bra{i} \ket{i}\bra{i} \rho- \frac{1}{2} \rho \ket{i}\bra{i} \ket{i}\bra{i}\Big)\\      
        &= \sum_{i} \gamma_{i} \rho_{ii} \ket{i}\bra{i} - \frac{1}{2} \sum_k \gamma_{i} \Big( \rho_{ik} \ket{i}\bra{k}  + \rho_{ki} \ket{k}\bra{i}\Big).
\end{aligned}
\end{equation}
Implying pure decoherence of the off-diagonal terms without affecting populations.

In both cases, the dissipator suppresses coherences in the energy eigenbasis. For a non-degenerate Hamiltonian, the unitary part does not couple populations and coherences, and therefore cannot regenerate off-diagonal elements. If, in addition, the dissipative dynamics has no decoherence-free subspaces, all coherences decay to zero, and the steady state is diagonal in the energy eigenbasis.

\section{Counting fields statistics}\label{appendix:A}

In this section, we derive heat currents using the formalism of full counting statistics, which can be applied irrespective of whether a secular approximation is performed. Full counting statistics has been extensively used to study heat exchange in open quantum systems~\cite{cuetara_stochastic_2015,esposito_nonequilibrium_2009,gasparinetti_heat-exchange_2014, murphy_laser_2022}. Our derivation of Eq.~(\ref{eq:Qcountingfields}) follows Ref.~\cite{gasparinetti_heat-exchange_2014}, however, instead of performing a full secular approximation, we retain the non-secular terms and apply the coarse-graining procedure.

We consider the situation described by the Hamiltonian Eq.~(\ref{eq:eq1}) in the main text, with system bath coupling $H_{SB} = \lambda(A \otimes B)$. The probability $\mathcal{P}(Q,t)$ of having an amount $Q$ of energy transferred from the system to the reservoir between a initial time $t_0 =0$ and $t$ is
\begin{equation}
   \mathcal{P}(Q,t) = \sum_{e, e'} \delta(e-e' -Q)p(e;e')p(e'), 
\end{equation}
where $p(e')$ and $p(e;e')$ are the probability of measuring $H_B$ at $t_0$ and obtaining $e'$, and the conditional probability of obtaining $e$ at time $t$ given $e'$. This product can be calculated by introducing the projectors $P_e$ into the measured energy and the unitary operator responsible for the evolution of the system and bath
\begin{equation}
   \mathcal{P}(Q,t) = \sum_{e, e'} \delta(e-e' -Q) \Tr[P_e U(t) P_{e'}\rho(o) P_{e'}U^\dagger(t)P_e].
\end{equation}

We are interested in the characteristic function of such a probability, given by
\begin{equation}
    G(u, t) = \int  dQ \mathcal{P}(Q,t) e^{i u Q} = \sum_{e, e'} \Tr[P_e U(t) P_{e'}\rho(0) P_{e'}U^\dagger(t)P_e] e^{i u (e-e')}.
\end{equation}
If the initial density matrix is in a factorized state $\rho_T(0) = \rho_s(0)\otimes\rho_B(0)$ and the environment in a thermal state, the projectors $P_e$ commute with $\rho_s$. We obtain
\begin{equation}
    G(u, t) =  \Tr[U^\dagger(t) e^{i u H_B} U(t) e^{-i u H_B}\rho_T(0)],
\end{equation}
where we used that $e^{\pm i u H_B} = \sum_e e^{\pm i u e}$. Defining an annotated density operator in terms of the characteristic function\begin{equation}
    G(u, t) = \mathrm{Tr}[\rho_{u,T}(t)],
\end{equation}
yields an evolution $\rho_u(t)=U_{u/2}\rho_u(0)U_{-u/2}^\dagger$, with annotated time-evolution operator $U_{u}=e^{iuH_b}Ue^{-iuH_b}$. The moments of the distribution follow from derivatives of the characteristic function; in particular, the average heat reads
\begin{equation}
    \langle Q  \rangle = -i \left.\frac{d G(u, t)}{d u}\right|_{u=0} = -i \mathrm{Tr}\left[ \frac{d \rho_{u,S}(t)}{d u}\right]_{u=0}.\label{eq:defQcf}
\end{equation}

The operator $U_{u}$ satisfies the equation of motion
\begin{equation}
i \frac{d}{dt} \mathcal{U}_u = H_u  \mathcal{U}_u, \qquad H_u = e^{i u H_B} H e^{-i u H_B},
\end{equation}
and the heat currents are obtained by evolving the annotated matrix $\rho_u(t)$, in a similar procedure as the one presented in Sec.~\ref{sec:ME}. Instead of the von Neumann equation, the total (system plus bath) annotated density matrix evolves according to 
\begin{equation}
\dot{\rho}_{u,T} = -i\left[H_S(t) + H_B, \rho_{u,T}\right] - i\left[H_{SB,u/2} \rho_{u,T} - \rho_{u,T}  H_{SB,-u/2}\right]
\end{equation} where $H_{SB\mu} = A \otimes e^{i u H_B}  B  e^{-i u H_B}$. We work in the Floquet-interaction picture introduced in the main text, in which the master equations simplifies to 
\begin{equation}
\dot{\Tilde{\rho}}_{u,T} = -i\left(\Tilde{H}_{SB,u/2}  \Tilde{\rho}_{u,T}- \Tilde{\rho}_{u,T} \Tilde{H}_{SB,-u/2}\right).
\end{equation}
Next, we integrate the above equation and substitute the result back into the evolution equation. Applying the Born--Markov approximation and using $\Tilde{H}_{SB,\pm u/2} =  \lambda(\Tilde{A}\otimes \Tilde{B}_{\pm u/2})$, the expression becomes\begin{equation}
\begin{aligned}
\dot{\Tilde{\rho}}_u = -\lambda^2 \int_0^t d\tau  \Tr_B \Big\{ &
\left(\Tilde{A}(t) \Tilde{A}(t-\tau) \Tilde{\rho}(t)\right)
\left(\Tilde{B}_{u/2}(t) \Tilde{B}_{u/2}(t-\tau) \rho_B\right) \\
&- \left(\Tilde{A}(t) \Tilde{\rho}(t) \Tilde{A}(t-\tau)\right)
\left(\Tilde{B}_{u/2}(t) \rho_B \Tilde{B}_{-u/2}(t-\tau)\right) \\
&- \left(\Tilde{A}(t-\tau) \Tilde{\rho}(t) \Tilde{A}(t)\right)
\left(\Tilde{B}_{u/2}(t-\tau) \rho_B \Tilde{B}_{-u/2}(t)\right) \\
&+ \left(\Tilde{\rho}(t) \Tilde{A}(t-\tau) \Tilde{A}(t)\right)
\left(\rho_B \Tilde{B}_{-u/2}(t-\tau) \Tilde{B}_{-u/2}(t)\right)
\Big\}
\end{aligned}
\end{equation}
We evaluate the four bath correlation functions in the bath energy eigenbasis:
\begin{equation}
    \begin{aligned}
\Tr_B\left[\Tilde{B}_{u/2}(t)\Tilde{B}_{u/2}(t-\tau)\rho_B\right]
&= \mathcal{B}(\tau) \\
\Tr_B\left[\Tilde{B}_{u/2}(t)\rho_B\Tilde{B}_{-u/2}(t-\tau)\right]
&= \sum_{jj^\prime} e^{iu(e_j-e_{j^\prime})}
\Tilde{B}_{jj^\prime}(t) p_{j^\prime} \Tilde{B}_{j^\prime j}(t-\tau) \\
\Tr_B\left[\Tilde{B}_{u/2}(t-\tau)\rho_B\Tilde{B}_{-u/2}(t)\right]
&= \sum_{jj^\prime} e^{iu(e_j-e_{j^\prime})}
\Tilde{B}_{jj^\prime}(t-\tau) p_{j^\prime} \Tilde{B}_{j^\prime j}(t) \\
\Tr_B\left[\rho_B\Tilde{B}_{-u/2}(t-\tau)\Tilde{B}_{-u/2}(t)\right]
&= \mathcal{B}(-\tau),
\end{aligned}
\end{equation} where $\mathcal{B}(\tau) =\Tr[\rho_B\Tilde{B}(t-\tau)\Tilde{B}(t)]$. After some algebraic manipulations, together with the decomposition in Eq.~(\ref{eq:jumpops}), we obtain
\begin{equation}
    \begin{aligned}
\dot{\Tilde{\rho}}_u =
-\lambda^2 \sum_{\alpha\alpha^\prime} e^{-i(\alpha-\alpha^\prime)t} \Big\{ &
\left[\int d\tau \, e^{-i\alpha^\prime\tau}\mathcal{B}(\tau)\right]
\left(A_\alpha A_{\alpha^\prime}^\dagger \Tilde{\rho}(t)\right) \\
&- \left[\int d\tau \, e^{-i\alpha^\prime\tau}
\sum_{jj^\prime} e^{iu(e_j-e_{j^\prime})}
\Tilde{B}_{jj^\prime}(t) p_{j^\prime} \Tilde{B}_{j^\prime j}(t-\tau)\right]
\left(A_\alpha \Tilde{\rho}(t) A_{\alpha^\prime}^\dagger\right) \\
&- \left[\int d\tau  e^{-i\alpha^\prime\tau}
\sum_{jj^\prime} e^{iu(e_j-e_{j^\prime})}
\Tilde{B}_{jj^\prime}(t-\tau) p_{j^\prime} \Tilde{B}_{j^\prime j}(t)\right]
\left(A_{\alpha^\prime}^\dagger \Tilde{\rho}(t) A_\alpha\right) \\
&+ \left[\int d\tau  e^{-i\alpha^\prime\tau}\mathcal{B}(-\tau)\right]
\left(\Tilde{\rho} A_{\alpha^\prime}^\dagger A_\alpha\right)
\Big\}.
\end{aligned}
\end{equation}
Exchanging the dummy indices $j$ and $j^\prime$ in the second line of the above expression and taking the derivative with respect to $u$, the first and last terms vanish. Furthermore, employing the approximation of Eq.~(\ref{eq:smoothbathapprox}), the heat current depends only on the real part of the rates $\Gamma$
\begin{equation}
\begin{aligned}
\frac{d\dot{\Tilde{\rho}}_u}{du}\bigg|_{u=0}= i \sum_{\alpha\alpha^\prime} e^{-i(\alpha-\alpha^\prime)t} \Big(
\alpha^\prime \mathbb{R}\text{e}[\Gamma(\alpha^\prime)]
A_\alpha \Tilde{\rho} A_{\alpha^\prime}^\dagger
-\alpha^\prime \mathbb{R}\text{e}[\Gamma(-\alpha^\prime)]
A_{\alpha^\prime}^\dagger \Tilde{\rho} A_\alpha
\Big).
\end{aligned}
\end{equation}
After coarse-graining
\begin{equation}
\begin{aligned}
\frac{d\dot{\Tilde{\rho}}_u}{du}\bigg|_{u=0}=i \sum_{\alpha\alpha^\prime} e^{-i(\alpha-\alpha^\prime)t}
 \mathrm{sinc}\!\left[(\alpha-\alpha^\prime)\frac{\Delta t}{2}\right]
\Big(
\alpha^\prime \mathbb{R}\text{e}[\Gamma(\alpha^\prime)]
A_\alpha \Tilde{\rho} A_{\alpha^\prime}^\dagger
-\alpha^\prime \mathbb{R}\text{e}[\Gamma(-\alpha^\prime)]
A_{\alpha^\prime}^\dagger \Tilde{\rho} A_\alpha
\Big).
\end{aligned}
\end{equation}
Applying Eq.~(\ref{eq:defQcf}) and exchanging the dummy indices $\alpha \leftrightarrow \alpha^\prime$ in the second term, we obtain Eq.~(\ref{eq:Qcountingfields})
\begin{equation}
\langle \dot{Q} \rangle
=
\sum_{\alpha\alpha^\prime}
\left(
\alpha \mathbb{R}\text{e}[\Gamma(\alpha)] + \alpha^\prime \mathbb{R}\text{e}[\Gamma(\alpha^\prime)]
\right)
e^{-i(\alpha-\alpha^\prime)t}
 \mathrm{sinc}\left[
(\alpha-\alpha^\prime)\frac{\Delta t}{2}
\right]\mathrm{Tr}\left[
A_\alpha \Tilde{\rho} A_{\alpha^\prime}^\dagger
\right]
\end{equation}
This expression can be further rewritten as
\begin{equation}
\langle \dot{Q} \rangle=
\sum_{\alpha} 2 \alpha \mathbb{R}\text{e}[\Gamma(\alpha)]\text{sinc}\left[
(\alpha-\alpha^\prime)\frac{\Delta t}{2}\right]\mathbb{R}\text{e}\left[e^{i(\alpha-\alpha^\prime)t} \Tr(A_{\alpha^\prime} \Tilde{\rho} A_{\alpha}^\dagger)\right]
\end{equation}
which shows that, despite the presence of the oscillatory term, the resulting expression is always real.

\section{System-bath rotating wave approximation}\label{sec:RWA}

A commonly used approach for eliminating the explicit time dependence of master equations in periodically driven systems is the rotating-wave approximation (RWA). The procedure can be illustrated by considering a system driven by an external field and coupled to a bosonic reservoir through the same transition via a transverse system operator (e.g. a $\sigma_x$-type coupling). In this case, the Hamiltonian in Eq.~(\ref{eq:eq1}) can be written as
\begin{equation}
    \begin{aligned}
        H_0 &= \sum_i \omega_i \ket{i}\bra{i}\\
        V &= g (e^{i \Omega t} \ket{j}\bra{k} + e^{-i \Omega t} \ket{k}\bra{j})\\
        H_{SB} &= \sum_q ( \ket{j}\bra{k} + \ket{k}\bra{j})(b_q^\dagger + b_q)\\
        H_B &= \sum_q b_q^\dagger b_q.
    \end{aligned}
\end{equation}
After applying the Floquet transformation P, the time dependence is shifted from the drive to the system-bath interaction
\begin{equation}
    \begin{aligned}
        H_0 &=  (\omega_j - \tfrac{\Omega}{2}) \ket{j}\bra{j} + (\omega_k + \tfrac{\Omega}{2}) \ket{k}\bra{k}+ \sum_{i \neq j,k} \omega_i \ket{i}\bra{i}, \\
        V &= g ( \ket{j}\bra{k} +  \ket{k}\bra{j})\\
        H_{SB} &= \sum_q (e^{-i \Omega t} \ket{j}\bra{k} + e^{i \Omega t}\ket{k}\bra{j})(b_q^\dagger + b_q)\\
        H_B &= \sum_q b_q^\dagger b_q.
    \end{aligned}
\end{equation}
The remaining time dependence can be removed by moving to an interaction picture with respect to the bath Hamiltonian 
\begin{equation}
    H_{\mathrm{rot}} = \Omega \sum_q b_q^\dagger b_q.\label{eq:intpicbath}
\end{equation}
In this frame, the bath operators transform as $b_q \rightarrow e^{-i \Omega t} b_q$.
As a result, the system-bath interaction contains both time-independent terms and terms oscillating at frequency $2\Omega$. Neglecting the rapidly oscillating contributions within a rotating-wave approximation, we obtain
\begin{equation}
    H_{SB} = \sum_q (b_q^\dagger \ket{k}\bra{j} + b_q \ket{j}\bra{k}).
\end{equation}
In this way, the fast oscillations are neglected prior to the derivation of the Born--Markov master equation. 

\begin{figure}[t]
    \centering
    \includegraphics[scale=0.35]{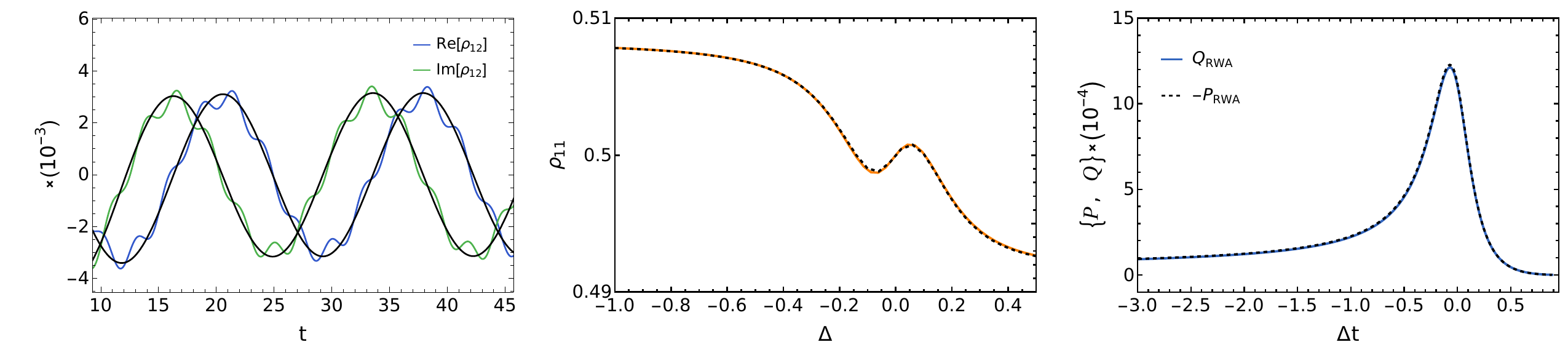}
    \caption{ Comparison between the rotating-wave approximation (RWA) and the exact dynamics. The left panel shows the time evolution of the off-diagonal elements of the density matrix in the interaction picture, with the RWA results shown in black and the exact results in color. The middle panel compares the period-averaged steady-state populations obtained from the two methods, in the Sch\"odinger picture. The right panel shows the period-averaged heat and work exchanged during a cycle as predicted by the RWA.}
    \label{fig:RWA}
\end{figure}

This approach is equivalent to applying a coarse-graining interval of $\Delta t = 2 \pi/ 2 \Omega$ directly to the master equation in the rotated–Schrödinger picture given by $P(t) \rho P^\dagger (t)$. The resulting averaging eliminates transitions arising from matrix elements of different Floquet zones, leading to a block-diagonal Kossakowski matrix. The master equation provides an averaged description of the dynamics and reproduces steady-state populations, coherences, and energy flow with good accuracy.  Nevertheless, like the Floquet-Redfield equation, it does not ensure complete positivity. Figure~\ref{fig:RWA} shows the results for the dynamics, steady-state, and energy currents in the case of the two-level system. The heat can be obtained from Eq.\eqref{eq:Qcountingfields} by taking $\Delta t = 0 $ and considering only the terms in the sum such that $q=q^\prime$.

\section{Other regularization procedures}\label{sec:otherreg}

For time-independent Hamiltonians, besides the coarse-graining approach~\cite{farina_open-quantum-system_2019}, other methods of regularization have been employed to obtain a completely positive equation from the Bloch-Redfield equation. 

Several works replace the arithmetic mean of the bath functions $\Gamma$ two Bohr frequencies by their geometric mean,
\begin{equation}
  \tfrac{1}{2}\big[\Gamma(\alpha)+\Gamma(\alpha')\big]
  = \sqrt{\Gamma(\alpha)\Gamma(\alpha')}
  + \tfrac{1}{2}\Big(\sqrt{\Gamma(\alpha)}-\sqrt{\Gamma(\alpha')}\Big)^{2},\label{eq:sqrtapprox}
\end{equation}
and discard the last term. We refer to this as the $\sqrt{\mathrm{SD}}$ approximation, where SD stands for the spectral density. It is used in Refs.~\cite{kirsanskas_phenomenological_2018,mccauley_accurate_2020,nathan_universal_2020,davidovic_completely_2020}, which differ in their justification and in their treatment of the Lamb shift. A thermodynamics of open systems described by this approximation has been analyzed in~\cite{soret_thermodynamic_2022,boubakour2026refined}. 

Another strategy is to directly regularize the time-dependent Kossakowski matrix $K(t)$~\cite{dabbruzzo_time-dependent_2023}. It consists in replacing it with the closest positive semidefinite matrix of the same dimension $K^{+}$, that is
\begin{equation}
  K^{+}(t)=\frac{K(t)+\sqrt{K^{\dagger}(t)K(t)}}{2}.\label{eq:colsestKpos}
\end{equation}

\begin{figure}[t]
    \centering
    \includegraphics[scale=0.4]{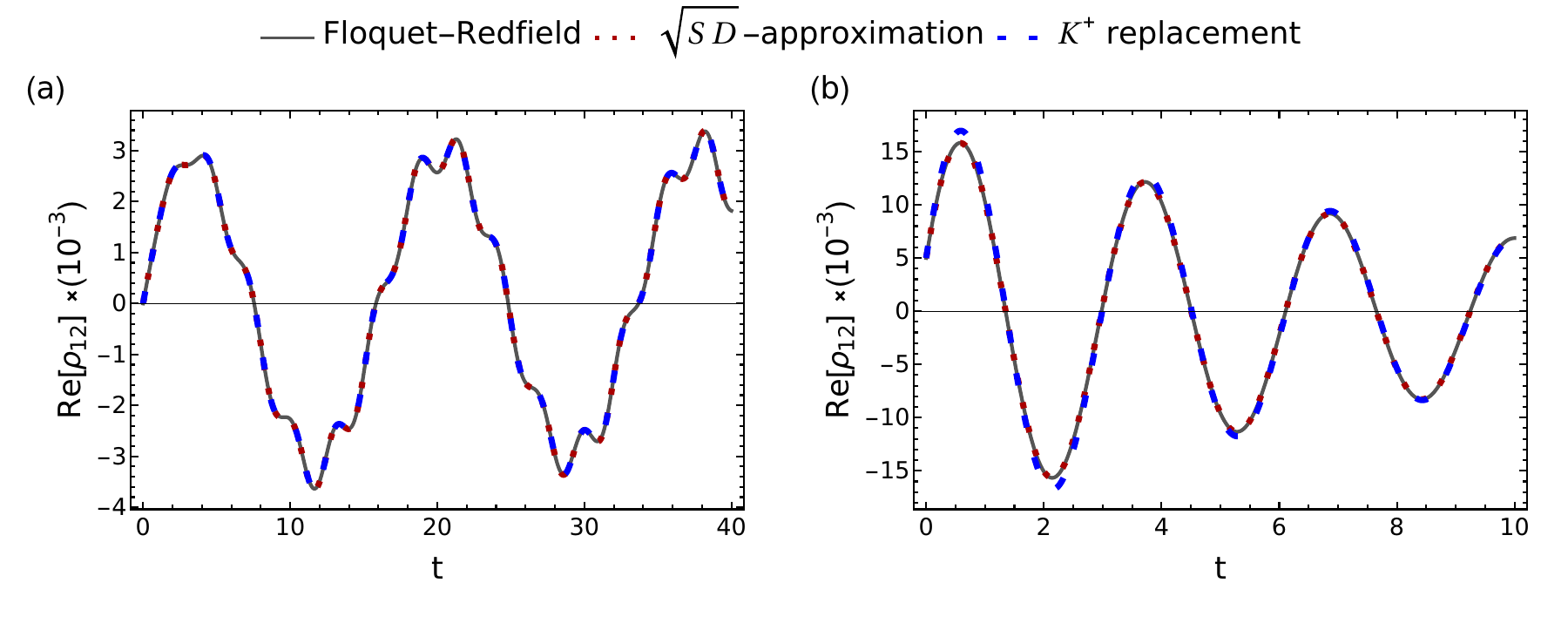}
    \caption{Comparison between the Floquet-Redfield equation, applying the $\sqrt{\mathrm{SD}}$ approximation Eq.\eqref{eq:sqrtapprox}, and replacing the Kossakowski matrix by $K^{+}$ Eq.\eqref{eq:colsestKpos}. The left and right panels show the dynamics of the real part of the coherence term of the density matrix for the driven two and three-level systems, respectively. The parameters used to simulate the dynamics as well as the pictures in which the desnsity matrix is depicted are the same as in Figs.~\ref{fig:Nonmarkov} and ~\ref{fig:dyn3levels}.}
    \label{fig:otherregularisations}
\end{figure}

Figure~\ref{fig:otherregularisations} compares the time evolution of the real part of the coherences obtained using the alternative regularization schemes discussed above with the Floquet–Redfield dynamics. Both regularizations reproduce the Redfield dynamics well, although replacing the Kossakowski matrix by $K^{+}$ leads to slight deviations at early times. It has been shown, however, that applying the $\sqrt{\mathrm{SD}}$ approximation to a system that couples to the bath through multiple operators, the Kossakowski matrix can exhibit small negative eigenvalues~\cite{fernandez_de_la_pradilla_recovering_2024}.

These regularization schemes may therefore provide viable alternatives for calculating the system dynamics. However, the heat current cannot be calculated directly from Eq.\eqref{eq:Qcountingfields}. The same approximations would have to be inserted into the derivation of the counting fields.

A third approach consists in realizing a partial secular approximation based on clustering the Bohr frequencies of the system~\cite{potts_thermodynamically_2021,trushechkin_unified_2021}. Transitions with nearly degenerate frequencies are grouped into clusters, terms coupling different clusters are dropped, and within each cluster the bath spectral density is evaluated at a single representative frequency.  This approach is closely related to coarse-graining, since it effectively sets a resolution at which the master equation distinguishes Bohr frequencies. However, this resolution is chosen based on the structure of the Born energies of the system instead of a coarse grain.

\section{Kossakowski matrix and dissipators of the driven two-level system}\label{appendix:B}

In this appendix, we provide the explicit expressions for the elements of the coarse-grained Kossakowski matrix associated with the driven-dissipative two-level system in the interaction picture. Throughout, we employ the shorthand notation $K^\text{c.g}_{ij} = K_{ij}$, so the dissipator Eq.\eqref{eq:BRdissipatorKdt} is
\begin{align}
     \mathcal{D}[\Tilde{\rho}] =  \sum_{ \alpha^\prime, \alpha \in \mathbb{F}} K_{\alpha^\prime \alpha}(t) \Big(A_{\alpha^\prime} \Tilde{\rho}(t) A^\dagger_\alpha - \frac{1}{2} \Big\{A^\dagger_\alpha A_{\alpha^\prime}, \Tilde{\rho}(t)\Big\}\Big).\label{eq:Dappendix}
\end{align}
We calculate each of the terms in the sum analytically and group them accordingly with their contribution to the dynamics 
\begin{equation*}
     \mathcal{D}[\Tilde{\rho}] =   \mathcal{D}[\Tilde{\rho}] \Big |_\text{sec}  +  \mathcal{D}[\Tilde{\rho}]\Big |_\text{coherence} +  \mathcal{D}[\Tilde{\rho}]\Big |_\text{coupling} +  \mathcal{D}[\Tilde{\rho}]\Big |_{2 \Omega}.\label{eq:termsdisspator}
\end{equation*}
For example, the secular contribution is given by the sum of all terms in Eq.~\eqref{eq:Dappendix}, such that $\alpha=\alpha^\prime$:
\begin{equation}
    \mathcal{D}[\Tilde{\rho}] \Big |_\text{sec} = \sum_{\alpha \in \mathbb{F}} K_{\alpha \alpha}(t) \Big(A_{\alpha} \Tilde{\rho}(t) A^\dagger_\alpha - \frac{1}{2} \Big\{A^\dagger_\alpha A_{\alpha}, \Tilde{\rho}(t)\Big\}\Big).
\end{equation}
The remaining contributions to the dissipator are calculated first and then grouped according to their structure, i.e., according to how they affect the different elements of $\Tilde{\rho}$.

As mentioned in the main text, we adopt the following ordering of frequency indices:
\[
\{ \Omega_R +\Omega, \Omega,-\Omega_R + \Omega ,\Omega_R -\Omega,-\Omega, -\Omega_R - \Omega \}.
\]
For clarity, in the following we express the dissipators in a simplified form by introducing the quantities $\phi_{ij}$, which compactly represent combinations of elements of the Kossakowski matrix $K_{ij}$. Their explicit definitions are given below. The diagonal terms of the Kossakowski matrix are the only terms retained under the full secular approximation (see Sec.~\ref{sec:FSA}), and are highlighted in gray in Fig.~\ref{fig:kmatrix}c. We refer to them as secular terms, and their contribution to the dissipator in Eq.~(\ref{eq:BRdissipatorK}) is given by
\begin{equation}
\mathcal{D}[\Tilde{\rho}]\Big |_\text{sec} =
\begin{pmatrix}
    \begin{aligned}
        &-\Phi_1^\text{sec}\ \Tilde{\rho}_{11}  + \Phi_2^\text{sec}\ \Tilde{\rho}_{22}
    \end{aligned}
    &
    \begin{aligned}
        &- \Phi_{12}^\text{sec}\ \Tilde{\rho}_{12}
    \end{aligned}
    \\
    \begin{aligned}
        &- \Phi_{12}^\text{sec}\ \Tilde{\rho}_{21}
    \end{aligned}
    &
    \begin{aligned}
        &\Phi_1^\text{sec}\ \Tilde{\rho}_{11}  - \Phi_2^\text{sec}\ \Tilde{\rho}_{22}
    \end{aligned}
\end{pmatrix}.
\end{equation}
As can be appreciated, they lead to independent evolution of the populations and coherences of the two-level system density matrix, with decoherence controlled by the real terms $\Phi_{12}^{\rm sec} \geq 0$. If the full secular approximation is performed, $\mathcal{D}[\Tilde{\rho}] = \mathcal{D}[\Tilde{\rho}] \Big |_\text{sec}$, the off-diagonal elements decay to zero in the long time limit, and the steady state populations become 
\begin{equation}
        \Tilde{\rho}_{11} = \frac{\Phi_2^{sec}}{\Phi_1^{sec} + \Phi_2^{sec}} = \frac{K_{11}+K_{44}}{K_{11}+K_{44} + K_{33}+K_{66}},\label{eq:ssfullsecular}
\end{equation}
and $\Tilde{\rho}_{22} = 1 - \Tilde{\rho}_{11}$. This is the standard result found in the literature~\cite{szczygielski_markovian_2013,gasparinetti_heat-exchange_2014,grifoni_driven_1998} which is well-justified in the fast driving regime $\Omega \gg \omega_0$; however, it exhibits the issues discussed in Sec.~\ref{sec:thermodconsistency}.

The off-diagonal terms of the Kossakowski matrix contribute to the evolution in three distinct ways. The terms oscillating at frequencies $\pm \{2 \Omega_R, 2 \Omega_R \pm 2 \Omega \}$ affect only of the coherence in the two level system and are highlighted in blue in Fig.~\ref{fig:kmatrix}. Their contribution to the dissipator is simply given by
\begin{equation}
\mathcal{D}[\Tilde{\rho}]\Big |_\text{coherence} =
\begin{pmatrix}
    \begin{aligned}
        &0
    \end{aligned}
    &
    \begin{aligned}
        & \Phi_{21}^\text{coh}\Tilde{\rho}_{21}
    \end{aligned}
    \\
    \begin{aligned}
        &\Phi_{12}^\text{coh}\Tilde{\rho}_{12}
    \end{aligned}
    &
    \begin{aligned}
        &0
    \end{aligned}
\end{pmatrix}.
\end{equation} These terms are generally complex in the interaction picture, and cause coherence oscillations. 

On the other hand, the terms highlighted in green in Fig.~\ref{fig:kmatrix} correspond to the frequencies $\pm\{\Omega_R,\Omega_R \pm 2 \Omega\}$. They couple the diagonal and off-diagonal elements of the density matrix, thereby driving the system away from a thermal evolution. Their contribution to the dissipator is essential to capture the dynamics of the coherences in the long-time run. Its contribution to the dissipator is given by
\begin{equation}
\mathcal{D}[\Tilde{\rho}]\Big |_\text{coupling} =
\begin{pmatrix}
     \Phi_{12}^{c}\ \Tilde{\rho}_{12} + \Phi_{21}^{c}\ \Tilde{\rho}_{21} & {\Phi'}_{1}^{c}\ \Tilde{\rho}_{11}+{\Phi'}_{2}^{c}\ \Tilde{\rho}_{22}\\
    {\Phi}_{1}^{c}\ \Tilde{\rho}_{11}+{\Phi}_{2}^{c}\ \Tilde{\rho}_{22} & -(\Phi_{12}^{c}\ \Tilde{\rho}_{12} + \Phi_{21}^{c}\ \Tilde{\rho}_{21}) 
\end{pmatrix}.
\end{equation}

Finally, the terms shown in red in Fig.~\ref{fig:kmatrix} oscillate at twice the driving frequency $\pm 2 \Omega$. Similarly to the secular terms, they lead to independent evolution of the coherences and populations:
\begin{equation}
\mathcal{D}[\Tilde{\rho}]\Big |_{2 \Omega} =
\begin{pmatrix}
    \begin{aligned}
        &- \Phi_{1}^{2\Omega}\ \Tilde{\rho}_{11} + \Phi_{2}^{2\Omega}\ \Tilde{\rho}_{22}
    \end{aligned}
    &
    \begin{aligned}
        &- \Phi_{12}^{2\Omega}\ \Tilde{\rho}_{12}
    \end{aligned}
    \\
    \begin{aligned}
        &- \Phi_{12}^{2\Omega}\ \Tilde{\rho}_{21}
    \end{aligned}
    &
    \begin{aligned}
        &\Phi_{1}^{2\Omega}\ \Tilde{\rho}_{11} -\Phi_{2}^{2\Omega}\ \Tilde{\rho}_{22}
    \end{aligned}
\end{pmatrix},\label{eq:D2om}
\end{equation}
with $\Phi_{12}^{2\Omega} \in \mathbb{I}\text{m}$. For strong driving, these contributions become important and may introduce modifications in the long-time limit values of the two-level system populations and coherence.

The full expressions of $\Phi$'s as a function of elements of the Kossakowski matrix are, starting with the secular contribution:
\begin{equation}
\begin{aligned}
\Phi_{1}^{\text{sec}} &= K_{11}+K_{44}\\
\Phi_{2}^{\text{sec}} &= K_{33}+K_{66}\\
\Phi_{12}^{\text{sec}} &= \frac{1}{2}(K_{11}+4K_{22}+K_{33} +K_{44}+4K_{55}+K_{66}).
\end{aligned}
\end{equation}
The contribution to the coherences is given by
\begin{equation}
\begin{aligned}
\Phi_{12}^{\text{coh}} &= K_{31}+ K_{34}+K_{61} +K_{64}\\
\Phi_{21}^{\text{coh}} &= K_{13}+ K_{16}+K_{43} +K_{46},
\end{aligned}
\end{equation}
and the coupling terms
\begin{equation}
\begin{aligned}
\Phi_{12}^{c}
&=\frac{1}{2}\left(K_{21}+K_{24}+K_{32}+K_{35}+K_{51}+K_{54}+K_{62}+K_{65}\right),\\
\Phi_{21}^{c}&=\frac{1}{2}\left(K_{12}+K_{15}+K_{23}+K_{26}+K_{42}+K_{45}+K_{53}+K_{56}
\right),
\\
{\Phi'}_{1}^{c}
&=\frac{1}{2}\left(3K_{12}+3K_{15}-K_{23}-K_{26}+3K_{42}+3K_{45}-K_{53}-K_{56}\right),
\\
{\Phi'}_{2}^{c}
&=
\frac{1}{2}\left(K_{12}+K_{15}-3K_{23}-3K_{26}+K_{42}+K_{45}-3K_{53}-3K_{56}\right),
\\
{\Phi}_{1}^{c}
&=\frac{1}{2}\left(3K_{21}+3K_{24}-K_{32}-K_{35}+3K_{51}+3K_{54}-K_{62}-K_{65}\right),
\\
{\Phi}_{2}^{c}&=\frac{1}{2}\left(K_{21}+K_{24}-3K_{32}-3K_{35}+K_{51}+K_{54}-3K_{62}-3K_{65}\right).
\end{aligned}
\end{equation}
And at lest the contribution of terms oscillating with $2 \Omega$ is
\begin{equation}
\begin{aligned}
\Phi_{1}^{2\Omega} &= K_{14}+K_{41}\\
\Phi_{2}^{2\Omega} &= K_{36}+K_{63} \\
\Phi_{12}^{2\Omega} &= \frac{1}{2}(K_{14}+4K_{25}+K_{36} + K_{41}+4K_{52}+K_{63}).
\end{aligned}
\end{equation}
 
The effect of the generalized Lamb shift Hamiltonian Eq.~(\ref{eq:Hls}) can be written by defining a matrix $X_{\alpha,\alpha^\prime} = e^{i(\alpha^\prime - \alpha)t} \chi(\alpha, \alpha^\prime)$, following the same indexing order as for the Kossakowski matrix we find
\begin{equation}
\begin{aligned}
   -i[H_{\mathrm{LS}}(t),\Tilde{\rho}(t)] = -&i \begin{pmatrix}
0
&
\phi^\text{ls}_\text{sec}\Tilde{\rho}_{12}
\\[1.5ex]
-\phi^\text{ls}_\text{sec}\Tilde{\rho}_{21}
&
0
\end{pmatrix}\\
-&i \begin{pmatrix}
\phi^\text{ls}_{1}\Tilde{\rho}_{12}+\phi^\text{ls}_{2}\Tilde{\rho}_{21}
&
-\phi^\text{ls}_{2}\Tilde{\rho}_{11}+\phi^\text{ls}_{3}\Tilde{\rho}_{12}+\phi^\text{ls}_{2}\Tilde{\rho}_{22}
\\[2ex]
-\phi^\text{ls}_{1}\Tilde{\rho}_{11}-\phi^\text{ls}_{3}\Tilde{\rho}_{21}+\phi^\text{ls}_{1}\Tilde{\rho}_{22}
&
-\phi^\text{ls}_{1}\Tilde{\rho}_{12}-\phi^\text{ls}_{2}\Tilde{\rho}_{21}
\end{pmatrix}.\end{aligned}\label{eq:LScontribution2ls}
\end{equation}
The first term above corresponds to the secular contribution to the Lamb shift, and is responsible for coherent oscillations. The second term stands for the non-secular contributions, which again couple population and coherence, and the $\phi$ factors correspond to

\begin{equation}
\begin{aligned}
\phi^\text{ls}_\text{sec} &= X_{11}-X_{33}+X_{44}-X_{66}\\
\phi^\text{ls}_{1} &=
X_{21}+X_{24}-X_{32}-X_{35}
+X_{51}+X_{54}-X_{62}-X_{65},
\\
\phi^\text{ls}_{2} &=
-X_{12}-X_{15}+X_{23}+X_{26}
-X_{42}-X_{45}+X_{53}+X_{56},
\\
\phi^\text{ls}_{3} &=
X_{14}+X_{41}-X_{46}-X_{63}.
\end{aligned}
\end{equation}

The elements of the Kossakowski matrix for the two-level system are given below. The diagonal elements are:
\begin{equation}
    \begin{aligned}
       K_{11}&= \left(\frac{\Delta + \Omega_R}{2 \Omega_R} \right)^2\gamma(\Omega_R +\Omega,\Omega_R +\Omega) &&
       K_{22} =  \left(\frac{g}{\Omega_R} \right)^2\gamma(\Omega,\Omega)\\
        K_{33}&= \left(\frac{\Delta - \Omega_R}{2 \Omega_R} \right)^2\gamma(-\Omega_R +\Omega,-\Omega_R +\Omega= &&
        K_{44} = \left(\frac{\Delta - \Omega_R}{2 \Omega_R} \right)^2\gamma(\Omega_R -\Omega,\Omega_R -\Omega)\\
        K_{55} &=  \left(\frac{g}{\Omega_R} \right)^2 \gamma( -\Omega,-\Omega) &&
        K_{66} =  \left(\frac{\Delta + \Omega_R}{2 \Omega_R} \right)^2 \gamma( -\Omega_R-\Omega,-\Omega_R-\Omega). 
    \end{aligned}
\end{equation}
The off diagonal terms responsible for decoherence, shown in blue in the diagram of Fig.~\ref{fig:kmatrix}, are
\begin{equation}
    \begin{aligned}
        K_{13}&=  -\left(\frac{g}{\Omega_R} \right)^2 \gamma(\Omega_R +\Omega, -\Omega_R + \Omega) \text{sinc}[\Omega_R \Delta t] e^{- i 2 \Omega_R t}\\ 
        K_{16}&= \left(\frac{\Delta + \Omega_R}{2 \Omega_R} \right)^2 \gamma(\Omega_R +\Omega,-\Omega_R-\Omega) \text{sinc}[(\Omega_R + \Omega)\Delta t] e^{- i 2 (\Omega_R +  \Omega) t}\\
        K_{34}&= \left(\frac{\Delta - \Omega_R}{2 \Omega_R} \right)^2 \gamma(-\Omega_R +\Omega, \Omega_R -\Omega) \text{sinc}[(\Omega_R - \Omega) \Delta t] e^{-i  2 (\Omega_R - \Omega)  t}\\ 
        K_{46}&=  - \left(\frac{g}{\Omega_R} \right)^2 \gamma(\Omega_R-\Omega,-\Omega_R-\Omega) \text{sinc}[\Omega_R \Delta t] e^{i 2 \Omega_R  t}\\
    \end{aligned}
\end{equation}
where we note that $\dfrac{\Delta^2 -\Omega_R^2}{4\Omega_R^2} = \left(\dfrac{g}{\Omega_R} \right)^2$. The coupling terms, represented in green, are
\begin{equation}
    \begin{aligned}
         K_{12}&= \frac{g(\Delta + \Omega_R)}{4 \Omega_R^2} \gamma(\Omega_R +\Omega,\Omega) \text{sinc}[\Omega_R \Delta t/2] e^{- i  \Omega_R t}\\
        K_{15}&= \frac{g(\Delta + \Omega_R)}{4 \Omega_R^2} \gamma(\Omega_R +\Omega,-\Omega) \text{sinc}[(\Omega_R + 2 \Omega) \Delta t/2] e^{- i (\Omega_R + 2 \Omega) t}\\
        K_{23}&= \frac{g(\Delta - \Omega_R)}{4 \Omega_R^2} \gamma(\Omega,-\Omega_R +\Omega) \text{sinc}[\Omega_R  \Delta t/2] e^{-i \Omega_R t}\\
        K_{24}&= \frac{g(\Delta - \Omega_R)}{4 \Omega_R^2} \gamma(\Omega,\Omega_R-\Omega) \text{sinc}[(\Omega_R - 2 \Omega) \Delta t/2] e^{i (\Omega_R - 2 \Omega)  t}\\
        K_{26}&= \frac{g(\Delta + \Omega_R)}{4 \Omega_R^2} \gamma(\Omega,-\Omega-\Omega_R) \text{sinc}[(\Omega_R + 2 \Omega) \Delta t/2] e^{-i [(\Omega_R + 2 \Omega)  t}\\
        K_{35}&= \frac{g(\Delta - \Omega_R)}{4 \Omega_R^2} \gamma(-\Omega_R +\Omega,-\Omega) \text{sinc}[(\Omega_R - 2 \Omega) \Delta t/2] e^{i (\Omega_R - 2 \Omega) t}\\ 
       K_{45}&= \frac{g(\Delta - \Omega_R)}{4 \Omega_R^2} \gamma(\Omega_R -\Omega, - \Omega) \text{sinc}[\Omega_R  \Delta t/2] e^{-i \Omega_R t}\\
        K_{56}&= \frac{g(\Delta + \Omega_R)}{4 \Omega_R^2} \gamma(-\Omega,-\Omega_R-\Omega) \text{sinc}[\Omega_R  \Delta t/2] e^{-i \Omega_R t}.
    \end{aligned}
\end{equation}
And the elements oscillating with frequency $2\Omega$, shown in red, are
\begin{equation}
    \begin{aligned}
        K_{14}&= - \left(\frac{g}{\Omega_R} \right)^2 \gamma(\Omega_R +\Omega,\Omega_R - \Omega) \text{sinc}[\Omega \Delta t] e^{-i  2 \Omega t}\\
        K_{25}&= \left(\frac{g}{\Omega_R} \right)^2 \gamma(\Omega, -\Omega) \text{sinc}[\Omega\Delta t] e^{-i  2 \Omega t}\\
       K_{36}&=- \left(\frac{g}{\Omega_R} \right)^2 \gamma(-\Omega_R +\Omega, -\Omega_R -\Omega) \text{sinc}[\Omega \Delta t] e^{-i  2 \Omega t}\\
    \end{aligned}
\end{equation}
with $K_{ji}$ given by similar expressions, but with entries of $\gamma$ inverted, that is, if $K_{ji} (\gamma(\alpha, \alpha^\prime))$, then $K_{ij} (\gamma(\alpha^\prime, \alpha))$. And we note that $\Tr[\mathcal{D}[\rho]] = 0$, which guarantees the preservation of the normalization of $\rho$ at all times. 

\section{Eigenvalue analysis of the Kossakowski Matrix}\label{appendix:positivity}

\begin{figure}[t]
    \centering
    \includegraphics[scale=0.35]{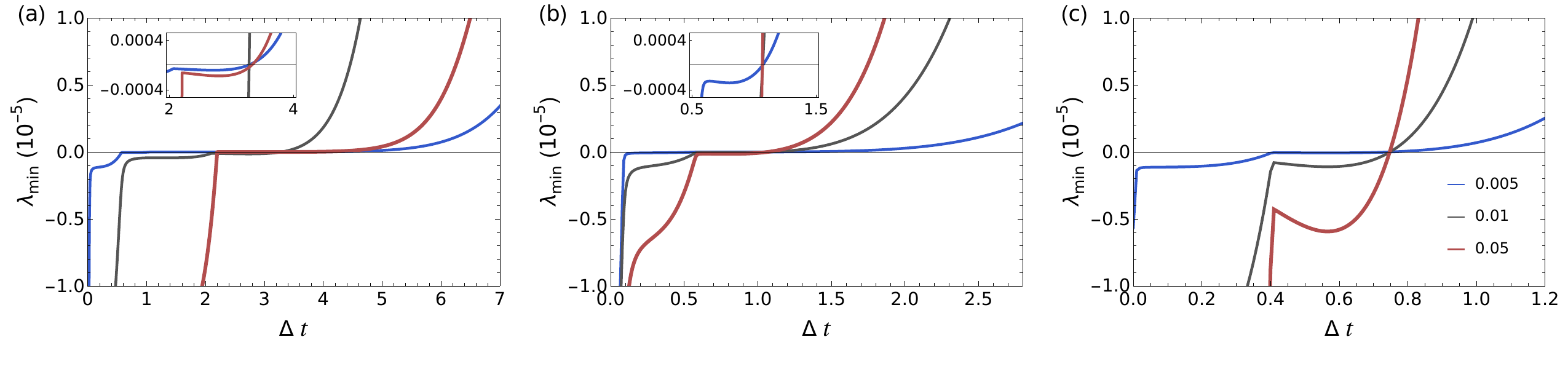}
    \caption{Minimum eigenvalue of the Kossakowski matrix as a function of the coarse-graining interval $\Delta t$, for the dissipator of (a) the driven two-level system, (b) the cold reservoir, and (c) the hot reservoir of the three-level system. Results are shown for drive strengths $g=0.01$ (blue), $0.2$ (gray), and $0.5$ (red). Insets: zoom around the zero crossing, showing that the minimum $\Delta t$ at which the minimum eigenvalue becomes non-negative is the same for all $g$, consistent with Fig.~\ref{fig:kmatrix}(b).}
    \label{fig:minEigenvalue}
\end{figure}

In this section we perform an eigenvalue analysis of the Kossakowski matrix of the two and three-level system. Fig.~\ref{fig:minEigenvalue} shows the minimum eigenvalue of $K$, $\lambda_{\min}$, as a function of the coarse-graining interval $\Delta t$ for three driving strengths $g$. Panel (a) corresponds to the dissipator of the driven two-level system (example I), and panels (b) and (c) to the dissipators associated with the cold and hot reservoirs of the three-level maser (example II), respectively. The dynamics is completely positive, only if $K$ is positive semidefinite, i.e., $\lambda_{\min} \geq 0$.

For small coarse-graining intervals, $\lambda_{\min}$ is negative, and for $\Delta t > \Delta t_{\min}$ it remains non-negative in all cases we examined, so $K$ stays positive semidefinite for all larger $\Delta t$. Below $\Delta t_{\min}$, however, $\lambda_{\min}$ is not a monotonic function of $\Delta t$. Since we have no analytical guarantee that $\lambda_{\min}$ stays non-negative beyond the first zero crossing, this must be checked numerically for each case. We stress, however, that this check is straightforward and computationally inexpensive. The insets show $\Delta t_{\min}$ for which $\lambda_{\min}>0$. The results are in agreement  with Fig~\ref{fig:kmatrix}(b), in which we show that $\Delta t_{\min}$ has weak dependence on $g$.

We also note that, in all three panels of Fig.~\ref{fig:minEigenvalue}, $\lambda_{\min}$ exhibits a plateau near zero, making $\Delta t_{\min}$ sensitive to the tolerance adopted to define positivity. Allowing $\lambda_{\min}$ to take slightly negative values would shift $\Delta t_{\min}$ to the beginning of the plateau, significantly reducing its value. This relaxed positivity criterion would yield better agreement with the Redfield equation.

Finally, we note that there may be situations in which the condition $\tau_B \ll \Delta t \ll \tau_S$ cannot be satisfied for any choice of coarse-graining interval. In such cases, it is likely that the system cannot be adequately described by a Markovian master equation, since the required separation of timescales, $\tau_B \ll \tau_S$, does not hold. One would then need to resort to non-Markovian methods, such as PT-TEMPO calculations, hierarchical equations of motion, or pseudomode approaches.

On the other hand, if $\tau_B \ll \tau_S$ holds, but no coarse-graining interval can be found that also satisfies $\Delta t \ll \tau_S$, the issue is not necessarily with the Markovian approximation itself. Rather, it may lie in the assumption that $\rho(t) \approx \rho$ over the coarse-graining interval, which is made when integrating Eq.\eqref{eq:intred}. In this case, other methods such a dynamical coarse-graining approach would be required, in which the integral in Eq.\eqref{eq:intred} is evaluated while explicitly accounting for the evolution of the system state. Another possibility would be to explore the other methods presented in Appendix~\ref{sec:otherreg}.

Our method, which has a simple numerical implementation (and provides analytical expressions), finds the minimum coarse-graining interval for which $\tau_B << \Delta t$. With that in hand, one should be able to easily evaluate whether this provides enough resolution to describe the system evolution. 

\section{Dissipator of the three-level system}\label{appendix:C}

We report below the dissipative contribution arising from the cold bath in the three-level heat engine Sec.~\ref{sec:3ls}. The corresponding expression for the hot bath is obtained analogously. The system Hamiltonian of the system-bath interaction in Eq.~(\ref{eq:H3ls}) is $A =\ket{3}\bra{2} + \ket{2}\bra{3}$. In the Floquet-interaction picture
\begin{equation}
    A(t) = U(\theta)^\dagger U_2^\dagger U_1^\dagger A U_1 U_2 U(\theta) =\begin{pmatrix}
    0 && 0 && \sqrt{\frac{\Omega_R - \Delta}{2 \Omega_R}}e^{i \epsilon_+ t} \\
    0 && 0 && \sqrt{\frac{\Omega_R + \Delta}{2 \Omega_R}}e^{i \epsilon_- t}\\
     \sqrt{\frac{\Omega_R - \Delta}{2 \Omega_R}}e^{-i \epsilon_+ t} && \sqrt{\frac{\Omega_R + \Delta}{2 \Omega_R}}e^{-i \epsilon_- t} && 0
\end{pmatrix}, 
\end{equation}

which can be decomposed into the operators Eq.~(\ref{eq:As3ls}). We adopt the following ordering of the frequency indices for the $K$ matrix
\[
\{ \epsilon_+, \epsilon_-, -\epsilon_+, - \epsilon_- \}.
\]
And group the contributions of the dissipator as explained in the main text and illustrated in Fig.~\ref{fig:diagrams3ls}
\begin{equation}
     \mathcal{D}[\Tilde{\rho}] =   \mathcal{D}[\Tilde{\rho}] \Big |_\text{sec}  +  \mathcal{D}[\Tilde{\rho}]\Big |_\text{coherence} +  \mathcal{D}[\Tilde{\rho}]\Big |_\text{coupling}. \label{eq:termsdisspator3ls}
\end{equation}

The secular contribution to the dissipator is 
\begin{equation}
\mathcal{D}[\Tilde{\rho}]\Big |_\text{sec} =
\begin{pmatrix}
     K_{33}\Tilde{\rho}_{33} - K_{11}\Tilde{\rho}_{11} & -(K_{11} +  K_{22})\frac{\Tilde{\rho}_{12}}{2} & -(K_{11} +  K_{33} +  K_{44})\frac{\Tilde{\rho}_{13}}{2} \\
    -(K_{11} +  K_{22})\frac{\Tilde{\rho}_{21}}{2}  & K_{44}\Tilde{\rho}_{33} - K_{22}\Tilde{\rho}_{22} & -(K_{22} +  K_{33}- K_{44})\frac{\Tilde{\rho}_{23}}{2}\\
    -(K_{11} +  K_{33} +  K_{44})\frac{\Tilde{\rho}_{31}}{2} & -(K_{22} +  K_{33} - K_{44})\frac{\Tilde{\rho}_{32}}{2} & \begin{aligned}
        &K_{11}\Tilde{\rho}_{11} + K_{22}\Tilde{\rho}_{22} \\
        &\quad - (K_{33} +K_{44})\Tilde{\rho}_{33}
    \end{aligned}
\end{pmatrix},
\end{equation}
with Kossakowski elements
\begin{equation}
    \begin{aligned}
       K_{11}&= \left(\frac{\Omega_R -\Delta }{2 \Omega_R} \right)\gamma(\epsilon_+,\epsilon_+)  &&
       K_{22} =  \left(\frac{\Omega_R +\Delta }{2 \Omega_R} \right)\gamma(\epsilon_-,\epsilon_-) \\
        K_{33}&= \left(\frac{\Omega_R -\Delta }{2 \Omega_R} \right)\gamma(-\epsilon_+,-\epsilon_+)&&
        K_{44} = \left(\frac{\Omega_R + \Delta }{2 \Omega_R} \right)\gamma(-\epsilon_-,-\epsilon_-) .
    \end{aligned}
\end{equation}

The contribution the coupling terms, shown in green in Fig.~\ref{fig:diagrams3ls}, to the dissipator is
\begin{equation}
\mathcal{D}[\Tilde{\rho}]\Big |_\text{coupling} =
\begin{pmatrix}
     -(K_{12} +  K_{21})\mathbb{R}\text{e}\Tilde{\rho}_{12} & K_{34} \Tilde{\rho}_{33} -K_{21}\frac{\Tilde{\rho}_{11} + \Tilde{\rho}_{22}}{2} & - K_{21}\frac{\Tilde{\rho}_{23}}{2}\\
    K_{43} \Tilde{\rho}_{33} -K_{12}\frac{\Tilde{\rho}_{11} + \Tilde{\rho}_{22}}{2} & - (K_{12} +  K_{21})\mathbb{R}\text{e}\Tilde{\rho}_{12} & -K_{12}\frac{\Tilde{\rho}_{13}}{2} \\
    -K_{12}\frac{\Tilde{\rho}_{31}}{2}& -K_{21}\frac{\Tilde{\rho}_{31}}{2} & 2 (K_{12} +  K_{21})\mathbb{R}\text{e}\Tilde{\rho}_{12}
\end{pmatrix},
\end{equation}
with Kossakowski elements
\begin{equation}
    \begin{aligned}
       K_{12}&= \left(\frac{\sqrt{ \Omega_R^2 -\Delta^2}}{2 \Omega_R} \right)\gamma(\epsilon_+,\epsilon_-) e^{-i(\epsilon_+-\epsilon_-)t}&&
       K_{21} =  \left(\frac{\sqrt{ \Omega_R^2 -\Delta^2}}{2 \Omega_R}  \right)\gamma(\epsilon_-,\epsilon_+) e^{i(\epsilon_+-\epsilon_-)t}\\
        K_{34}&= \left(\frac{ \sqrt{ \Omega_R^2 -\Delta^2}}{2 \Omega_R}  \right)\gamma(-\epsilon_+,-\epsilon_-) e^{i(\epsilon_+ +\epsilon_-)t}&&
        K_{43} = \left(\frac{\sqrt{ \Omega_R^2 -\Delta^2}}{2 \Omega_R}  \right)\gamma(-\epsilon_-,-\epsilon_+)e^{-i(\epsilon_+  +\epsilon_-)t}.
    \end{aligned}
\end{equation}
And the contribution affecting only coherence terms, shown in blue, in the dissipator is
\begin{equation}
\mathcal{D}[\Tilde{\rho}]\Big |_\text{coherence} =
\begin{pmatrix}
    0 & 0 & K_{31} \Tilde{\rho}_{31} + K_{32} \Tilde{\rho}_{32}\\
    0 & 0 & K_{41} \Tilde{\rho}_{31} + K_{42} \Tilde{\rho}_{32}\\
    K_{13} \Tilde{\rho}_{13} + K_{23} \Tilde{\rho}_{23} & K_{14} \Tilde{\rho}_{13} + K_{24} \Tilde{\rho}_{23} & 0
\end{pmatrix},
\end{equation}
with Kossakowski elements
\begin{equation}
    \begin{aligned}
       K_{31}&= \left(\frac{\Omega_R -\Delta }{2 \Omega_R} \right)\gamma(-\epsilon_+,\epsilon_+) e^{i \epsilon_+t}&&
       K_{13} =  \left(\frac{\Omega_R -\Delta }{2 \Omega_R}  \right)\gamma(\epsilon_+,-\epsilon_+) e^{-2i\epsilon_+t}\\
        K_{42}&= \left(\frac{\Omega_R +\Delta }{2 \Omega_R}  \right)\gamma(-\epsilon_-,\epsilon_-) e^{2 i \epsilon_- t}&&
        K_{24} = \left(\frac{\Omega_R +\Delta }{2 \Omega_R}  \right)\gamma(\epsilon_-,-\epsilon_-)e^{-2 i \epsilon_- t}\\
        K_{14}&= \left(\frac{\sqrt{ \Omega_R^2 -\Delta^2}}{2 \Omega_R} \right)\gamma(\epsilon_+,-\epsilon_-) e^{-i(\epsilon_+ + \epsilon_-)t}&&
       K_{41} =  \left(\frac{ \sqrt{ \Omega_R^2 -\Delta^2}}{2 \Omega_R}  \right)\gamma(-\epsilon_-,\epsilon_+)e^{i(\epsilon_+ + \epsilon_-)t}\\
        K_{23}&= \left(\frac{\sqrt{ \Omega_R^2 -\Delta^2}}{2 \Omega_R}  \right)\gamma(\epsilon_-, -\epsilon_+) e^{-i(\epsilon_+ + \epsilon_-)t}&&
        K_{32} = \left(\frac{\sqrt{ \Omega_R^2 -\Delta^2}}{2 \Omega_R}  \right)\gamma(-\epsilon_+,\epsilon_-)e^{i(\epsilon_+ + \epsilon_-) t}.
    \end{aligned}
\end{equation}
We note that $\Tr[\mathcal{D}[\Tilde{\rho}]] = 0$, which guarantees the preservation of the state normalization at all times. 


\bibliography{Biblio} 

\end{document}